\documentclass[a4paper,twoside]{article}
\def\HI{\hbox{\rm H\,{\sc i}}}
\newcommand{\affil}[1]{$^{\rm #1}$}
\usepackage[authoryear]{natbib}
\bibpunct{(}{)}{;}{a}{}{,}
\usepackage{graphicx}
\date{} 
\title{\large\bf\flushleft Comparison of potential ASKAP HI survey source finders }
\author{\parbox{\textwidth}{\flushleft
\vspace{-0.5cm}
{\it Attila Popping\affil{A,E}, Russell Jurek\affil{B}, Tobias Westmeier\affil{A}, Paolo Serra\affil{C}, Lars Fl$\ddot{o}$er \affil{D}, Martin Meyer\affil{A} and
  B$\ddot{a}$rbel Koribalski\affil{B}}\\
\vspace{0.4cm}
{\small \affil{A}\,International Centre for Radio Astronomy Research (ICRAR), The University of Western Australia, 35 Stirling Hwy, Crawley, WA 6009, Australia}\\
{\small \affil{B}\,Australia Telescope National Facility, CSIRO Astronomy and Space Science, PO Box 76, Epping NSW 1710, Australia}\\
{\small \affil{C}\,Netherlands Institute for Radio Astronomy (ASTRON), Postbus 2, 7990 AA Dwingeloo, The Netherlands}\\
{\small \affil{D}\,Argelander-Institut f\"ur Astronomie, Auf dem H\"ugel 71, 53121 Bonn, Germany}\\
{\small \affil{E}\,Email: attila.popping@icrar.org}}}
\begin{document}
\twocolumn[
\begin{changemargin}{.8cm}{.5cm}
\begin{minipage}{.9\textwidth}
\vspace{-1cm}
\maketitle
%
%
\small{\bf Abstract}\\
The large size of the ASKAP {\HI} surveys DINGO and WALLABY
necessitates automated 3D source finding. A performance difference of
a few percent corresponds to a significant number of galaxies being
detected or undetected. As such, the performance of the automated
source finding is of paramount importance to both of these surveys. We
have analysed the performance of various source finders to determine
which will allow us to meet our survey goals during the DINGO and
WALLABY design studies. Here we present a comparison of the
performance of five different methods of automated source
finding. These source finders are {\sc Duchamp}, the {\sc
  Gamma-finder}, {\sc CNHI}, a 2D-1D Wavelet Reconstruction and {\sc
  S+C finder}, a sigma clipping method. Each source finder was applied
on the same three-dimensional data cubes containing (a) point sources
with a Gaussian velocity profile and (b) spatially extended
model-galaxies with inclinations and rotation profiles. We
focus on the completeness and reliability of each algorithm when
comparing the performance of the different source finders.


\medskip
\medskip
\end{minipage}
\end{changemargin}
]
\small

\section{Introduction}

Radio astronomy is facing a new era, acquiring extremely large data
volumes with the coming of the Square Kilometre Array (SKA)
\citep{dewdney2009} and precursors such as MeerKAT \citep{jonas2009}
in South Africa, APERTIF \citep{verheijen2008} in the Netherlands and
the Australian SKA Pathfinder (ASKAP) \citep{deboer2009} in
Australia. Various continuum (2D) and spectral line (3D) surveys,
which cover large fractions of the sky, will be conducted with these
telescopes. The surveys are expected to detect millions of objects,
accelerating the need for reliable automated source finders.

A good source finder should have high {\em completeness} and high {\em
  reliability}, ie. a low rate of false detections. Choosing a suitable
trade-off between both parameters is necessary and depends on both
the algorithm and the rms uniformity of the data. Detecting objects is
relatively easy in the case of (strong) point sources, but becomes
more complicated in the case of irregular shapes and diffuse or
extended emission in one or more dimensions and at low signal to noise
ratios. The work presented in this paper aims to highlight the
strengths and weaknesses of potential 3D source finders for the Deep
Investigations of Neutral Gas Origins (DINGO) survey \citep{meyer2009}
and the Widefield ASKAP L-band Legacy All-sky Blind Survey (WALLABY)
\citep{koribalski2009}. These are two of the large {\HI} survey science
projects for ASKAP \citep{johnston2008}. To achieve the respective
science goals, we aim to develop source finding algorithms which
reliably and efficiently recover 3D {\HI} sources.

We have identified five different source finders that will be
subjected to testing and comparison; 1) the {\sc Duchamp} source
finder \citep{whiting2011a}, 2) the {\sc Gamma-finder}
\citep{boyce2003} 3) the CNHI source finder \citep{jurek2011}, 4)
the 2D-1D Wavelet Reconstruction source finder \citep{floer2011} and
5) the {\sc S+C finder} \citep{serra2011a}.

Testing of each algorithm was done on the same set of data cubes. The
first containing 961 point sources with varying peak flux and a
Gaussian velocity profile. The second cube contains 1024 modelled
galaxies with more realistic properties such as extended disks,
inclinations and rotation profiles. Here we compare their performance
in terms of completeness and reliability.

In section 2 we briefly summarise the main properties of the
source finding algorithms and in section 3 we describe the testing
method and the two model cubes that have been used for the
testing. The test results are presented in Section 4, followed by a
discussion in Section 5. We compare in detail the performance and
reliability of the source finders, to understand where the strong and
weak points of the different source finders are and to highlight
possible improvements. We finish with a short conclusion in the final
section.

\section{Source Finders}
Here we provide a short description of the five source finders
compared throughout the paper. For a more extended review of the
individual algorithm we refer to the reference papers describing each
method in detail.

\subsection{Duchamp source finder} {\sc Duchamp} \citep{whiting2011a}
is a source finder designed for 3D data, although it can be used for
2D and even 1D datasets. The source finder has been developed by
Matthew Whiting at CSIRO.\footnote{Duchamp website:
  http://www.atnf.csiro.au/people/Matthew.Whiting/Duchamp/}
{\sc Duchamp} identifies sources by simply applying a specified flux
or signal-to-noise threshold and searching for signals above that
threshold. In a second step, detections are merged or rejected
based on several criteria specified by the user. To improve its
performance, {\sc Duchamp} offers several methods of preconditioning
and filtering of the input data, including spatial and spectral
smoothing as well as wavelet reconstruction of the entire image or
cube. In a final step, {\sc Duchamp} measures several basic parameters
for each detected source, including position, radial velocity, size,
line width, and integrated flux.  The performance of the {\sc Duchamp}
source finder is tested in \citet{westmeier2011}.

\subsection{CNHI source finder}
The Characterised Noise {\HI} (CNHI) source finder \citep{jurek2011}
is being developed as part of the WALLABY design study.  The
CNHI source finder treats spectral datacubes as a collection of
spectra, using the Kuiper test, which is a variant of the
Kolmogorov-Smirnov test, to identify regions in each spectrum that do
not look like noise. The Kuiper test is used to calculate the
probability that the test region and the rest of the spectrum come
from the same distribution of voxel flux values. If the probability is
sufficiently low, then the test region is flagged as an object
section. The probability threshold is specified by the user. Once all
of the spectra have been processed, the object sections are combined
into objects. Object sections are combined using a variant of Lutz's
one pass algorithm.

There are two caveats to using the CNHI source finder. Firstly, the
CNHI source finder assumes that each spectrum is dominated by
noise. This is a safe assumption as spectral datacubes are generally
sparsely populated by sources. The presence of ripples, artifacts and
continuum signal will potentially invalidate this assumption
though. The second caveat is that the test region needs to be at least
four channels wide for the Kuiper test to be reliable. This matches
the smallest channel extent expected of WALLABY {\HI}
sources. Spectral datacubes with a poorer velocity resolution than
WALLABY will be affected by this. For a more detailed description of
the CNHI source finder see \citet{jurek2011}.

\subsection{Gamma-finder} {\sc Gamma-finder} is a {\sc Java}
application developed by \cite{boyce2003} which automatically searches
for objects in 3-dimensional data cubes. The searching algorithm of
{\sc Gamma-finder} is based on the {\sc Gamma-test}
\citep{Stefansson1997}, and a full description can be found in
\cite{jones2002}. The {\sc Gamma-test} is a near-neighbour data
analysis routine which estimates the noise variance in a continuous
dataset. This estimate is known as the {\it Gamma Statistic}, denoted
by $\Gamma$. When using the {\sc Gamma-finder} a Gamma signal-to-noise
ratio can be defined which is used as a clipping for objects to be
qualified as a detection. The output of the {\sc Gamma-finder} is
limited compared to other source finders (eg. {\sc Duchamp} and CNHI),
because it does not do any parametrisation, but only gives the three
dimensional position of a detection and the sigma level.

\subsection{2D-1D Wavelet Reconstruction source finder}
The 2D-1D Wavelet Reconstruction source finder is described in detail
in \citet{floer2011}, they have adapted a multi-dimensional wavelet
denoising scheme first used by \citet{starck2009}. It takes into
account that 3D data from spectroscopic surveys have two angular
dimensions and one spectral dimension, in which the shape of the
sources is vastly different than in the angular dimensions. The
algorithm therefore performs a two-dimensional wavelet transform
in all planes of the cubes and a subsequent one-dimensional wavelet
transform along each line of sight, i.e. each pixel.

Once the image has been de-noised by thresholding of the wavelet
coefficients, reconstructing the data from only the significant
coefficients yields a noise-free cube. The latter can be used to create
a mask for the sources in the original data.

\subsection{Smooth plus clip (S+C) finder}
\citet{serra2011a} developed a source finder which uses a limited
number of filters in order to optimise the signal-to-noise ratio of
objects present in a data cube. For each dataset, the finder looks for
sources in the original {\HI} cube and in the cubes obtained by
smoothing the original cube either on the sky, or in velocity, or
along all three axes. In this study we use a Gaussian filter of
FWHM=60 arcsec for smoothing on the sky, and a box filter of width 2,
4, 8, 16, and 32 channels for smoothing in velocity.  For each
smoothed cube a mask is built including all voxels brighter (in
absolute value) than a chosen threshold. The final mask is the
  union of all masks (i.e., a voxel is included in the total mask if
it is included in at least one of the individual masks), a value
  of 1 is allocated to all masked voxels and 0 to all unmasked
  voxels. A size filter is applied to the final binary mask by
convolving it with a 30 arcsec Gaussian kernel, equal to the original
angular resolution of the cube and to 3 channels in
velocity. Subsequently the mask is shrunk again by taking only
voxels in the convolved mask brighter than 0.5. This procedure removes
a large number of noise peaks included in the mask whose size is of
the order of the cube resolution.

\section{Testing method}
When comparing the five 3D source finders, we concentrate on two main
parameters, the completeness and the reliability of a source
finder. Completeness is defined as the number of detected sources
divided by the total number of sources. While this number is known for
simulated cubes, in reality we usually have a much harder problem: we
neither know the number of detectable sources in a cube, nor their
shape, size or velocity extent. There are a few examples of real
datacubes where there is a much deeper datacube of the same region of
sky, for example the HIPASS region that is covered by the HIDEEP
survey \citep{minchin2003}. The completeness can be given as a single
number, but can also be measured as a function of a certain parameter
such as integrated flux or velocity-width. Raw reliability is defined
as the number of true detections divided by the total number of
detections. In a good scenario the number of false detections is very
low, so the reliability is close to 100\%.

We have to stress that although completeness is a general parameter
for a simulation, reliability is highly dependent on the size of a
cube. When making a cube twice as large but keeping the number of
sources constant, the completeness will not change. However as the
noise voxels approximately double, so do the number of false
detections. In practise this is complicated by the non-linear steps
used by some source finders, and the number of false detections does
not necessarily scale linearly with the size of a data cube. The
reliability of different source finders can only be compared if the
finders are applied on exactly the same data sets. In many cases the
reliability of a source finder can be improved upon by applying a
threshold for one or more measured parameters like integrated flux.

We only concentrate on the capability of source finders to
determine detections. Not all source finders
have the capability of parametrizing detections, this however is a
different problem that can be addressed in the post-processing of
detections once they have been identified.

\subsection{Input Models}
For the testing and comparison of the different source finders we have
used 2 data cubes containing: 1) 961 artificial point sources with
Gaussian spectra and 2) 1024 artifical model galaxies with a range of
orientation parameters.

{\sc ASKAP}-specific noise has been added to the cubes, which was
generated by the {\sc Uvgen} task within {\sc Miriad} and is based on
the {\sc ASKAP} telescope configuration, a system temperature of
$T_{sys}$=50K and an integration time of 8 hours. The {\sc rms} in the
cubes is 1.95 mJy/beam ($30''$) per channel (3.9 km/s). The cubes are
similar to the cubes that have been used for the testing of the {\sc
  Duchamp} source finder by \cite{westmeier2011}. 

In the first cube with point sources each source was randomly assigned
a peak flux in the range of 1 to 20 $\sigma$, spectral line widths
(FWHM) range from approximately 0.4 to 40 km s$^{-1}$. While in
reality sources with line widths as small as 0.4 km s$^{-1}$ do not
occur, they are included to test the performance of source finders on
objects that are spectrally unresolved. In the second cube with model
galaxies all sources have an infinitely thin discs with varying
inclination (0$^{\circ}$ to 89$^{\circ}$), position angle (0$^{\circ}$
to 180$^{\circ}$), and rotation velocity (20 to 300 km s$^{−1}$).  For
a more detailed description of the cubes and the input parameters we
refer to the paper describing the {\sc Duchamp} testing
\citep{westmeier2011}.

\begin{figure*}[t]
\begin{center}
\includegraphics[width=0.48\textwidth, angle=0]{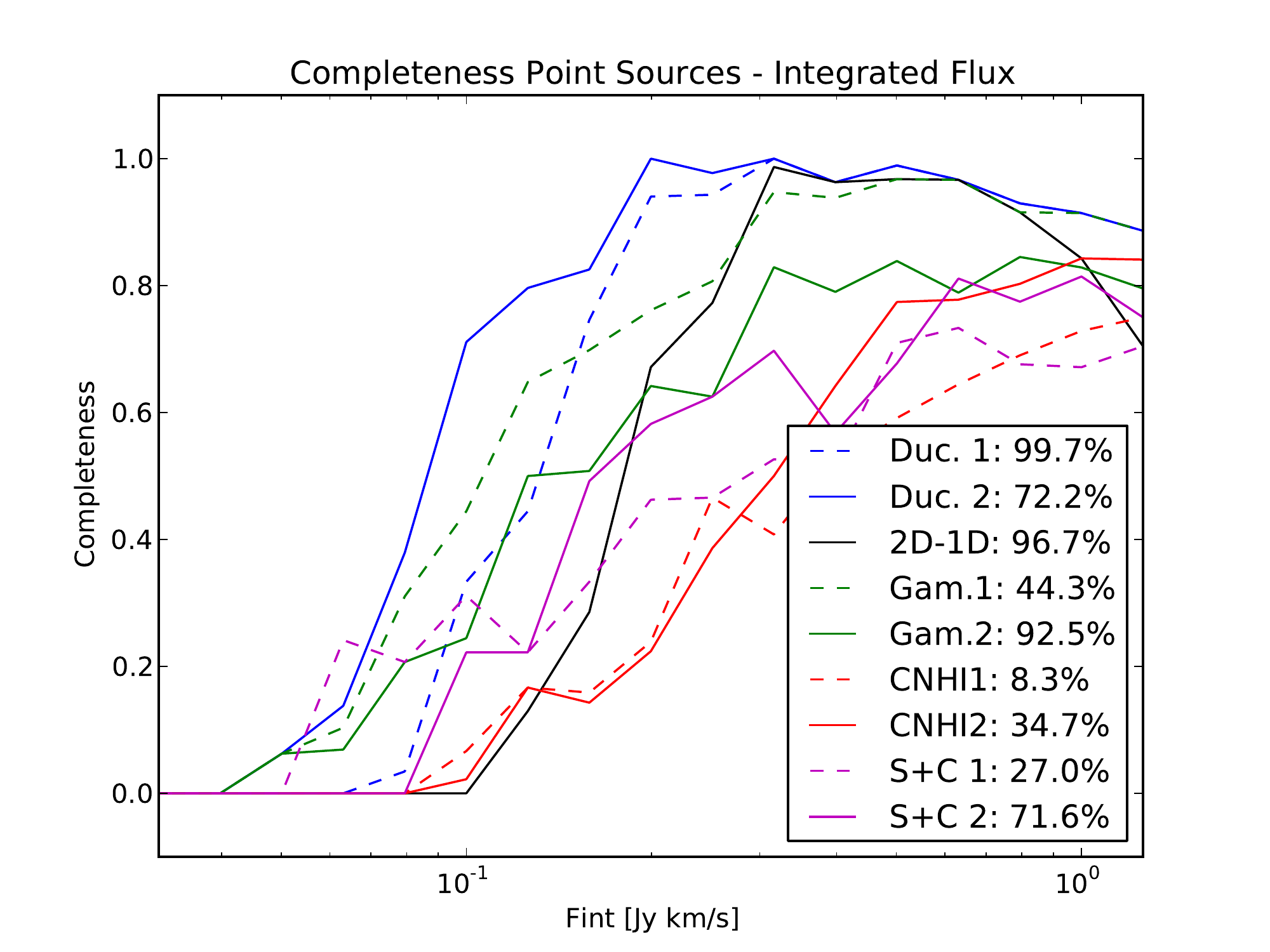}
\includegraphics[width=0.48\textwidth, angle=0]{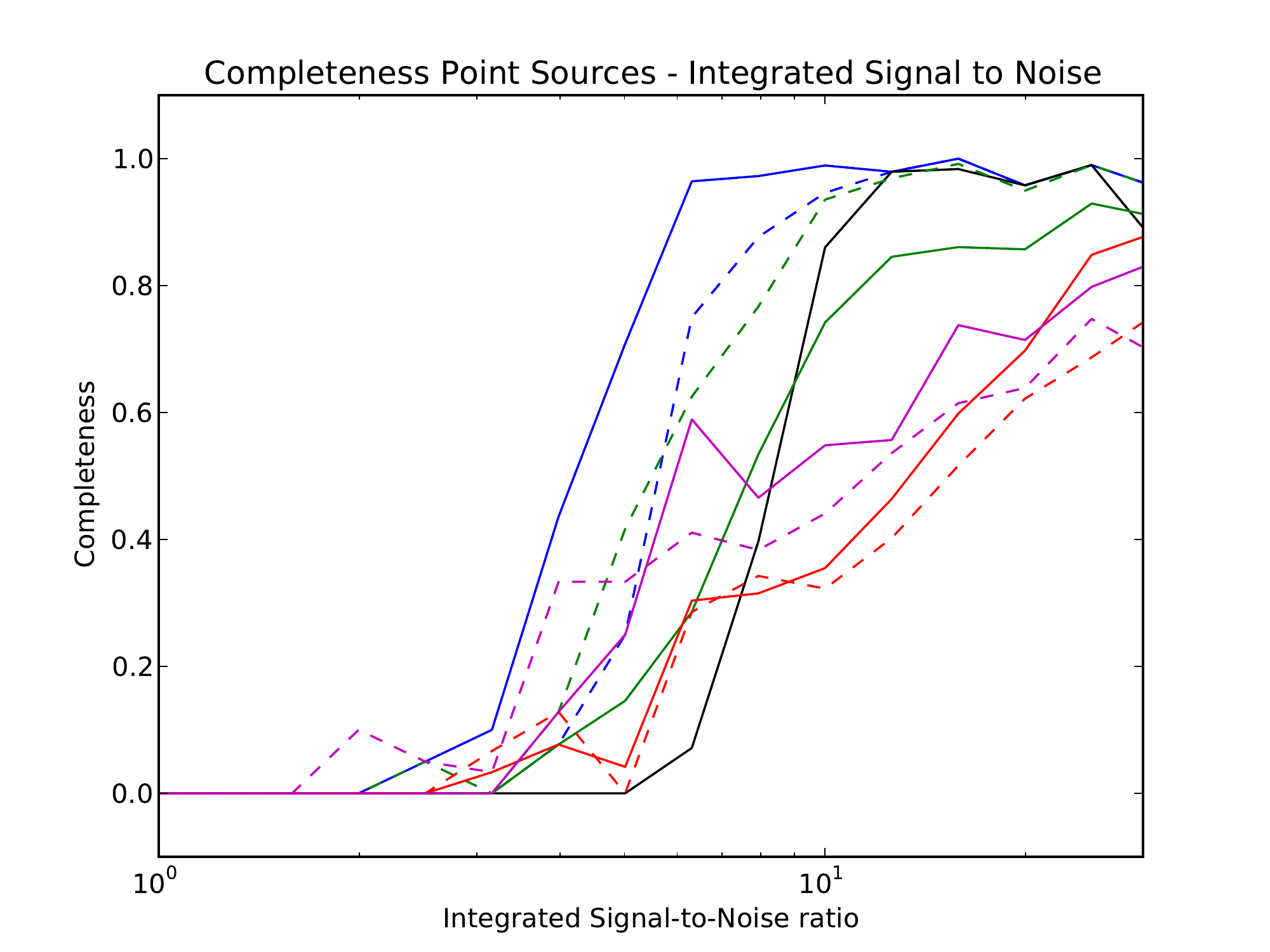}
\includegraphics[width=0.48\textwidth, angle=0]{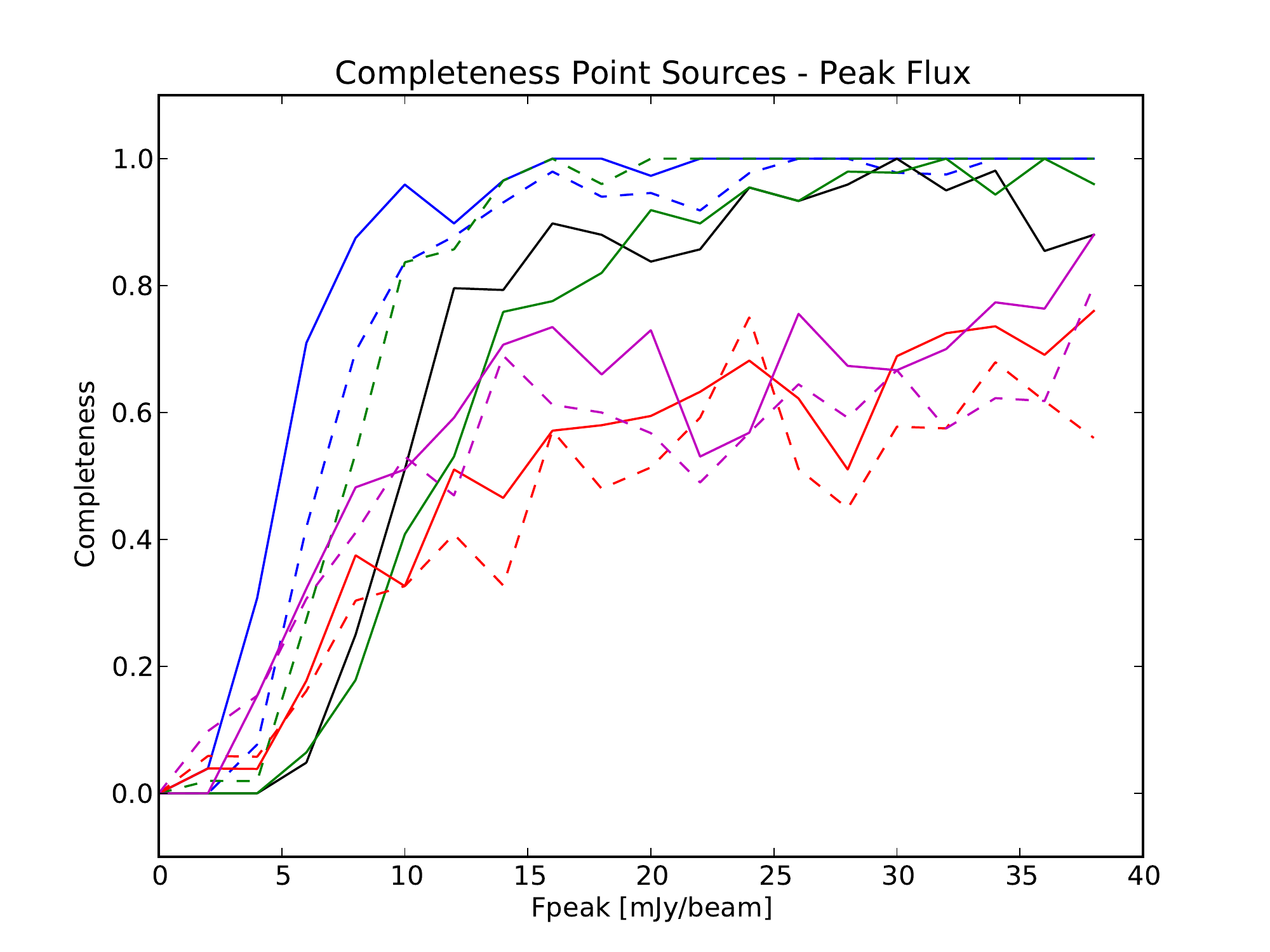}
\includegraphics[width=0.48\textwidth, angle=0]{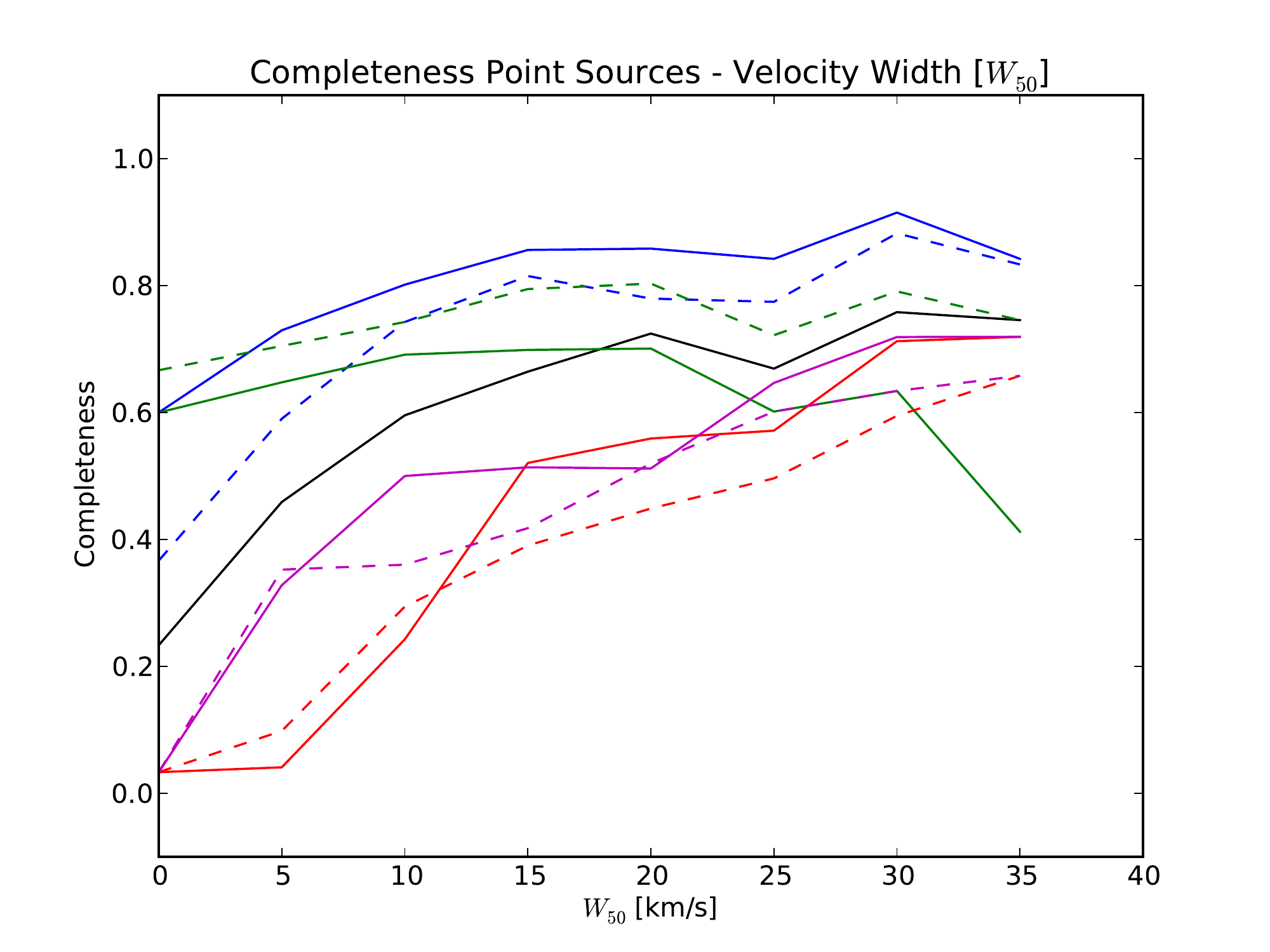}

\caption{Completeness of point sources plotted as a function of
  integrated {\HI} flux [Jy km/s] (top left), integrated signal to
  noise (top right), peak flux [mJy/beam] (bottom left) $W_{50}$ [km/s]
  (bottom right) for the different source finders. The legend
  gives the reliability of each source finder.}
\label{point_comp_flux}
\end{center}
\end{figure*}

\subsection{Cross-Matching}
To properly compare the five source finders, they have to be
analysed in exactly the same manner to exclude any discrepancies based
on different methods or interpretations.

Apart from the {\sc Gamma-Finder}, all source finders produce a
3-dimensional mask containing all the voxels that belong to a
detection. Although some source finders such as {\sc Duchamp} have the
capability to determine source parameters, we have chosen to extract
the source parameters from the produced masks, using a separate
script. In this way the results of all source finders are treated in
exactly the same manner and we are able to make an objective
comparison of the results. Using the mask, we have merged all
detections that were separated by one pixel in the two spatial
dimensions and seven channels in the spectral dimension. Furthermore
we required detections to be apparent in at least three channels
of the cube to reject spurious detections.

The way in which detections are merged can effect the results
significantly. For example double-horned, unresolved sources are often
split up into two separate sources. They can be recovered as one
source, however this depends on the scale that is used for the
merging, and it is inevitable that in the merging process not all
split sources are recovered properly.

 Some basic object parameters that have been
extracted are the position of the source, the velocity width, the peak
flux and the integrated flux.

Crucial but not trivial is how the cross-matching is done between the
implemented input catalogues and the results of the different source
finders. 

Measuring the central position of a source can be difficult, however in the
case of the model cube with point sources the position of the objects
is very well determined, both in spatial and velocity direction. The
list of input objects is compared with the detections of the source
finders, and pairs are sought within $\pm$1 pixel in the spatial
direction and $\pm$2 pixels in the spectral direction. As the
synthesised beam at FWHM of the used models is described by only three
pixels, this is a very robust method.

For the cube with disk galaxies the measured centre of a certain
object is not always trivial to determine as the sources can be very
extended. Due to the rotation, for many objects several components are
detected, without detecting emission in the actual centre of the
object. As the objects can have line widths of up to several hundred
km/s, the central velocity is difficult to estimate and might differ
significantly amongst the different source finders.

To do the cross-matching we have used a Python script that is used and
described in the paper on testing of the {\sc Duchamp} source
finder \citep{westmeier2011}. We created a three-dimensional mask
containing all voxels containing emission from the model galaxies. For
each detection we assess whether the central position $\pm$1 pixel
overlaps with one of the voxels in the mask and then determine to
which object from the input model catalogue it belongs.

\section{Results}
The range of {\HI} source properties is large and well documented in
many published galaxy catalogs (e.g.  \citet{koribalski2004},
\citet{meyer2004}, \citet{springob2005}, \citet{haynes2011}) as well as catalogs of high velocity clouds (HVCs;
e.g. \citet{putman2002}) and peculiar {\HI} features
(\citet{hibbard2001}, Rogues Gallery).  The shape of {\HI} spectra ranges
from simple Gaussian profiles to steep double-horn profiles and almost
everything in between. The distribution of {\HI} in disk galaxies is
often symmetric and regular, but many irregular {\HI} sources exist,
from peculiar dwarf galaxies and {\HI} rings to {\HI} plumes/filaments
and clouds. As typically only the highest column density gas is
detected, it is likely that the low column density gas is more
pervasive and irregular.

In the following we present a comparison of source finding algorithms
applied to the two cubes described in Section 3.1. We start with the
simple point sources with Gaussian profiles, then progress to extended
disks with more complex {\HI} profiles.

\subsection{Point sources}
Point sources with a Gaussian velocity profile are ideal sources in
the sense that they do not have any complicated structures and are
relatively easy to detect. Fig.~\ref{point_comp_flux} shows the
completeness as a function of integrated flux ($F_{int}$), integrated
signal-to-noise ratio, peak flux ($F_{peak}$) and 50\% velocity width
($W_{50}$). The integrated flux and integrated signal-to-noise
ratio are plotted on a logarithmic scale, to highlight the differences
between the source finders. All parameters are the true parameters
determined from the input models. For $F_{int}$ we use the same
definition as \cite{westmeier2011} (their equation 4).

\begin{table}[t]
\label{duchamp_setup}
\begin{tabular}{lcc}
\hline
Parameter & Value & Comment \\
\hline
threshold (test 1)  & 0.0039 & $2\times ${\sc rms} \\
threshold (test 2)  & 0.0029 & $1.5\times ${\sc rms} \\
minPix & 5 & \\
minChannels & 3 & \\
flagAdjacent & true & \\
flagATrous & true & Wavelet reconstr.\\
reconDim & 3 & in 3 dimensions \\
snrRecon   & 3 & \\
scaleMin (test 1) & 1 & \\
scaleMin (test 2) & 2 & \\
\hline
\end{tabular}
\caption{{\sc Duchamp} input parameters for the data cube with point sources.}
\end{table}

We have plotted two results for each of the individual source finders
on this particular cube, apart from the 2D-1D wavelet reconstruction
method which only produced one output. For {\sc Duchamp} the input
parameters are given in Table.~\ref{duchamp_setup}. For the {\sc
  Gamma-finder} we use a 3$\sigma$ and a 4$\sigma$ clipping threshold
and for the CNHI source finder we use a probability of $10^{-3}$ and
$3\cdot10^{-4}$. The {\sc S+C finder} has been tested using clipping
levels of 3$\sigma$ and 4$\sigma$. For each test, the raw reliability
is given as a percentage in the legend of the figure. Here the
  completeness is the principal value to compare the source
  finders as the single value for raw reliability can be a misleading
  number.

  The number of possible settings or input parameters for each source
  finder is very large and we experimented with each source finder
  until we found a set of parameters that was representative for its
  performance. We emphasise that the scope of this paper is to compare
  the results of the different source finders, rather than to test
  them individually which has been done in other papers in this
  special issue.

{\sc Duchamp} performs very well on point sources, and the
completeness is superior to the other source finders for all plotted
parameters. The completeness starts at very low values, but rapidly
increases to a completeness of about $\sim$50\% at an integrated flux
of $\sim$0.08 Jy km s$^{-1}$. There is a turnover in the plot reaching
full completeness around $\sim$0.2 Jy km s$^{-1}$. The completeness
does not stay at 100\% as some of the bright sources become merged due
to the wavelet reconstruction and multiple objects are counted as one.
The other source finders show a very similar behaviour however the
completeness levels are lower. There is a large variation in the
reliability numbers, but apart from CNHI the reliabilities for all
source finders have values above 70\%. We have to stress here again
that the raw reliability is an initial estimate of the quality of a
source finder, but is likely to be improved upon in post-processing of
the data.  We will explain this in more detail in the discussion. The
reliability will go down however with more realistic noise containing
unpredictable features such as e.g. continuum sources and solar
interference.

In the top-right panel of Fig.~\ref{point_comp_flux} the
completeness is plotted as a function of integrated signal-to-noise
ratio $(F_{int}/{\sigma}_{int})$. The integrated noise is calculated
as:

\begin{equation}
\sigma_{int} =  rms \cdot dV \cdot \sqrt{2.35 \cdot W_{50}/dV}. 
\end{equation}

where $dV$ is the spectral resolution of the cube and the coefficient
is used to convert the $W_{50}$ value to the line width of a Gaussian.
The general trend is very similar, here for the better performing
source finders in terms of completeness, about 50\% completeness can
be achieved at an integrated signal-to-noise ratio of $\sim4-5$. The
completeness increases very rapidly and for several source finders
100\% completeness is achieved at an integrated signal-to-noise ratio
around 10 while for the best {\sc Duchamp} run this result is achieved
at an integrated signal-to-noise ratio close to 6.

\begin{figure*}[t]
\begin{center}
\includegraphics[width=0.48\textwidth, angle=0]{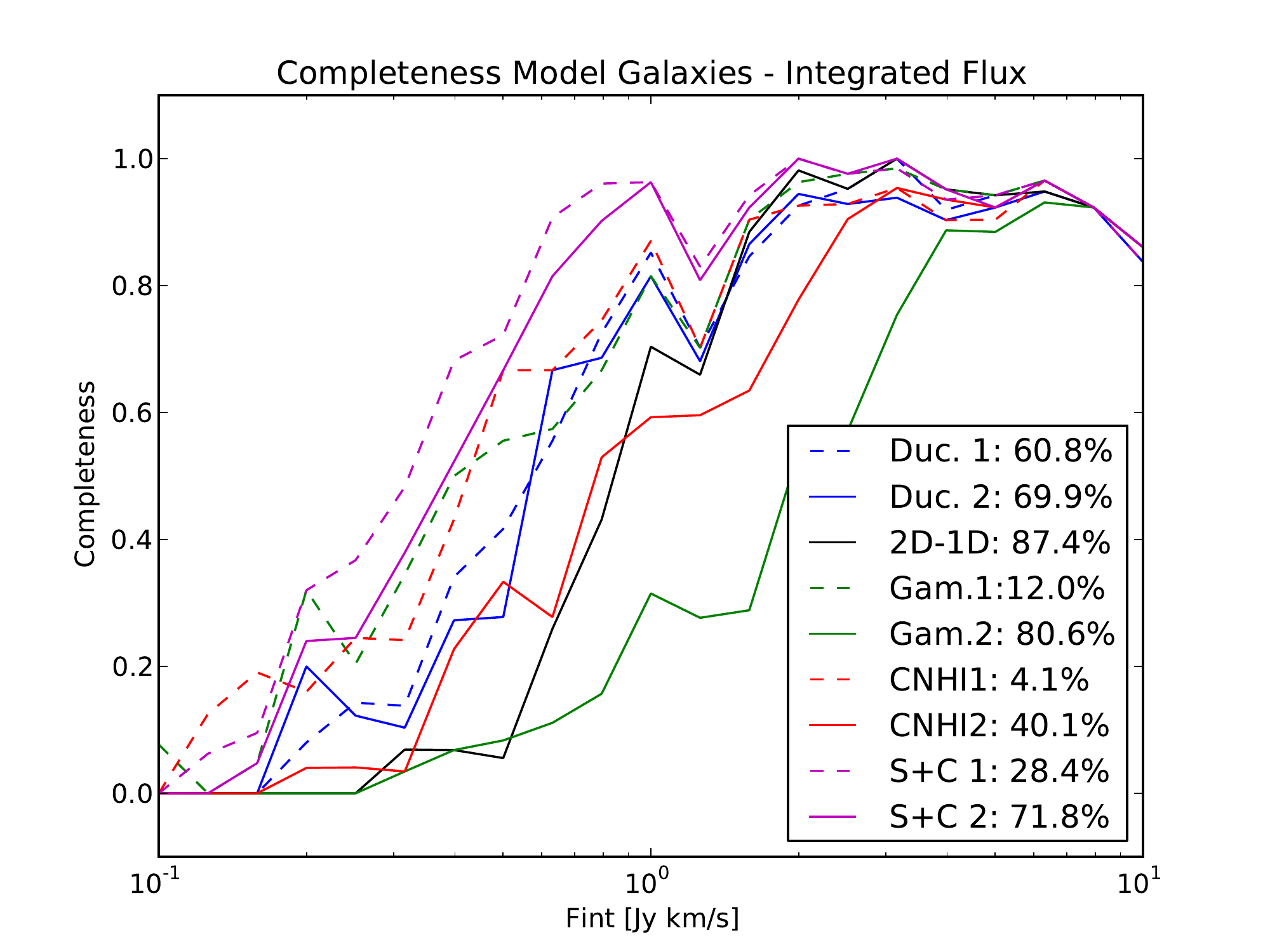}
\includegraphics[width=0.48\textwidth, angle=0]{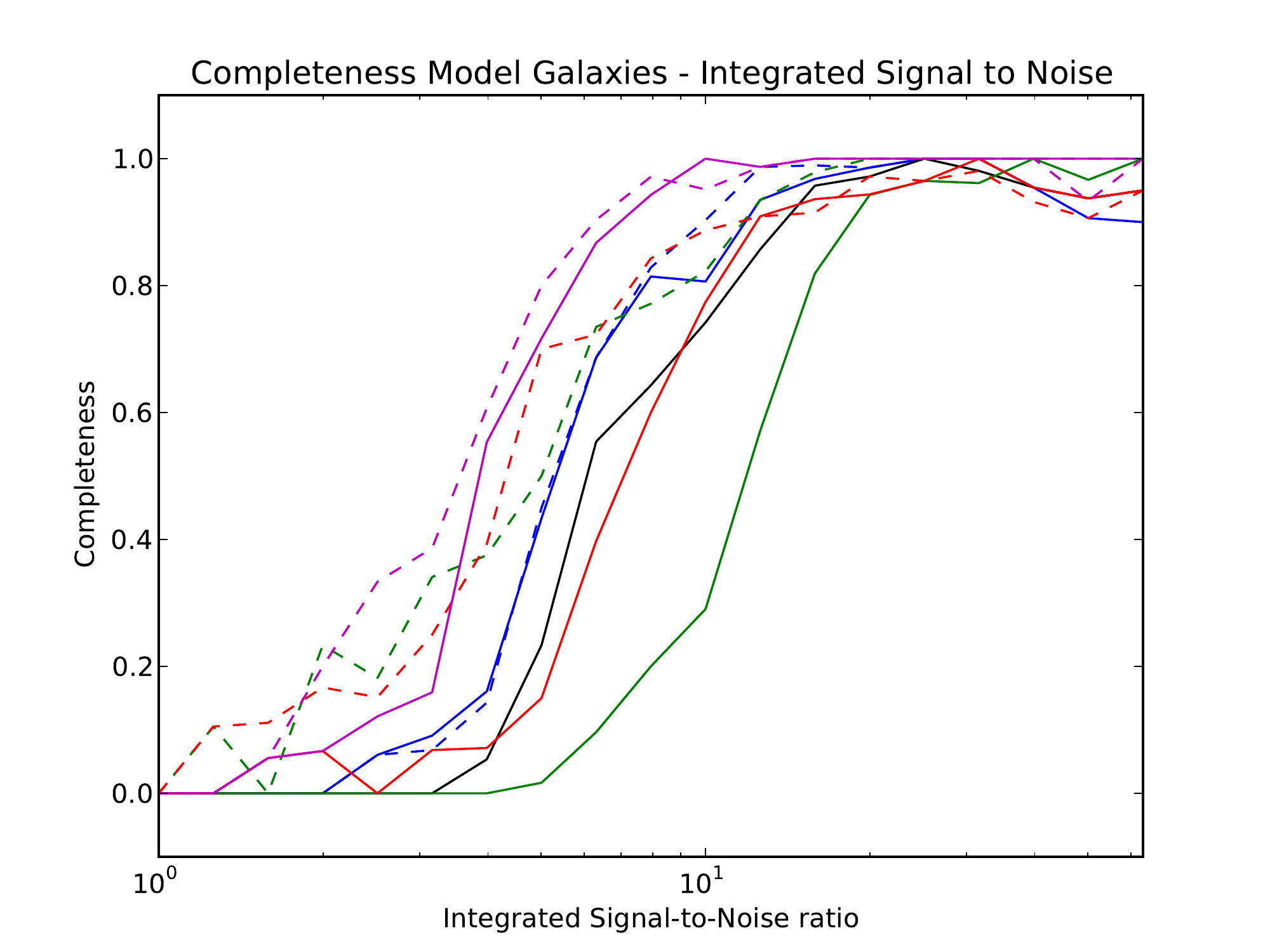}
\includegraphics[width=0.48\textwidth, angle=0]{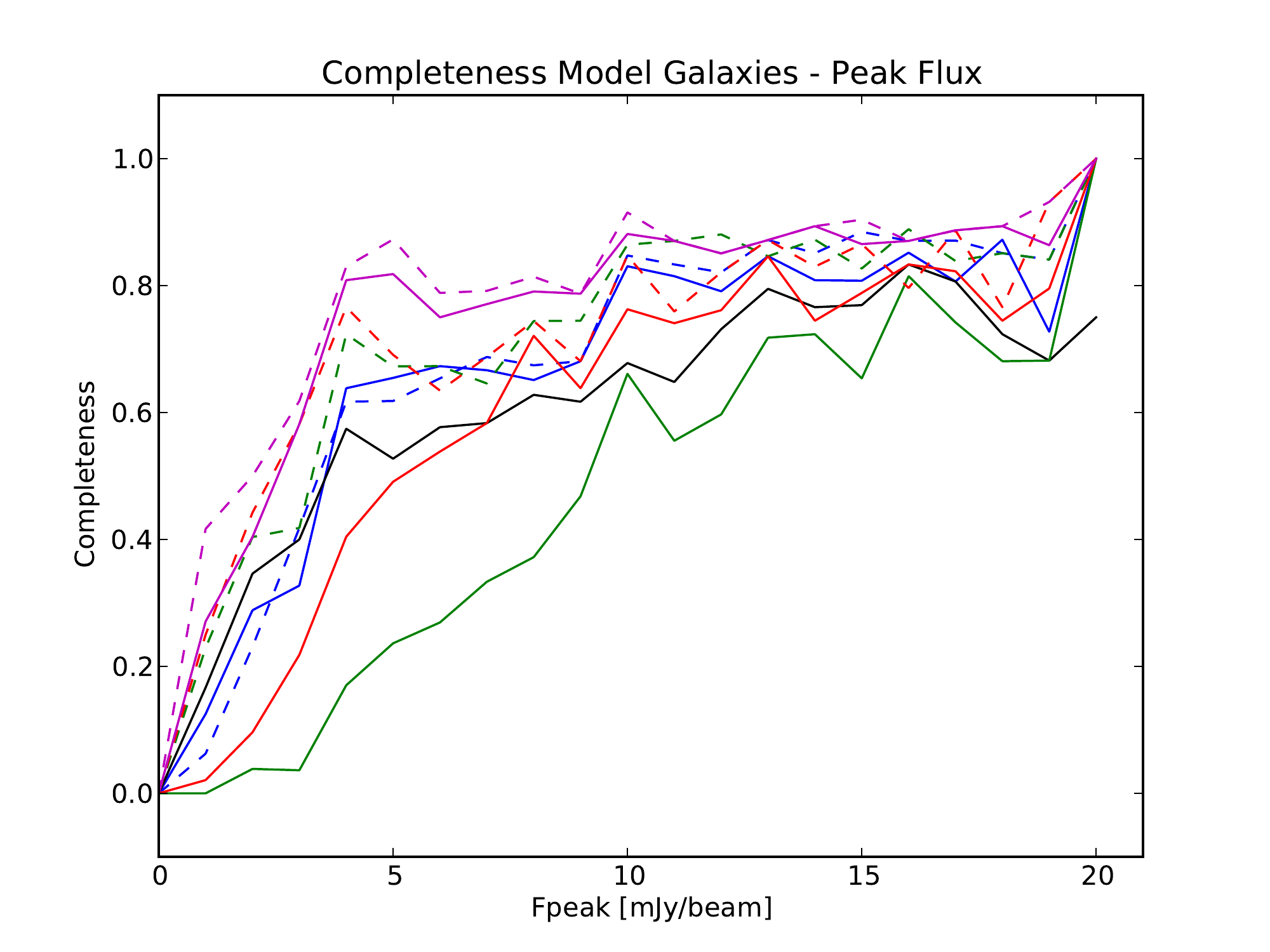}
\includegraphics[width=0.48\textwidth, angle=0]{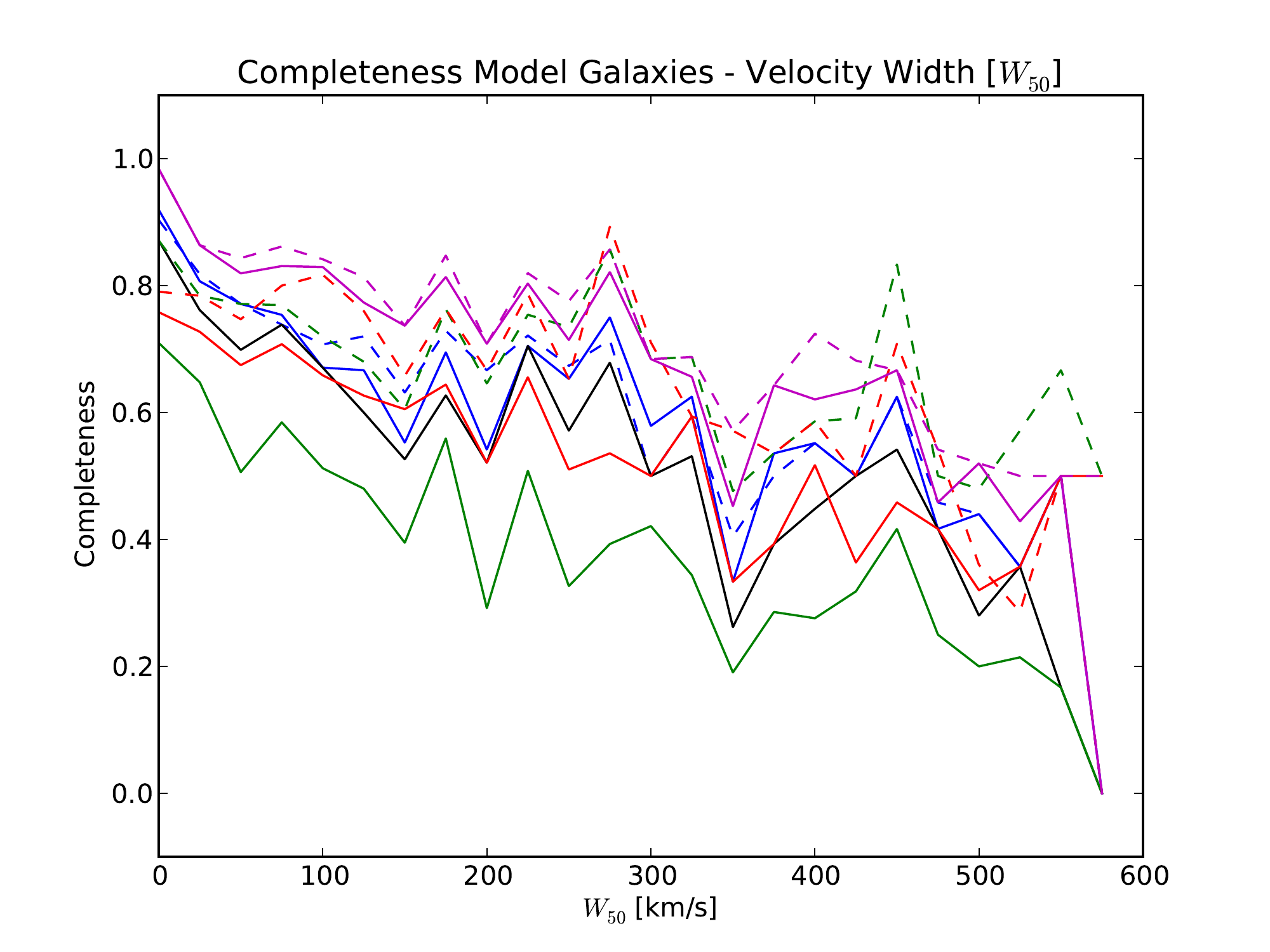}

\caption{Completeness of model galaxies plotted as a function of
  integrated {\HI} flux [Jy km/s] (top left), integrated
  signal-to-noise ratio (top right), peak flux [mJy/beam] (bottom
  left) and velocity width $W_{50}$ [km/s] (bottom right) for the
  different source finders. The legend gives the reliability of each
  source finder.}
\label{model_completeness}
\end{center}
\end{figure*}

\subsection{Model galaxies}
For the testing of the source finders on the cube with model galaxies,
we analyse again two different runs for each of the source finders
apart from the 2D-1D wavelet source finder. The tested parameters for {\sc
  Duchamp} are almost identical to two results as presented in
\cite{westmeier2011}, in table~\ref{model_duchamp} we summarise the
parameters that were used. The only difference between the two runs is
that in the second run the objects are {\it "grown"} to a lower
threshold once detected. When doing this, objects that are broken up
into multiple detections can get merged. The {\sc Gamma-finder} has
been used with a 3$\sigma$ and a $5\sigma$ clipping level, while for
the CNHI source finder we have used probability thresholds of
$5\cdot10^{-4}$ and $5\cdot10^{-5}$ respectively. In the case of the
{\sc S+C finder} clipping levels of 3.5$\sigma$ and 4$\sigma$ have
been used.

\begin{table*}
\begin{center}
\begin{tabular}{lrrr}
\hline
\hline
Parameter & Run 1 & Run 2 & Comment \\
\hline
threshold             & 0.00186  & 0.00186 & $1.0 \times${\sc rms}\\
minPix                &      10  &      10 & \\
minChannels           &       3  &       3 & \\
flagAdjacent          &    true  &    true & \\
flagGrowth            &   false  &    true & \\
growthThreshold       &      --  & 0.00093 & $0.5 \times${\sc rms} \\
flagRejectBeforeMerge &   false  &    true & \\
flagATrous            &    true  &    true & Wavelet reconstruction \\
reconDim              &       3  &       3 & in 3 dimensions \\
snrRecon              &       2  &       2 & \\
scaleMin              &       3  &       3 & \\

\hline
\hline
\end{tabular}
\caption{{\sc Duchamp} parameters that have been used for the cube
  with model galaxies}
\label{model_duchamp}
\end{center}
\end{table*}

In Fig.~\ref{model_completeness} we plot again the completeness of the
source finders as function of integrated flux, integrated signal-to-noise ratio, peak flux and velocity
width ($W_{50}$). The integrated
flux of the model galaxies is defined as:

\begin{equation}
F_{int} [\textrm{Jy km s$^{-1}$}] = F_{peak} \cdot (2\pi)^{1.5} \cdot disp \cdot B_{maj} \cdot B_{min}
\end{equation}
where $F_{peak}$ is the peak flux, $disp$ is the velocity dispersion and
$B_{maj}$ and $B_{min}$ are the FWHM major and minor axis respectively of
the 2-dimensional Gaussian describing the galaxy. The integrated noise is given by:

\begin{equation}
\sigma_{int} = \sqrt{\frac{2.35 \cdot W_{50}}{dV}} \cdot \sqrt{\frac{1.13 \cdot B_{maj} \cdot B_{min}}{b_{min} \cdot b_{maj}}} \cdot rms \cdot dV  \cdot 2.35
\end{equation}

where $W_{50}$ is the velocity width FWHM given by the model catalogue,
$dV$ is the channel separation, $b_{maj}$ and $b_{min}$ are the major
and minor axis of the synthesised beam and {\sc rms} is the noise in the
cube.

\ \\ 

The general results are slightly different to the results as obtained
from the cube with point sources. The performance of the different
source finders is quite comparable, however in general both
completeness and reliability levels are slightly lower than for the
point sources. Sources that are extended in space or velocity can be
almost hidden in the noise and hard to detect. For the better
performing source finders, we reach 50\% completeness around an
integrated signal-to-noise ratio between 4 and 6 and 100\%
completeness for a signal-to-noise ratio between 10 and 15. These are
very promising results given that the achieved completeness values are
very close to the completeness of the point sources which should be
much easier to detect. Compared to the point sources the S+C finder is
performing much better and seems the best algorithm here in terms
  of completeness. This is due to the fact that with smoothing to
different spatial or spectral scales the real shape of an object is
matched as close as possible. In the case of point sources smoothing
to a larger scale does not increase the signal to noise and hence the
S+C finder does not benefit as much. The {\sc Gamma-finder} performs
much worse for model galaxies as this source finder is most sensitive
to sudden changes in the spectrum, which are not as apparent in the
case of extended sources.

\begin{figure*}[t]
\begin{center}
\includegraphics[width=0.48\textwidth, angle=0]{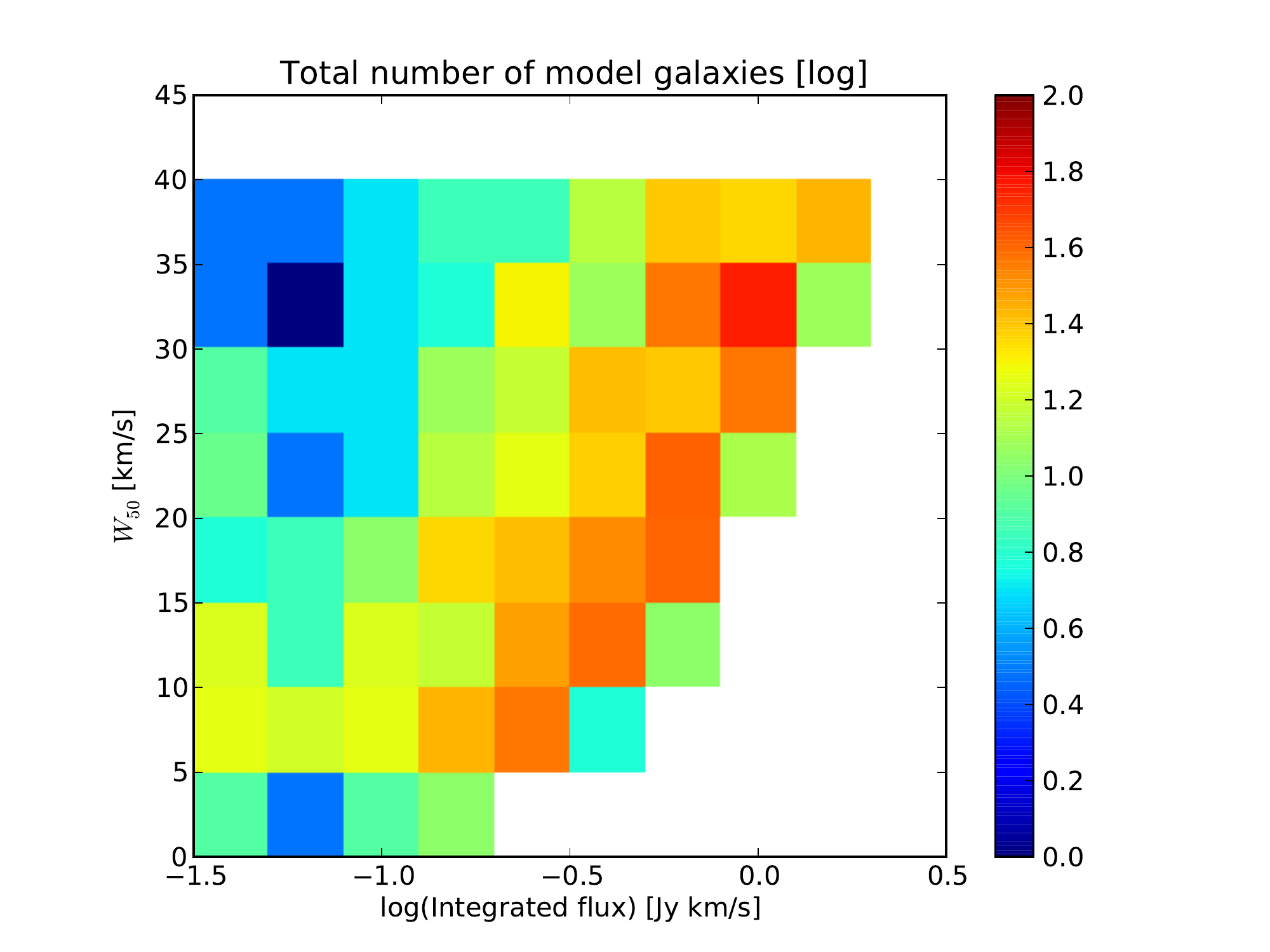}
\includegraphics[width=0.48\textwidth, angle=0]{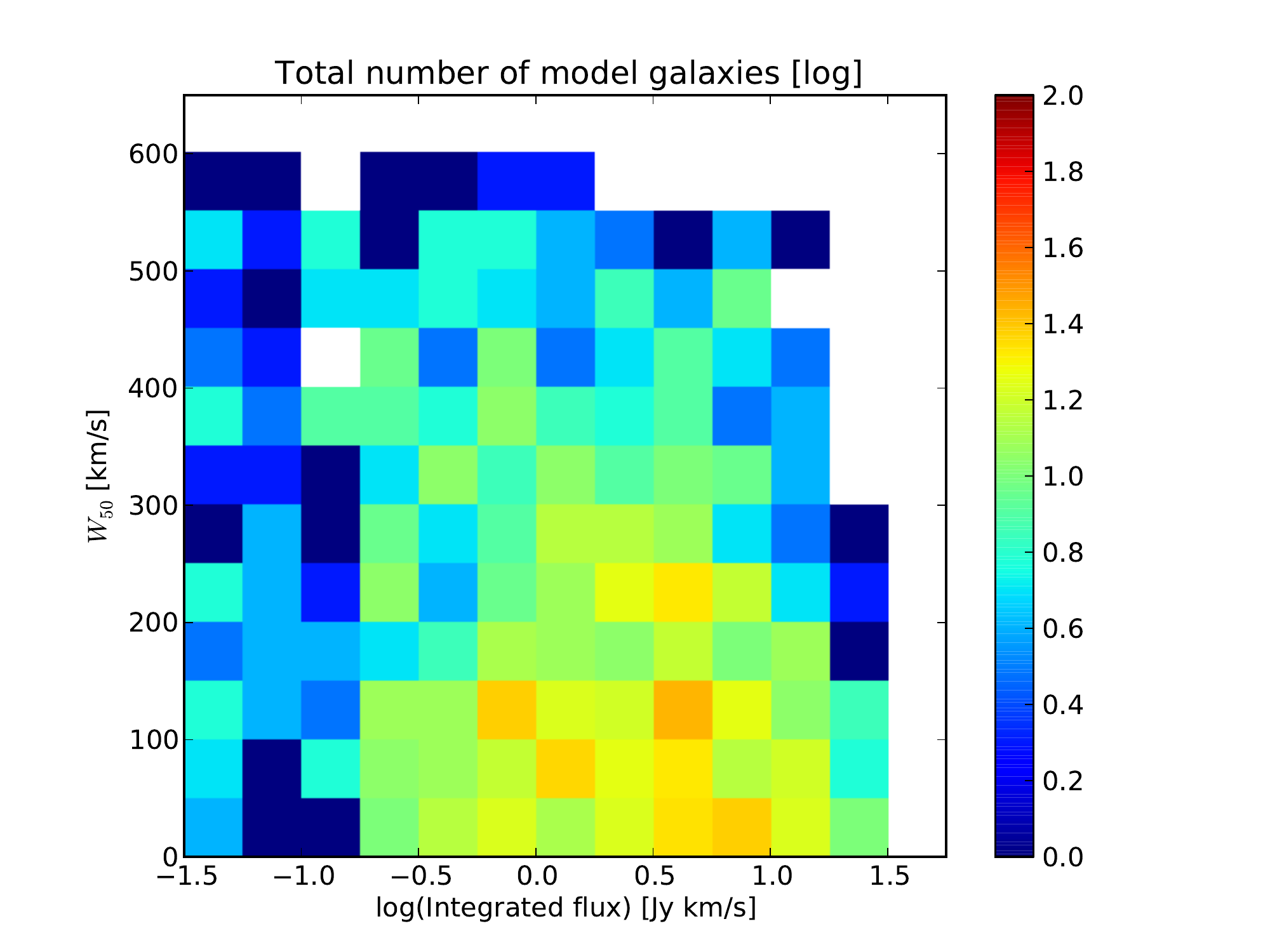}
\caption{Total number of objects in the cube with point sources (left)
  and model galaxies (right) is plotted on a two dimensional
  logarithmic scale as function of line width $(W_{50})$ [km/s] and integrated flux [Jy km/s].}
\label{total_detections_2D}
\end{center}
\end{figure*}


\begin{figure*}[t]
\begin{center}

\includegraphics[width=0.48\textwidth, angle=0]{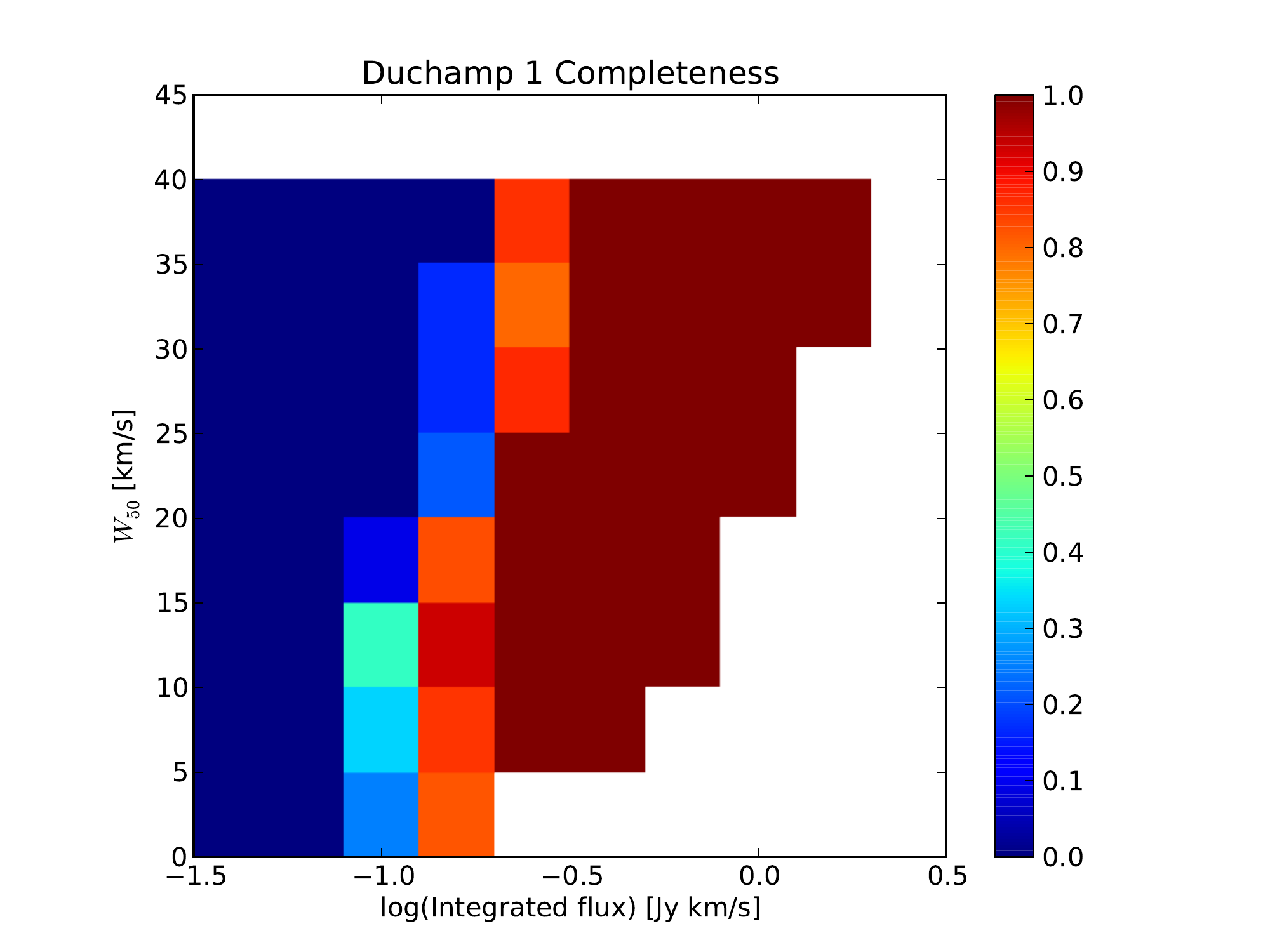}
\includegraphics[width=0.48\textwidth, angle=0]{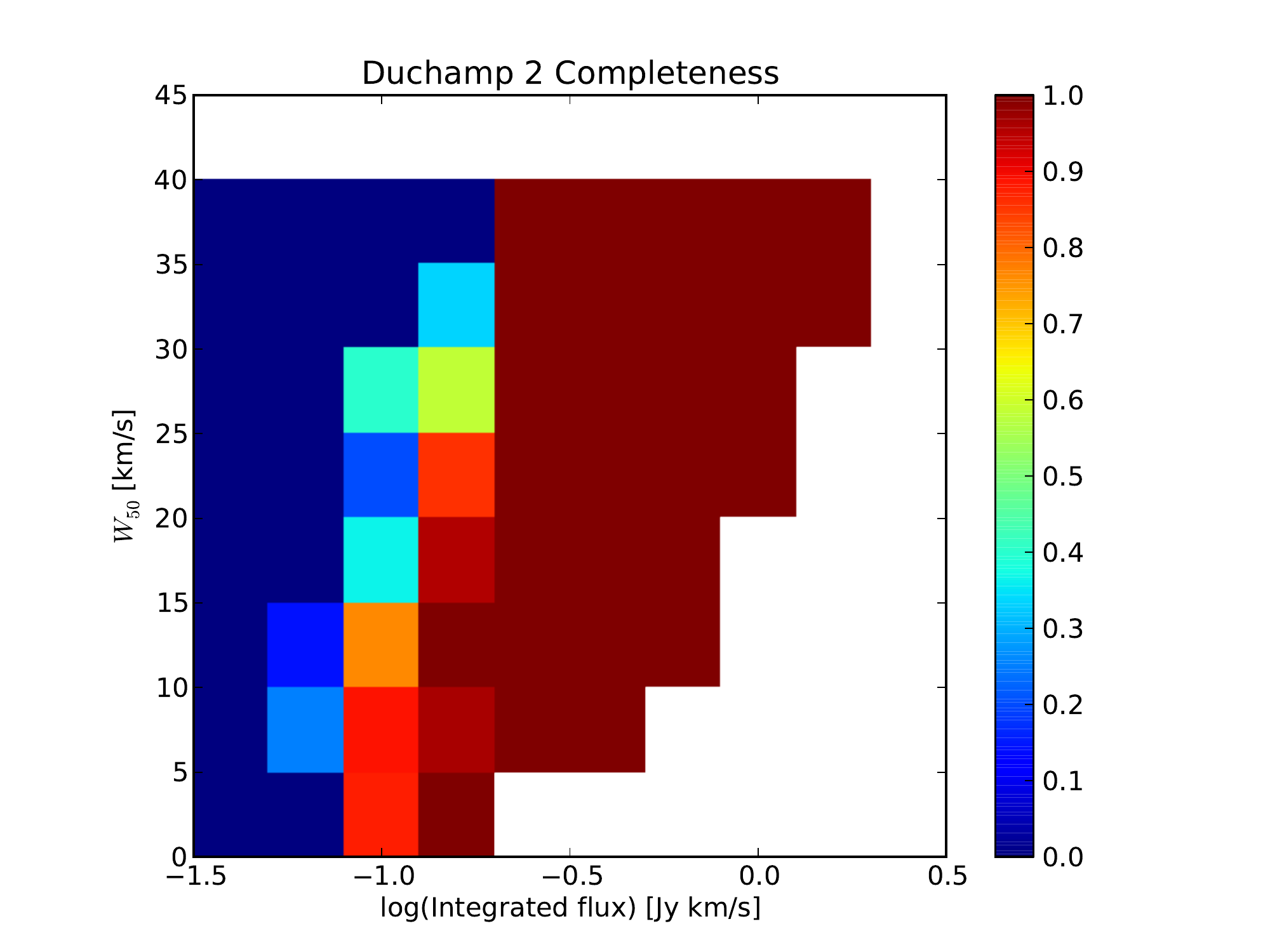}
\includegraphics[width=0.48\textwidth, angle=0]{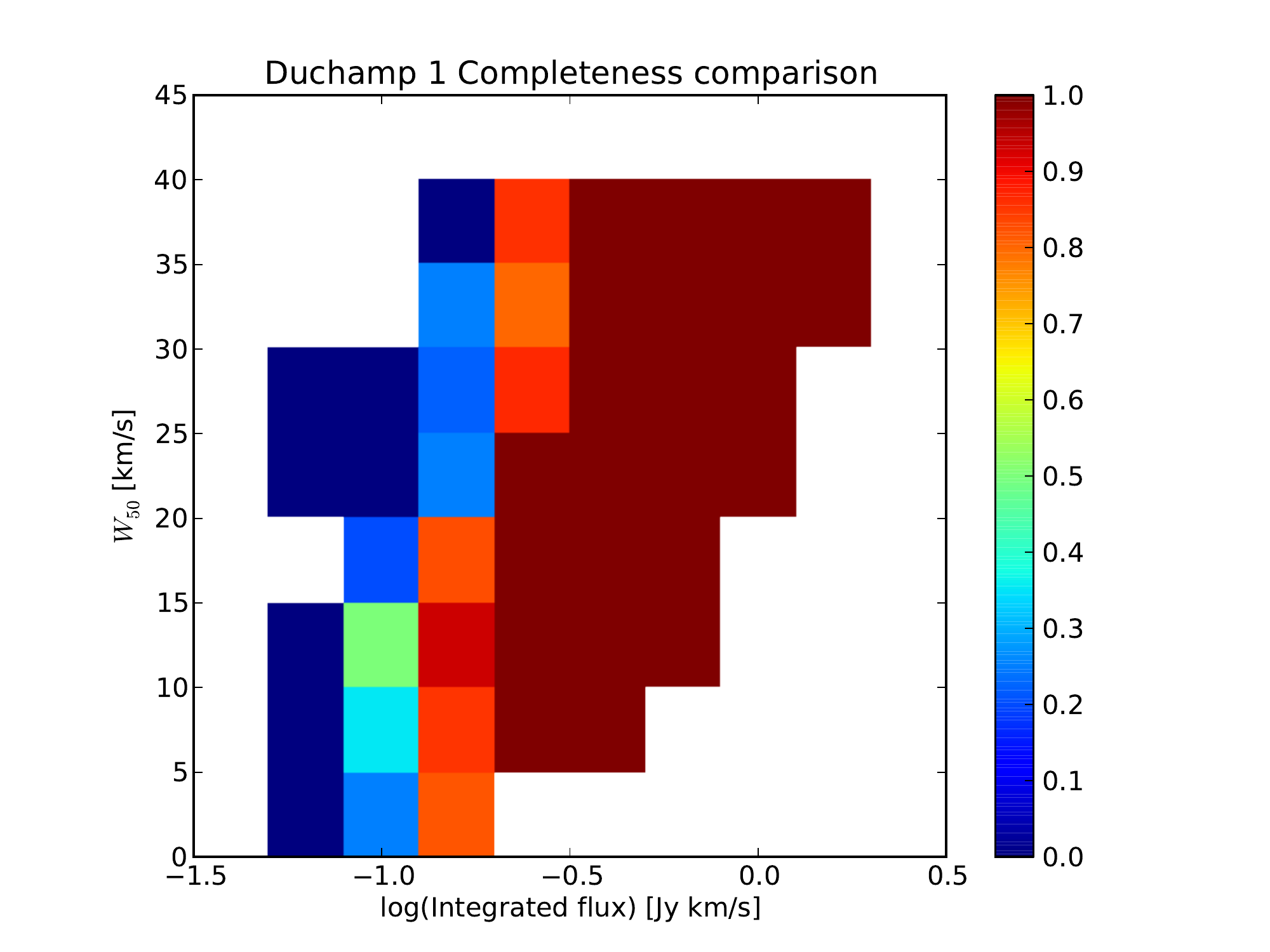}
\includegraphics[width=0.48\textwidth, angle=0]{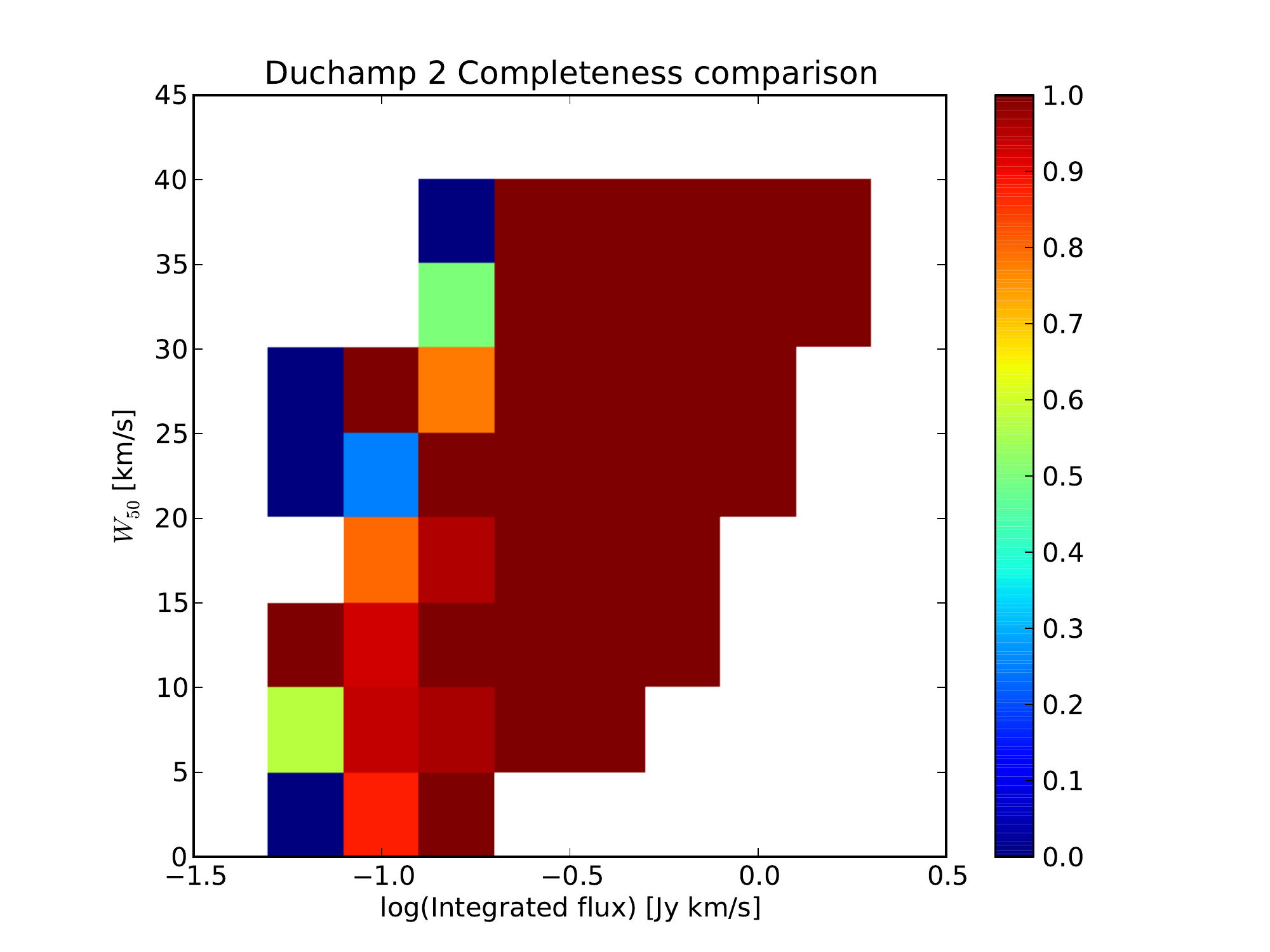}
\includegraphics[width=0.48\textwidth, angle=0]{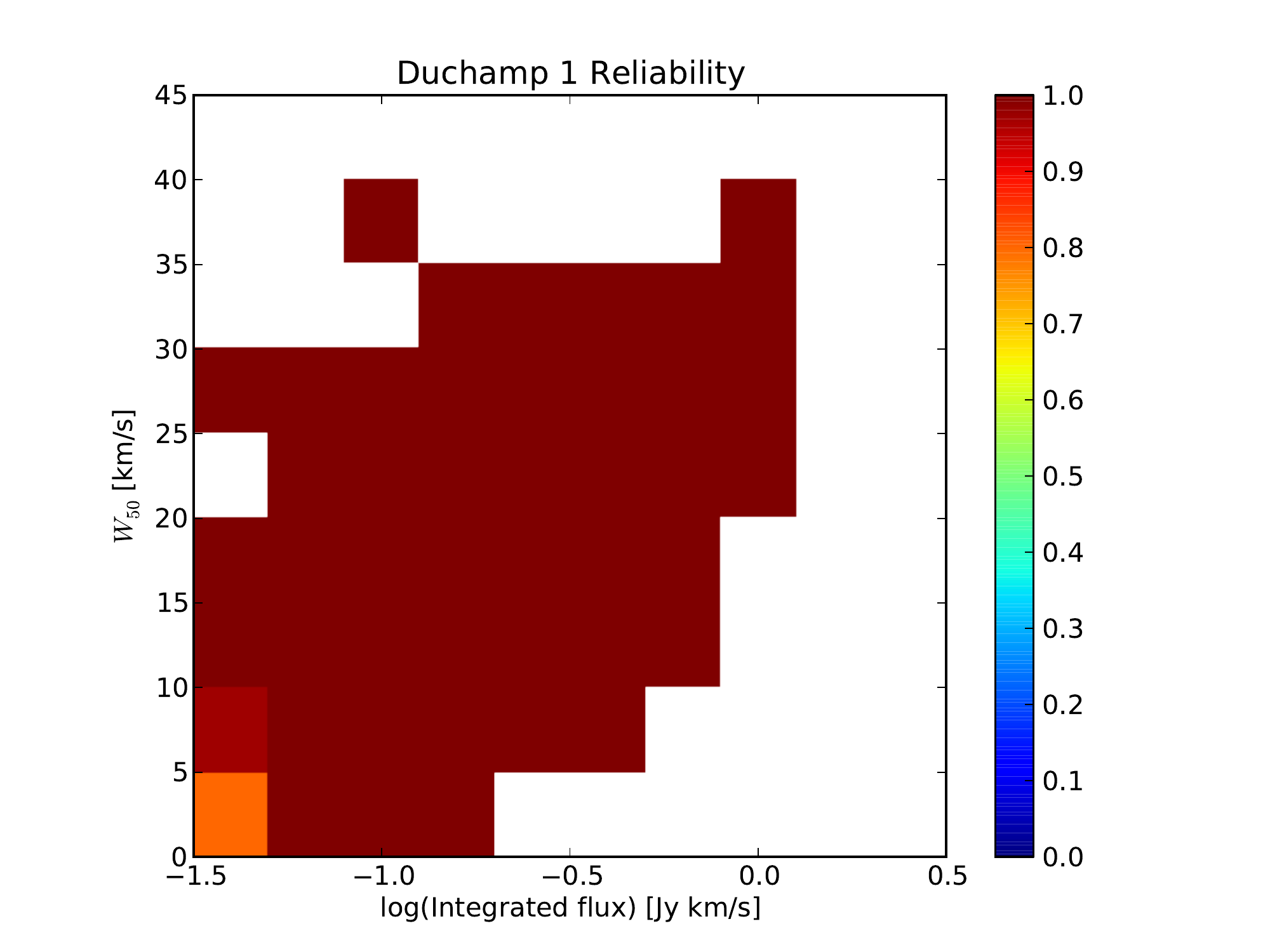}
\includegraphics[width=0.48\textwidth, angle=0]{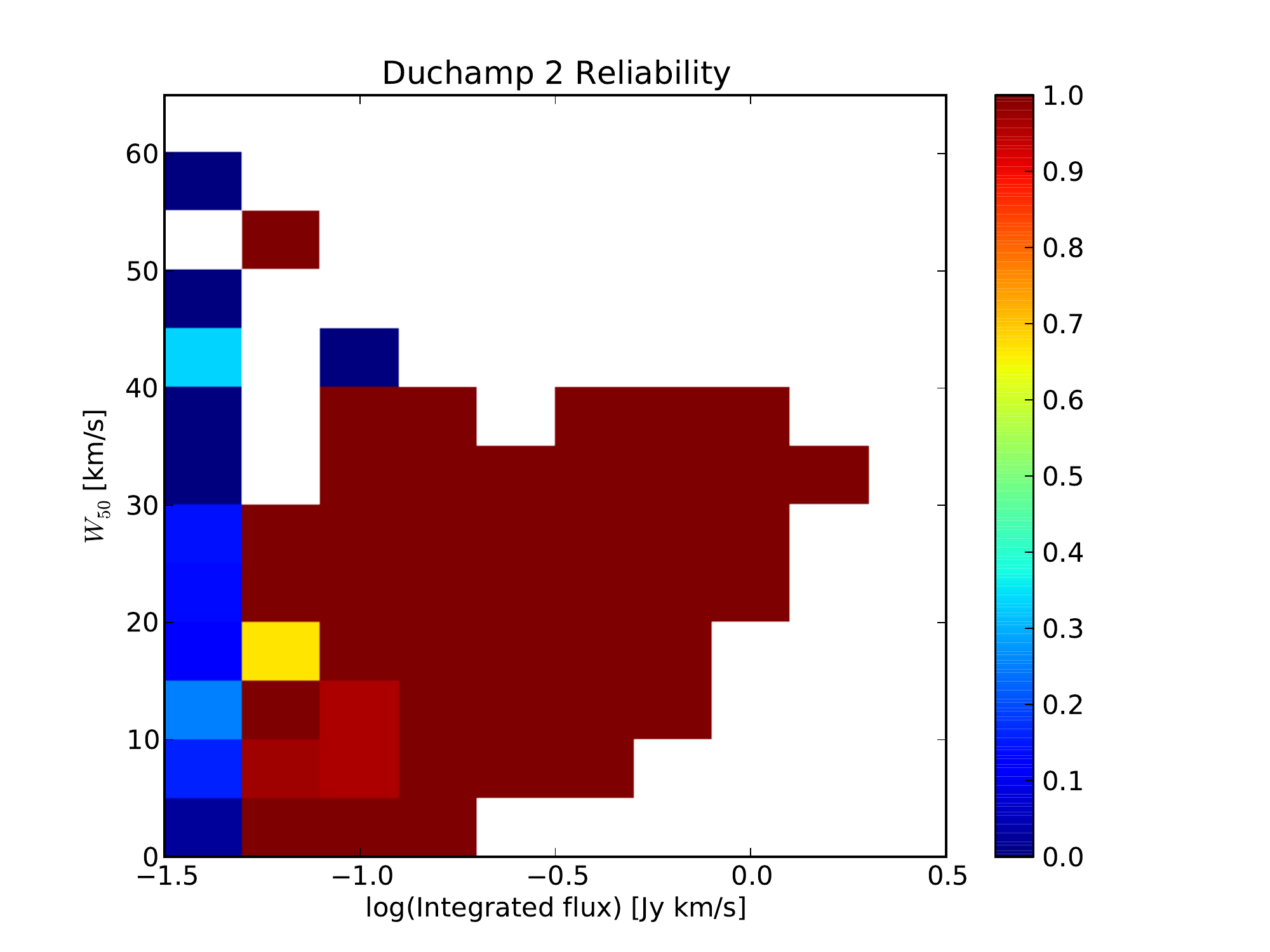}

\caption{2-Dimensional completeness and reliability of the
  source-finding tests on point sources is plotted as function of line
  width $(W_{50})$ [km/s] and integrated flux [log(Jy km/s)]. In the
  top panels completeness is plotted, while the middel panel shows a
  comparison where the relative completeness is plotted which is
  defined by the number of detections of a single source finder over
  the number of detections by any of the source finders.  In the
  bottom panel the reliability of each source finding result is
  plotted. For the completeness plots the source parameters are
  determined from the input catalogues are similar for each result,
  for the reliability plots the source parameters are measured, hence
  the scaling is different for each source.}
\label{point_2D}
\end{center}
\end{figure*}

\begin{figure*}[t]
\begin{center}

\includegraphics[width=0.48\textwidth, angle=0]{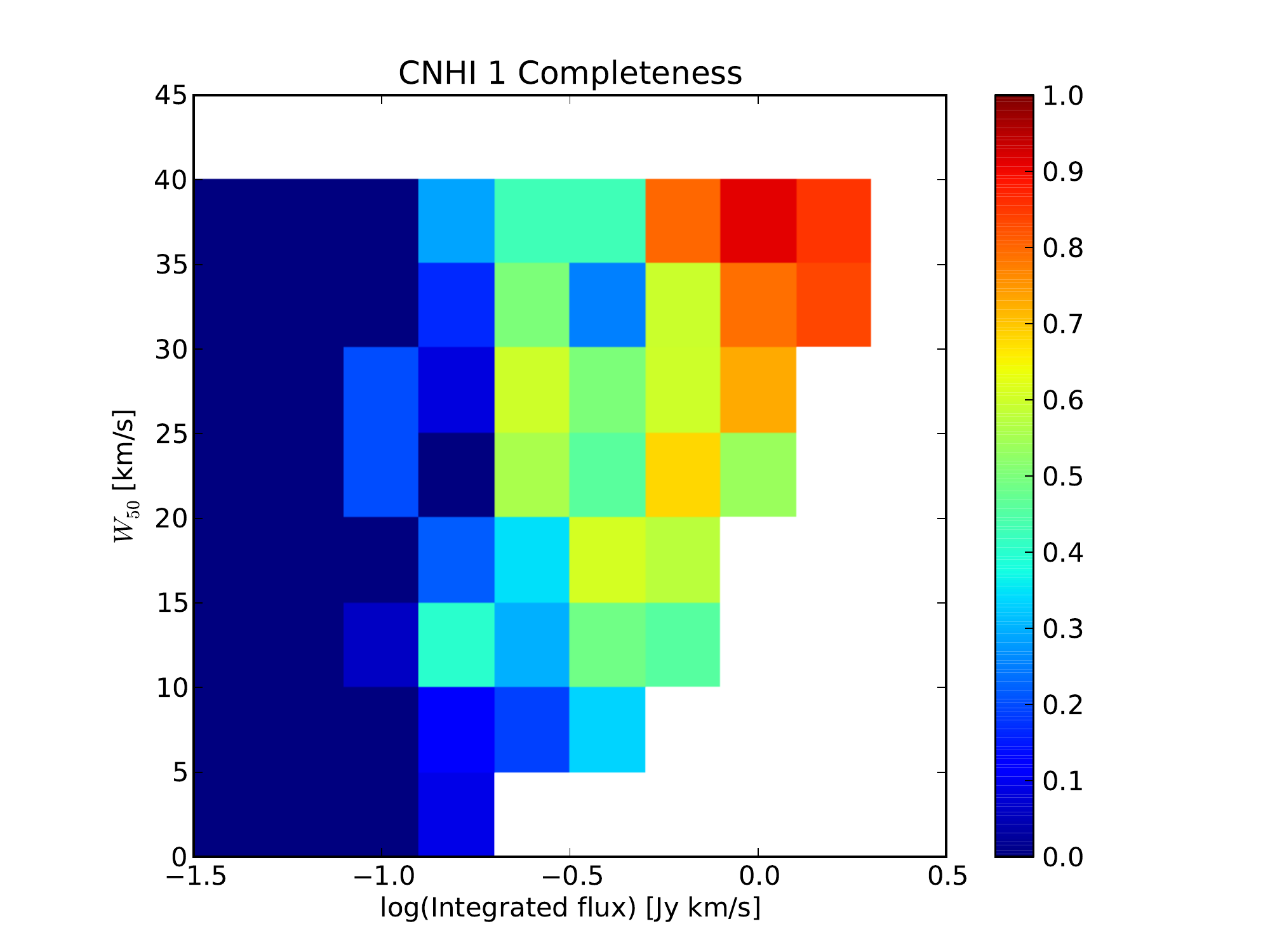}
\includegraphics[width=0.48\textwidth, angle=0]{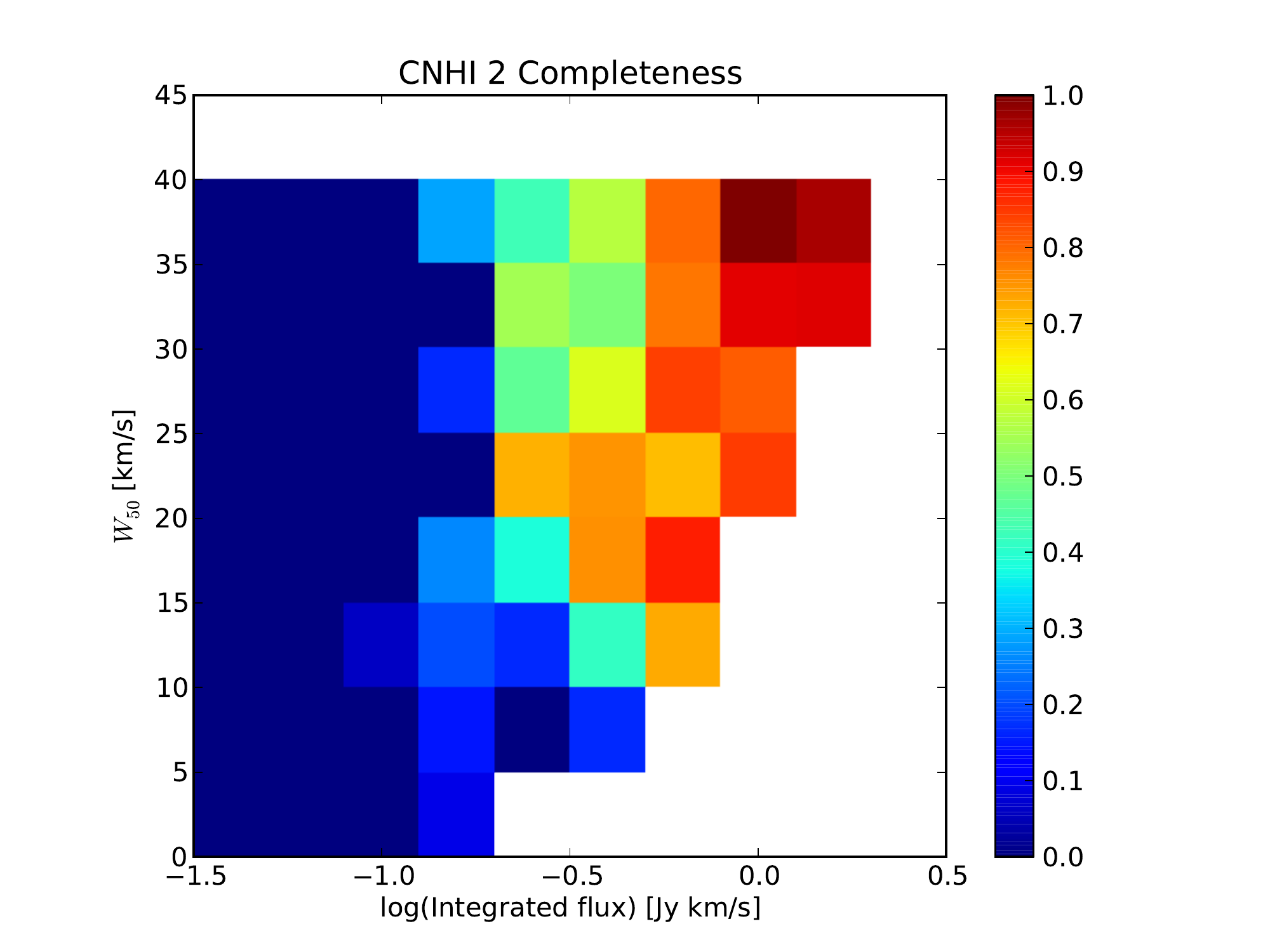}
\includegraphics[width=0.48\textwidth, angle=0]{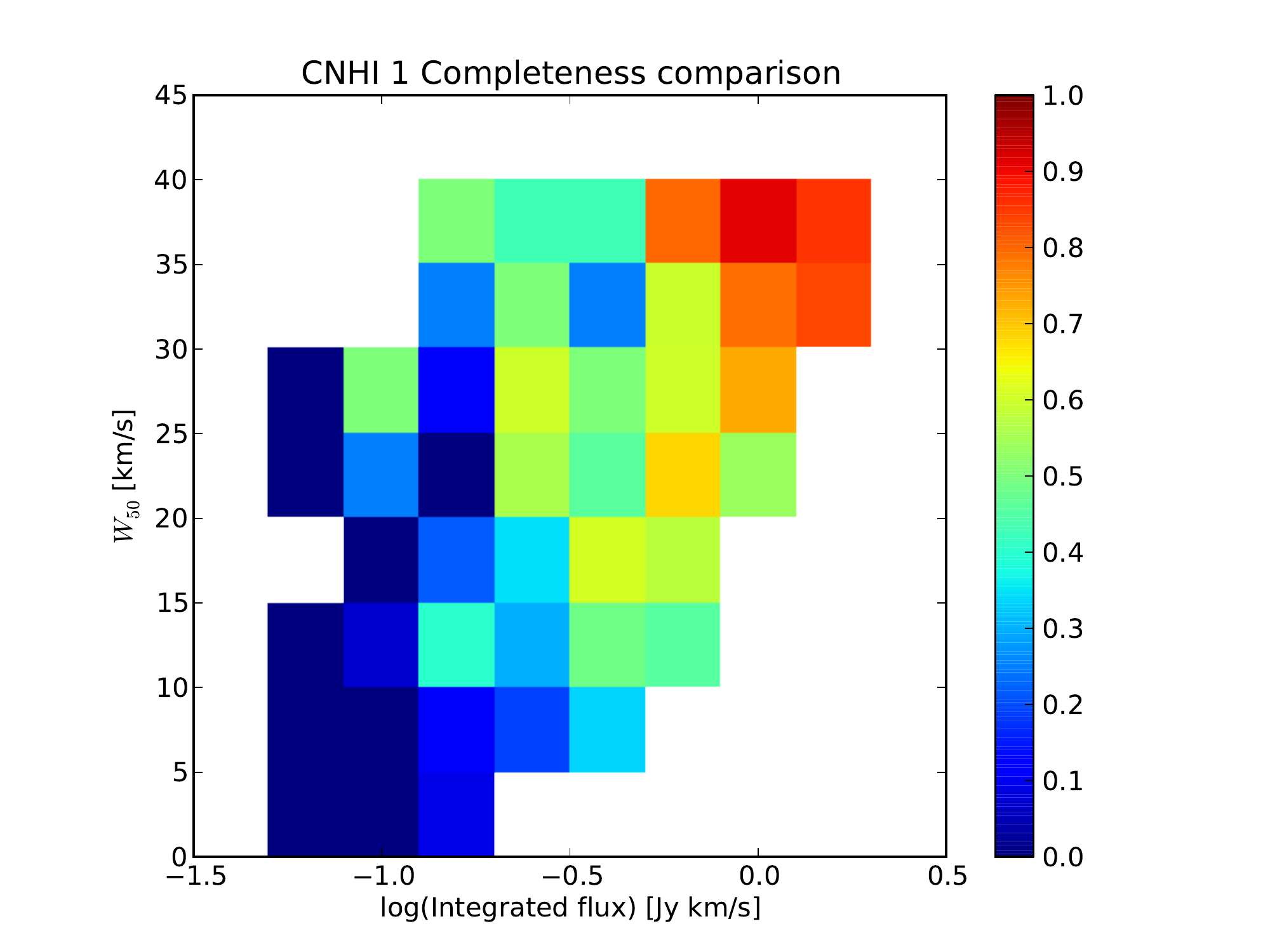}
\includegraphics[width=0.48\textwidth, angle=0]{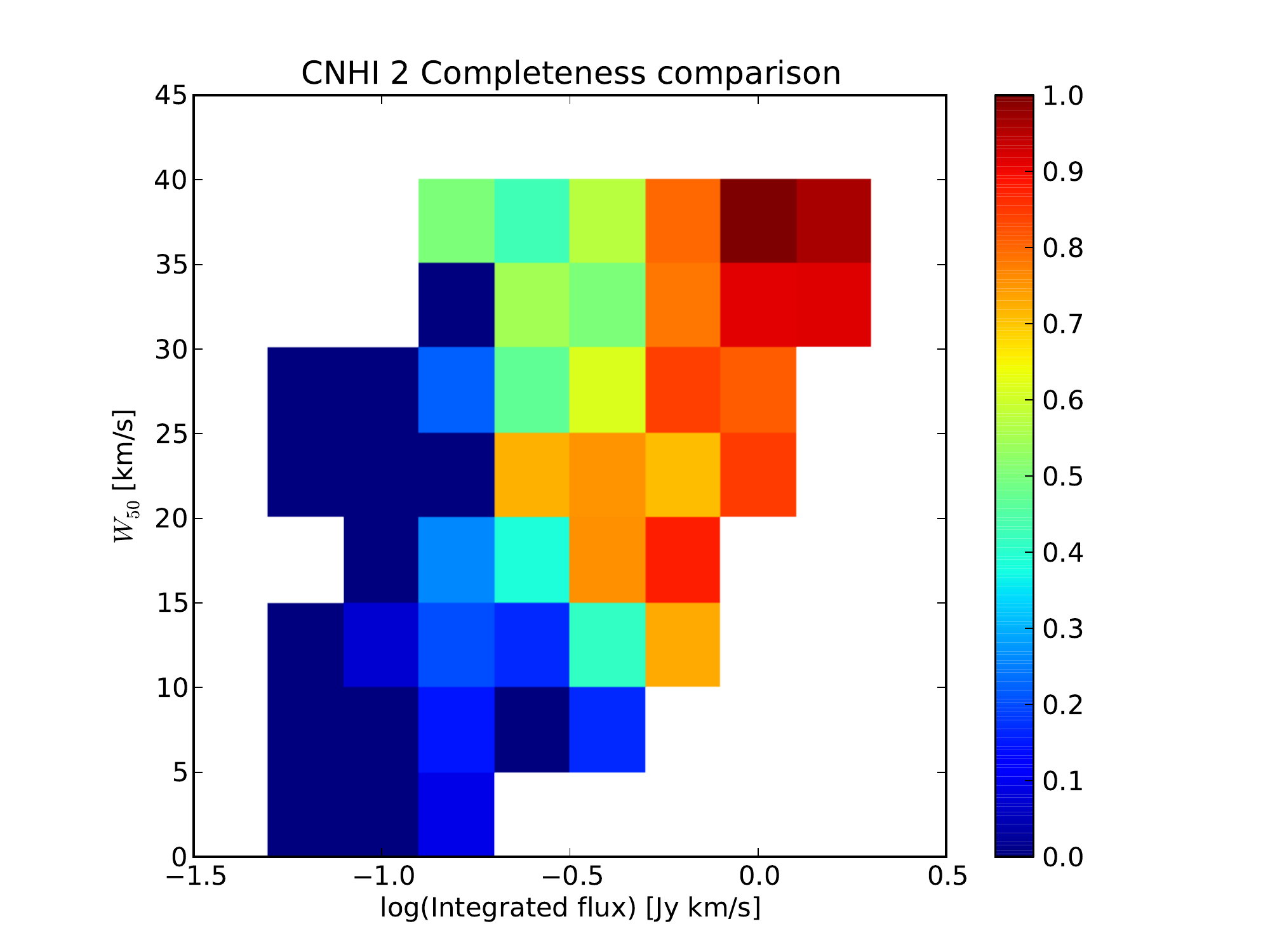}
\includegraphics[width=0.48\textwidth, angle=0]{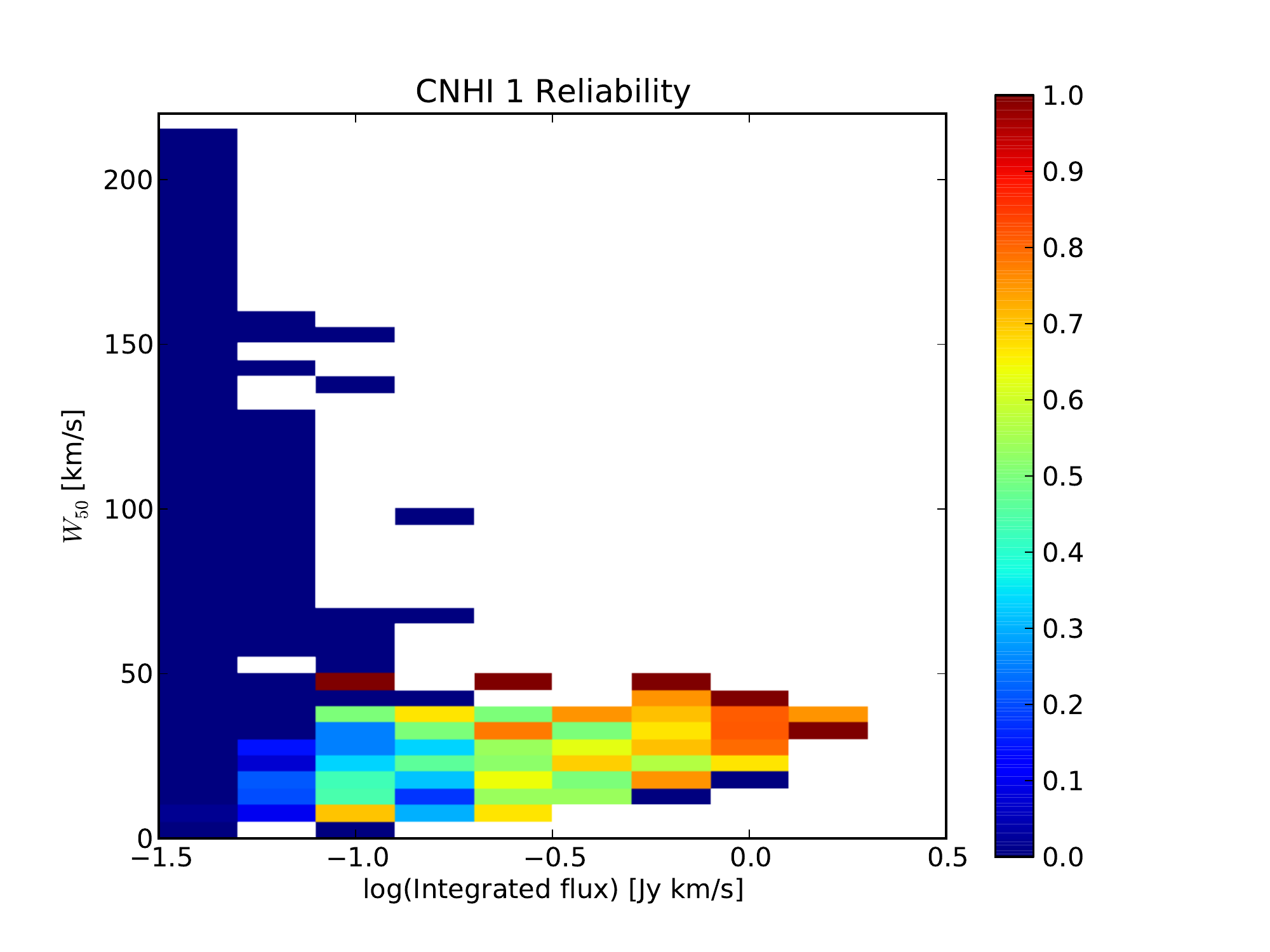}
\includegraphics[width=0.48\textwidth, angle=0]{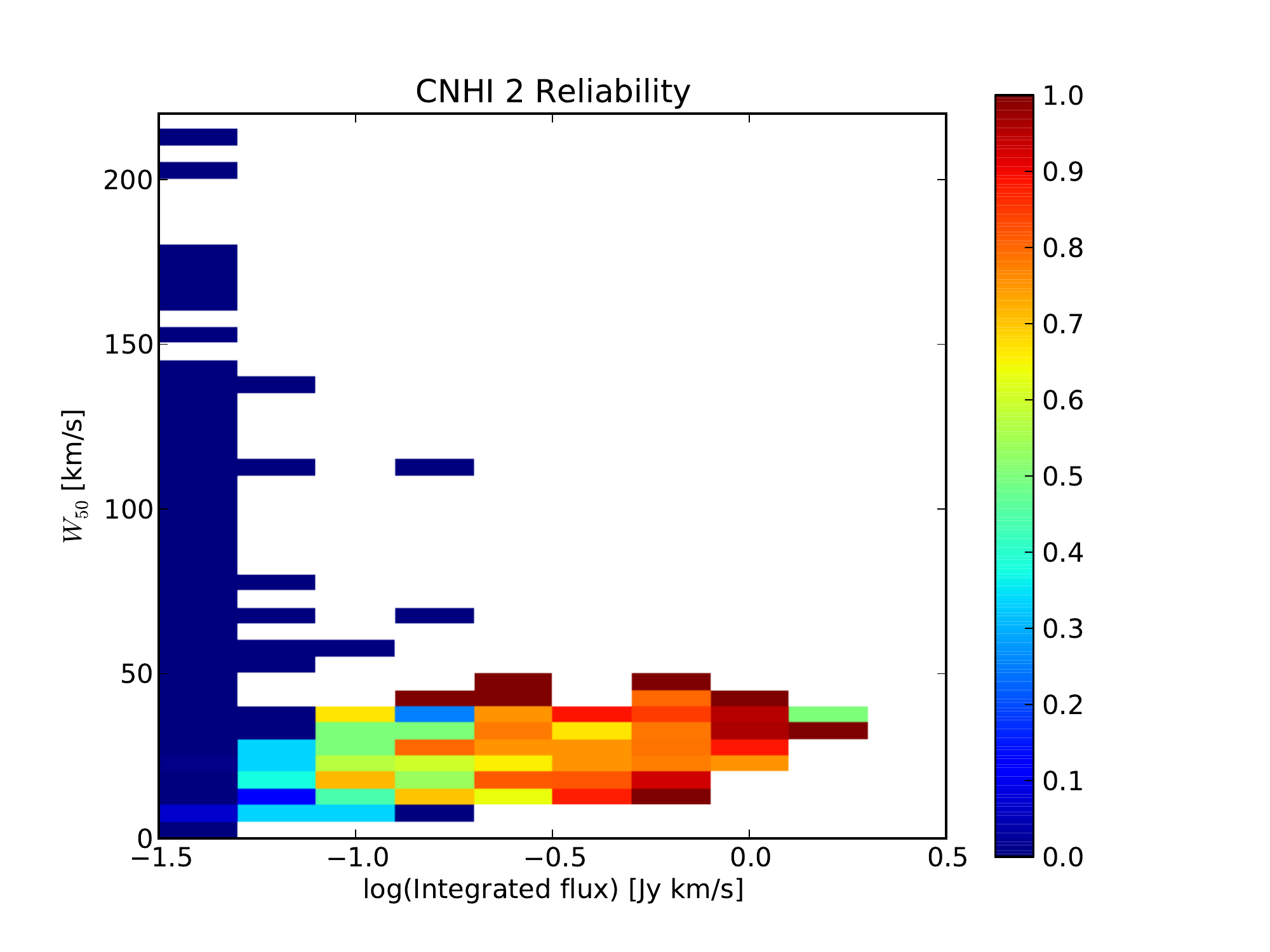}

\end{center}
\end{figure*}

\begin{figure*}[t]
\begin{center}

\includegraphics[width=0.48\textwidth, angle=0]{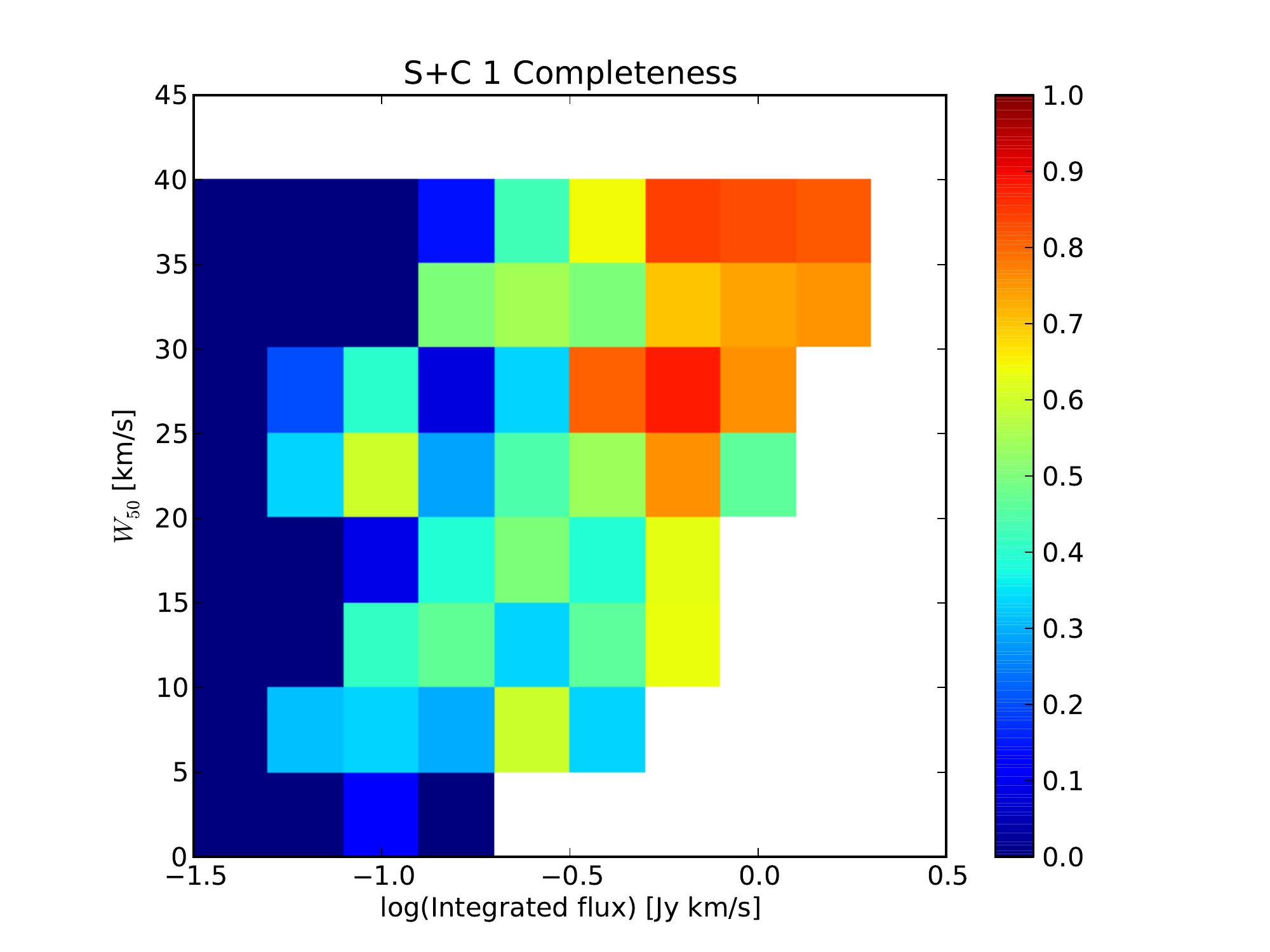}
\includegraphics[width=0.48\textwidth, angle=0]{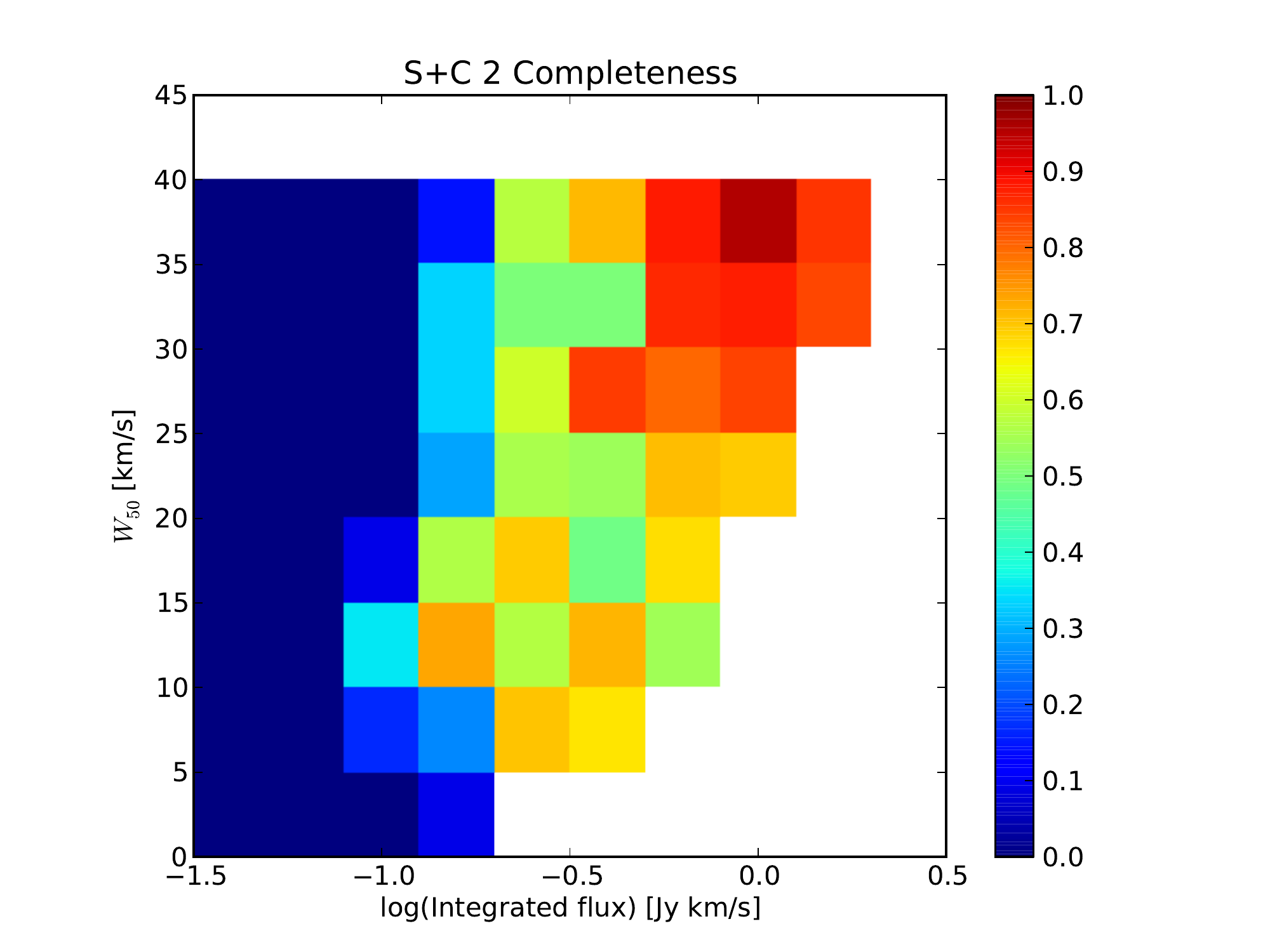}
\includegraphics[width=0.48\textwidth, angle=0]{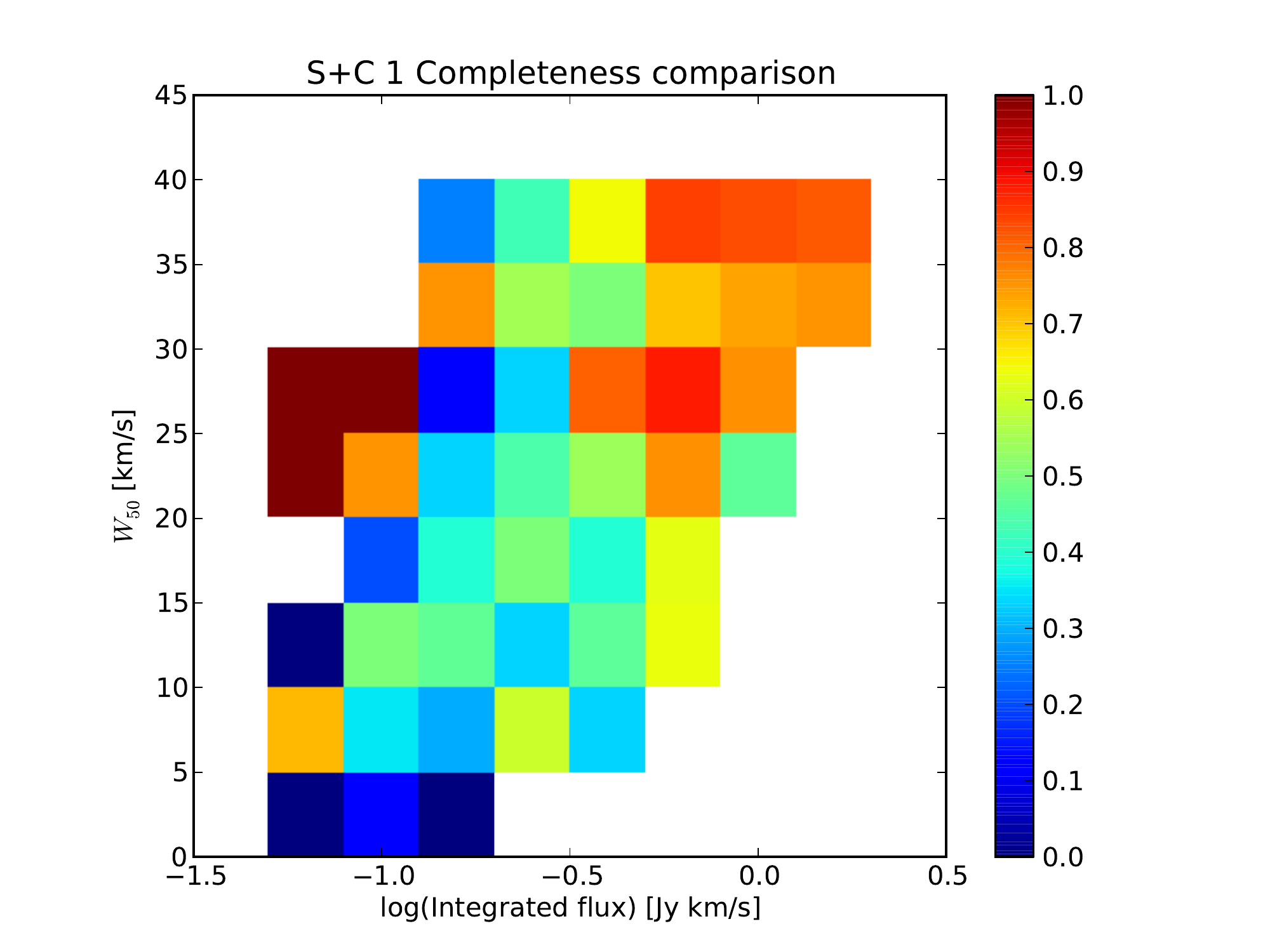}
\includegraphics[width=0.48\textwidth, angle=0]{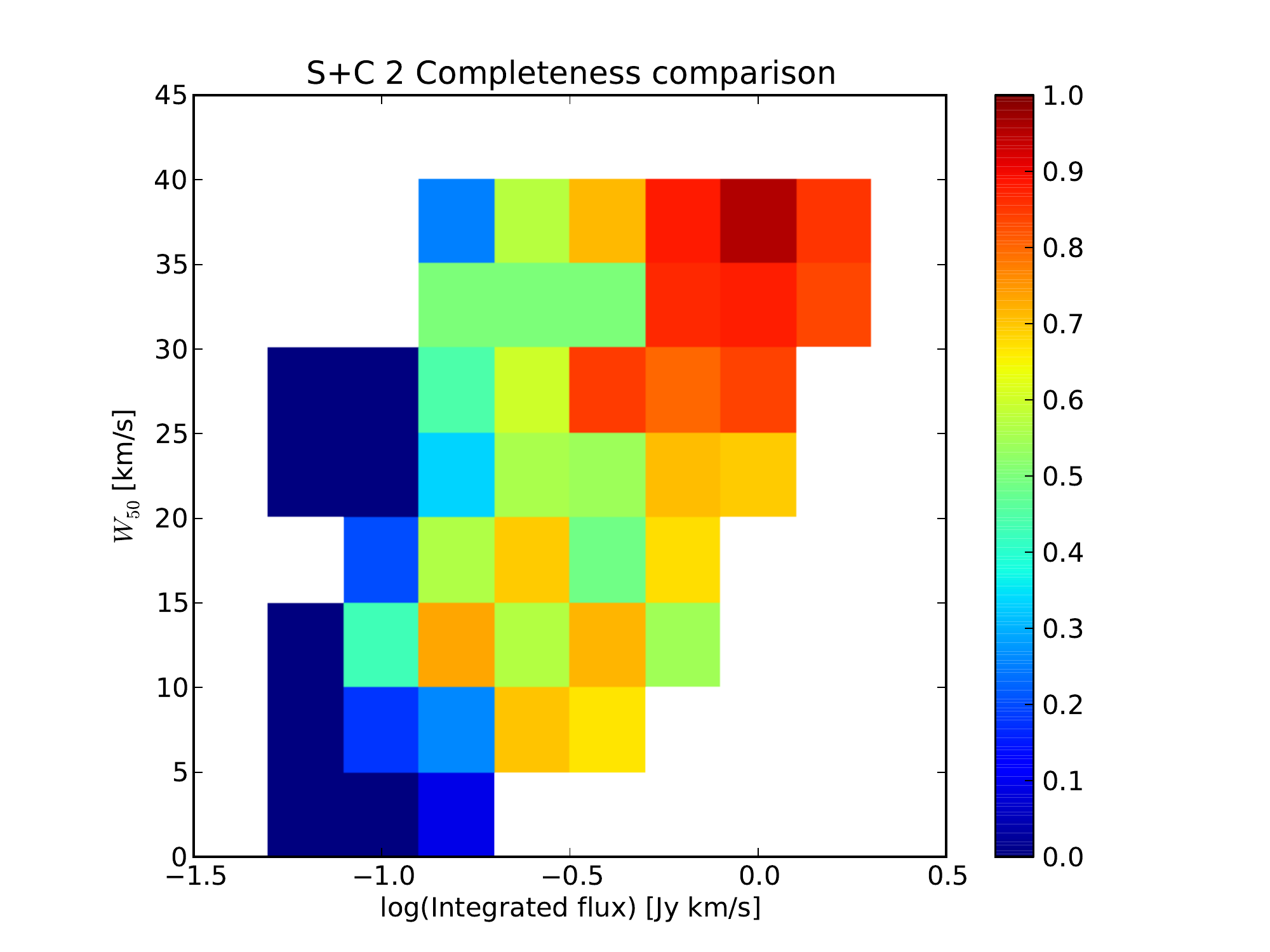}
\includegraphics[width=0.48\textwidth, angle=0]{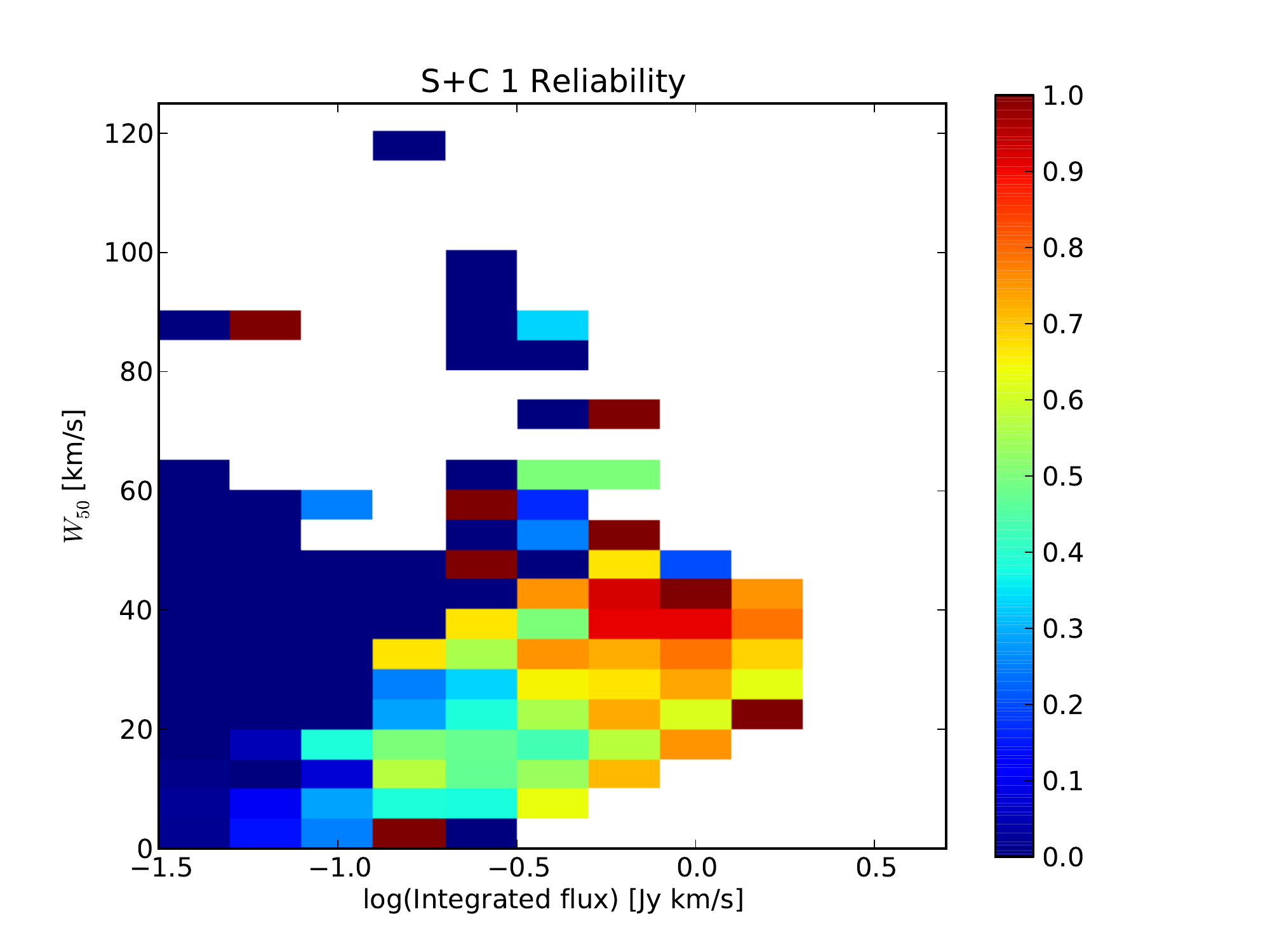}
\includegraphics[width=0.48\textwidth, angle=0]{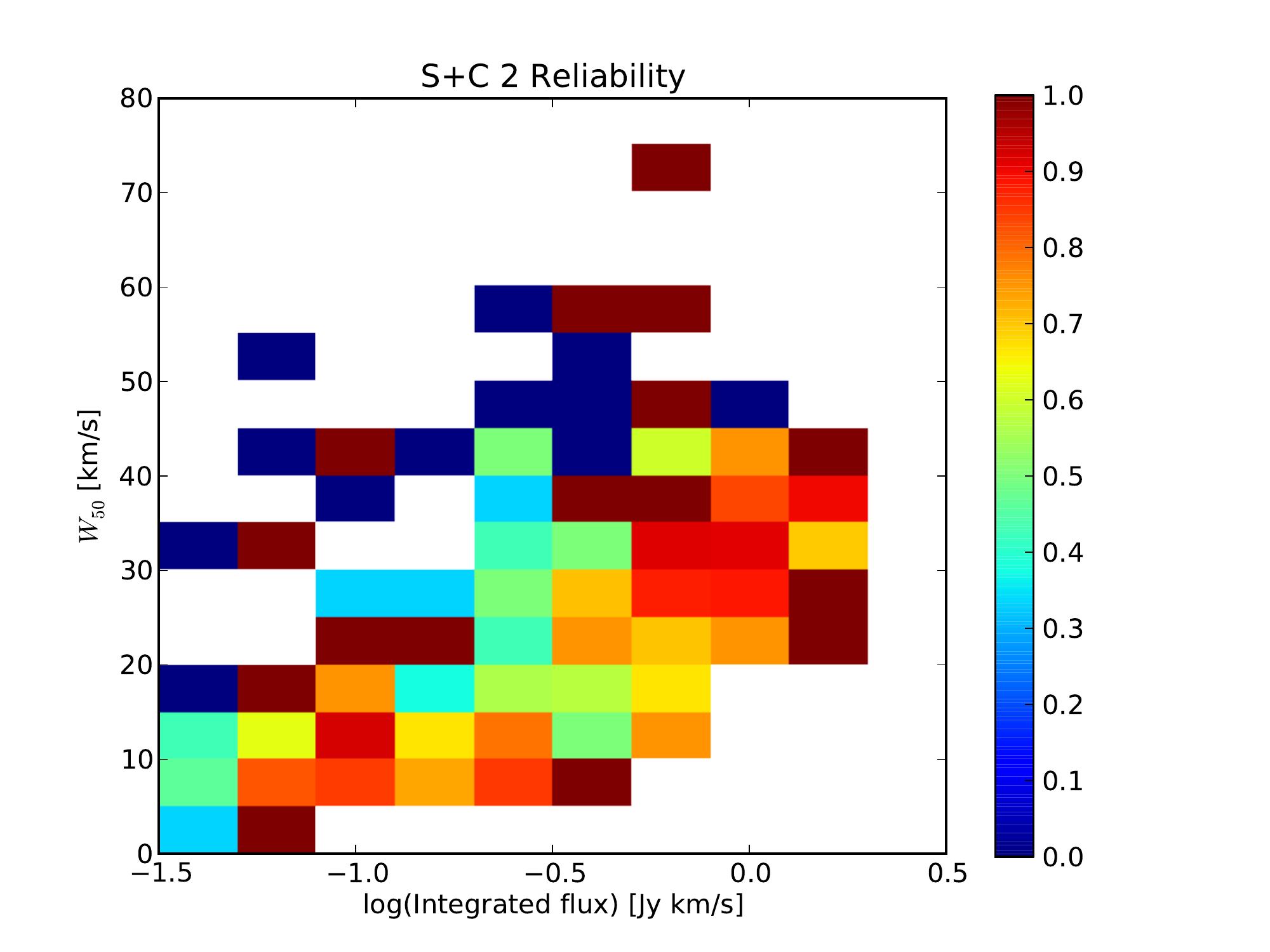}

\end{center}
\end{figure*}

\begin{figure*}[t]
\begin{center}

\includegraphics[width=0.48\textwidth, angle=0]{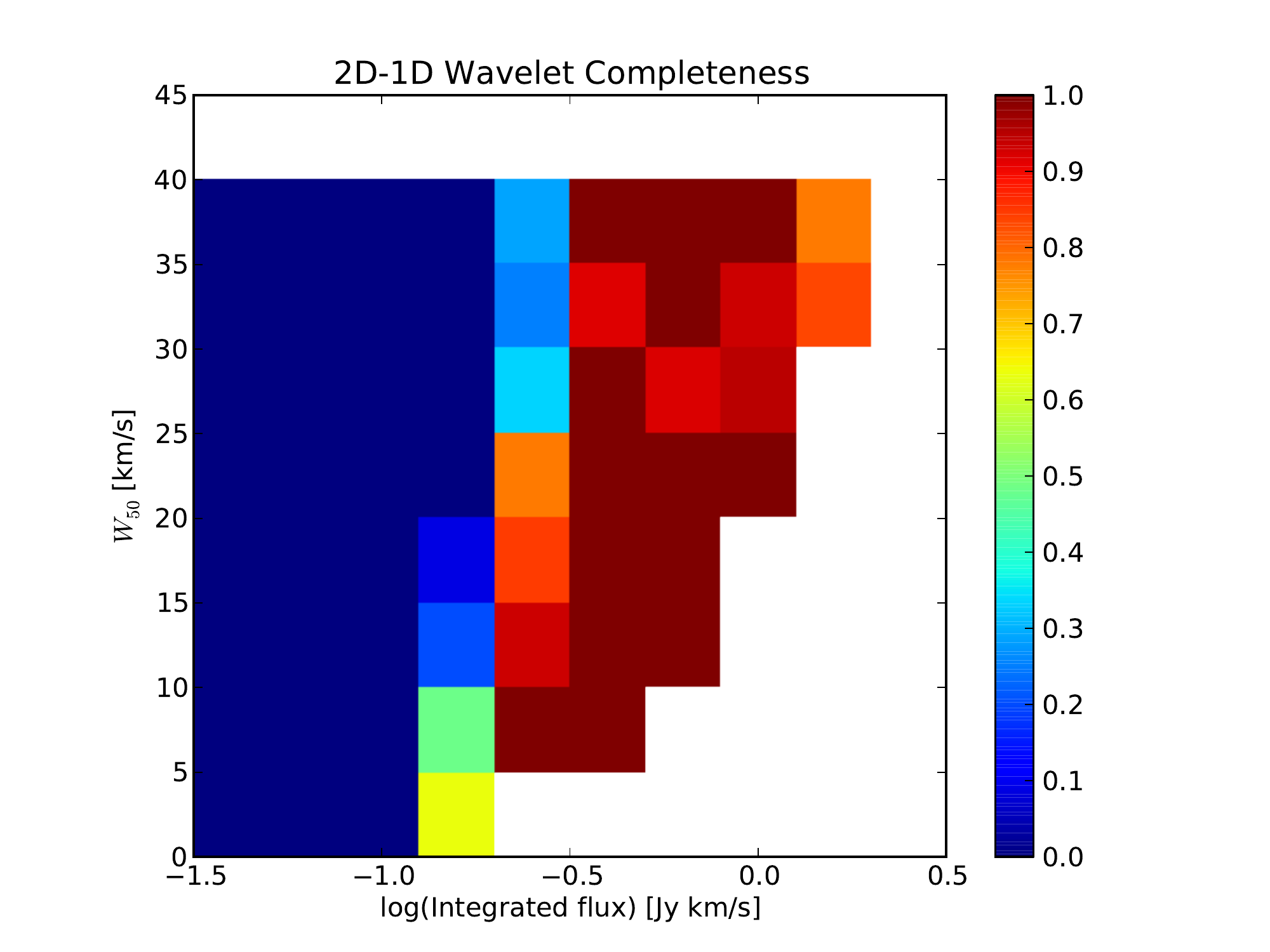}
\\
\includegraphics[width=0.48\textwidth, angle=0]{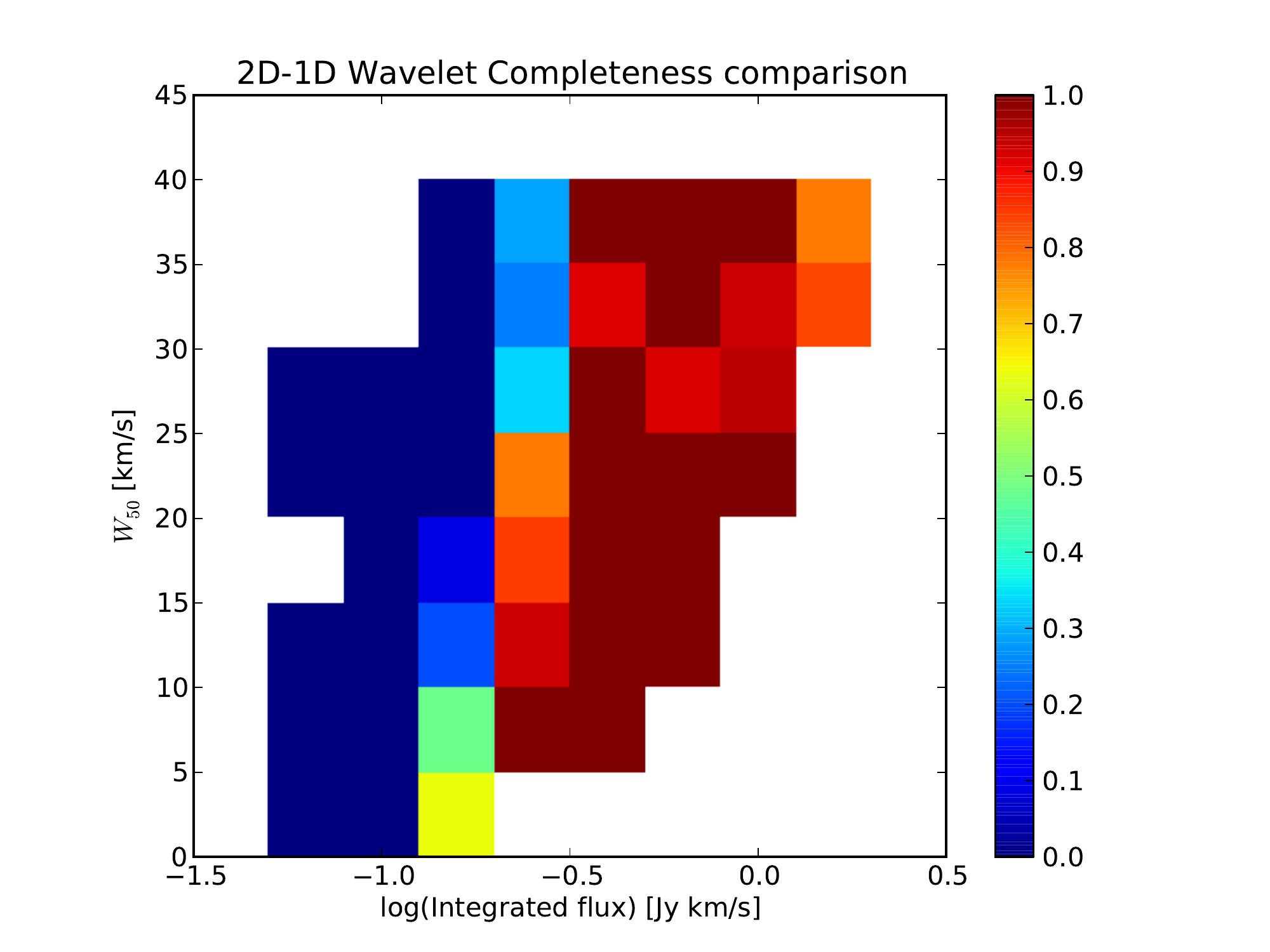}
\\
\includegraphics[width=0.48\textwidth, angle=0]{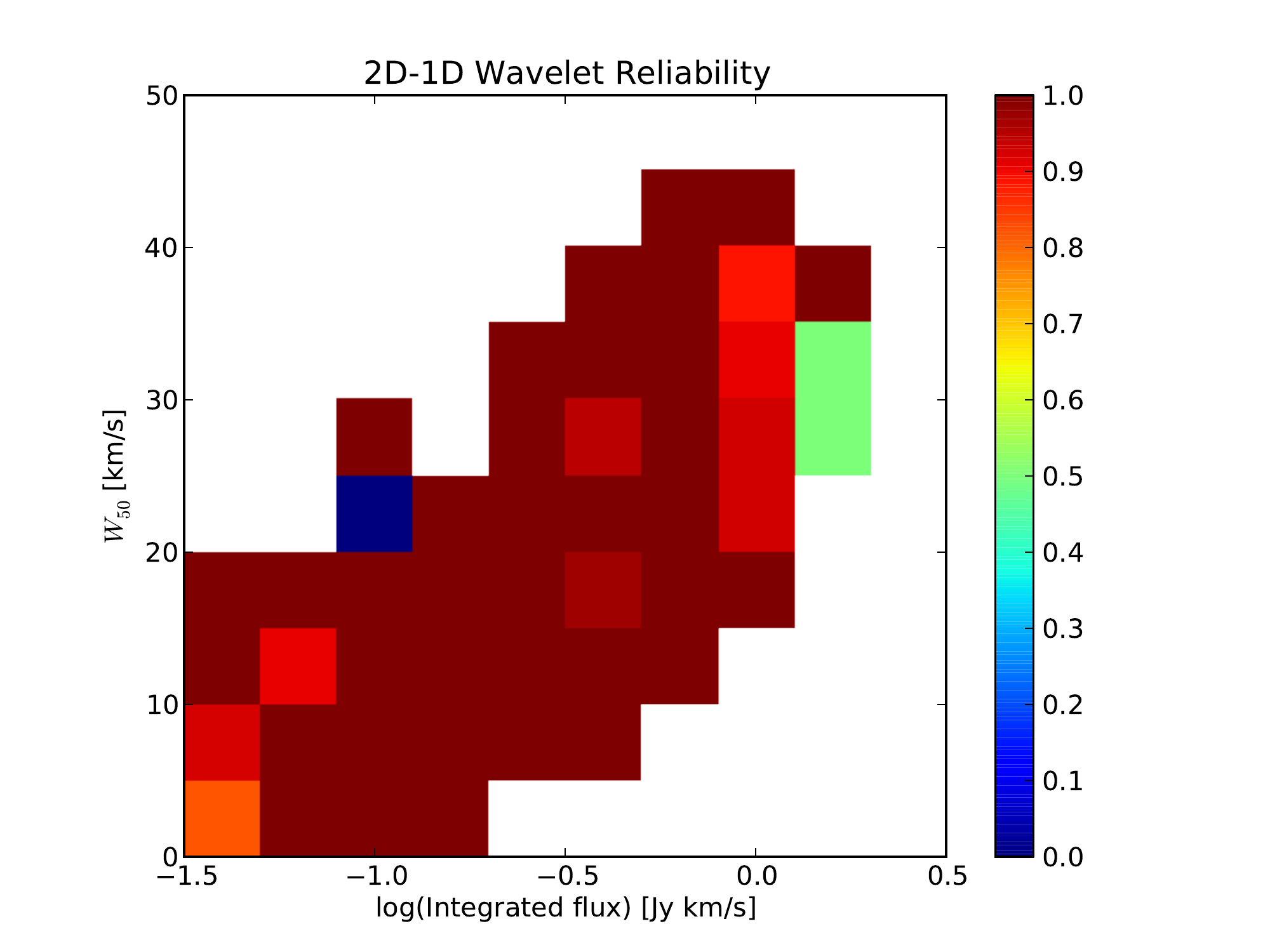}
\\

\end{center}
\end{figure*}

\begin{figure*}[t]
\begin{center}

\includegraphics[width=0.48\textwidth, angle=0]{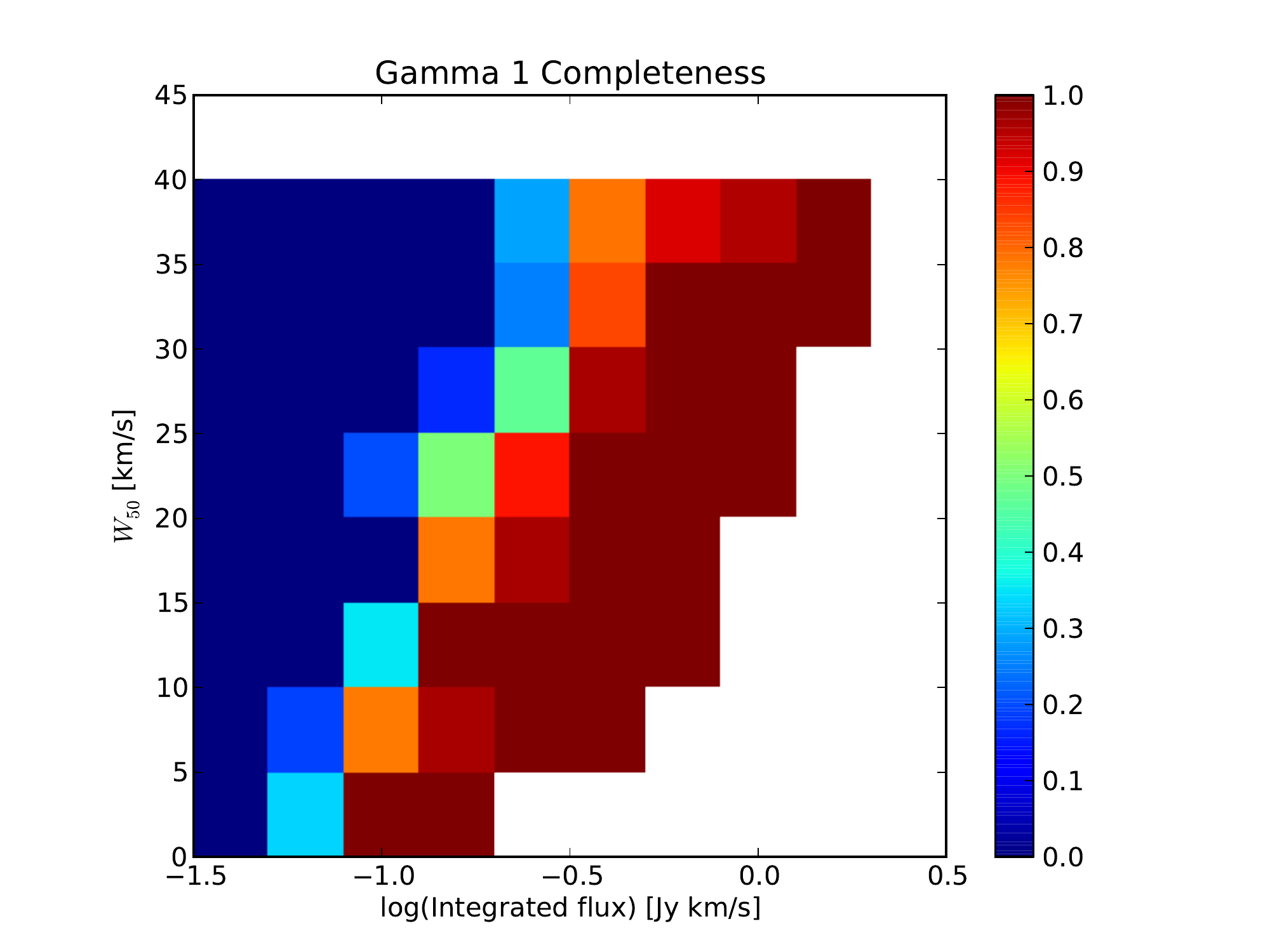}
\includegraphics[width=0.48\textwidth, angle=0]{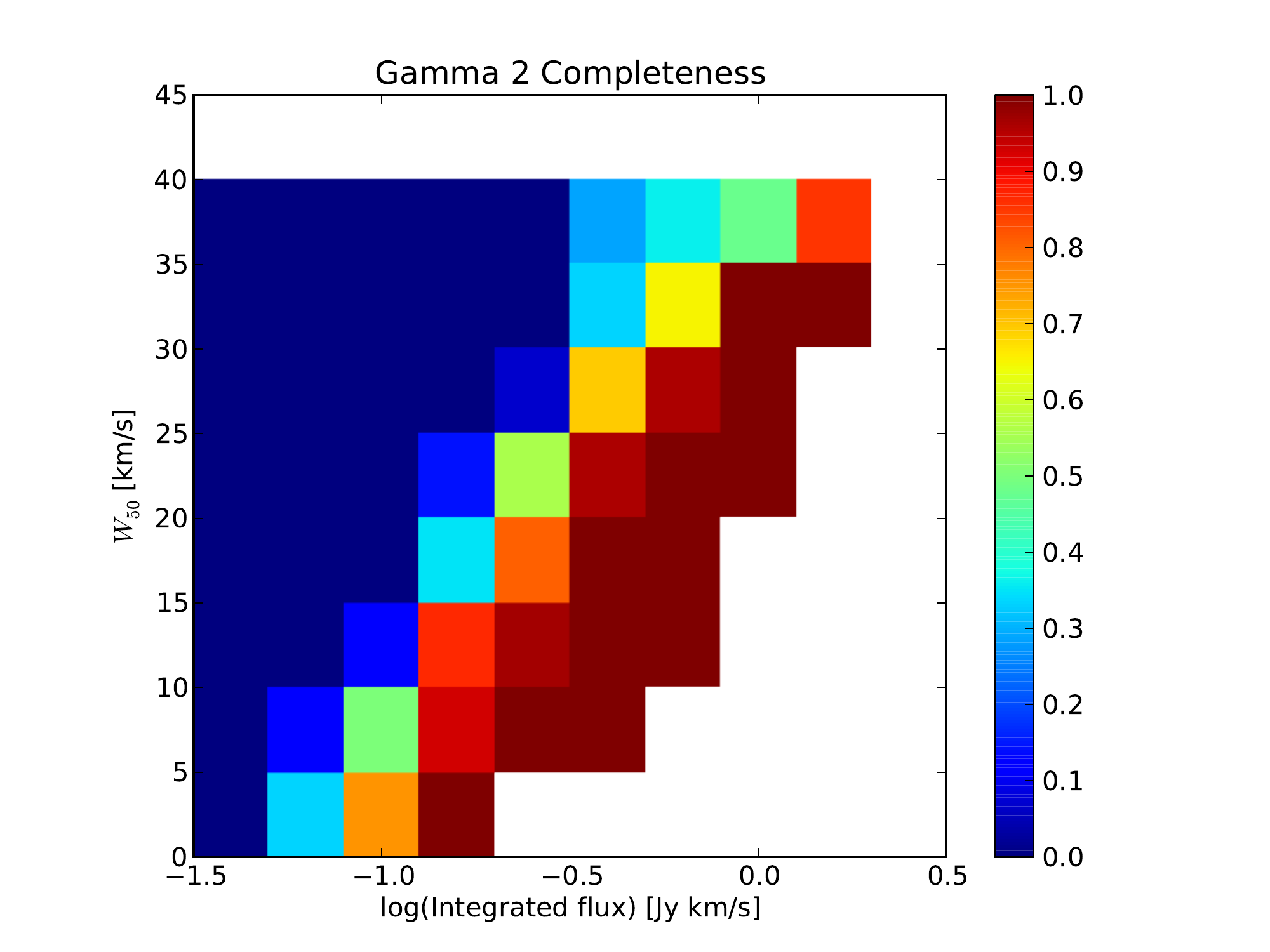}
\includegraphics[width=0.48\textwidth, angle=0]{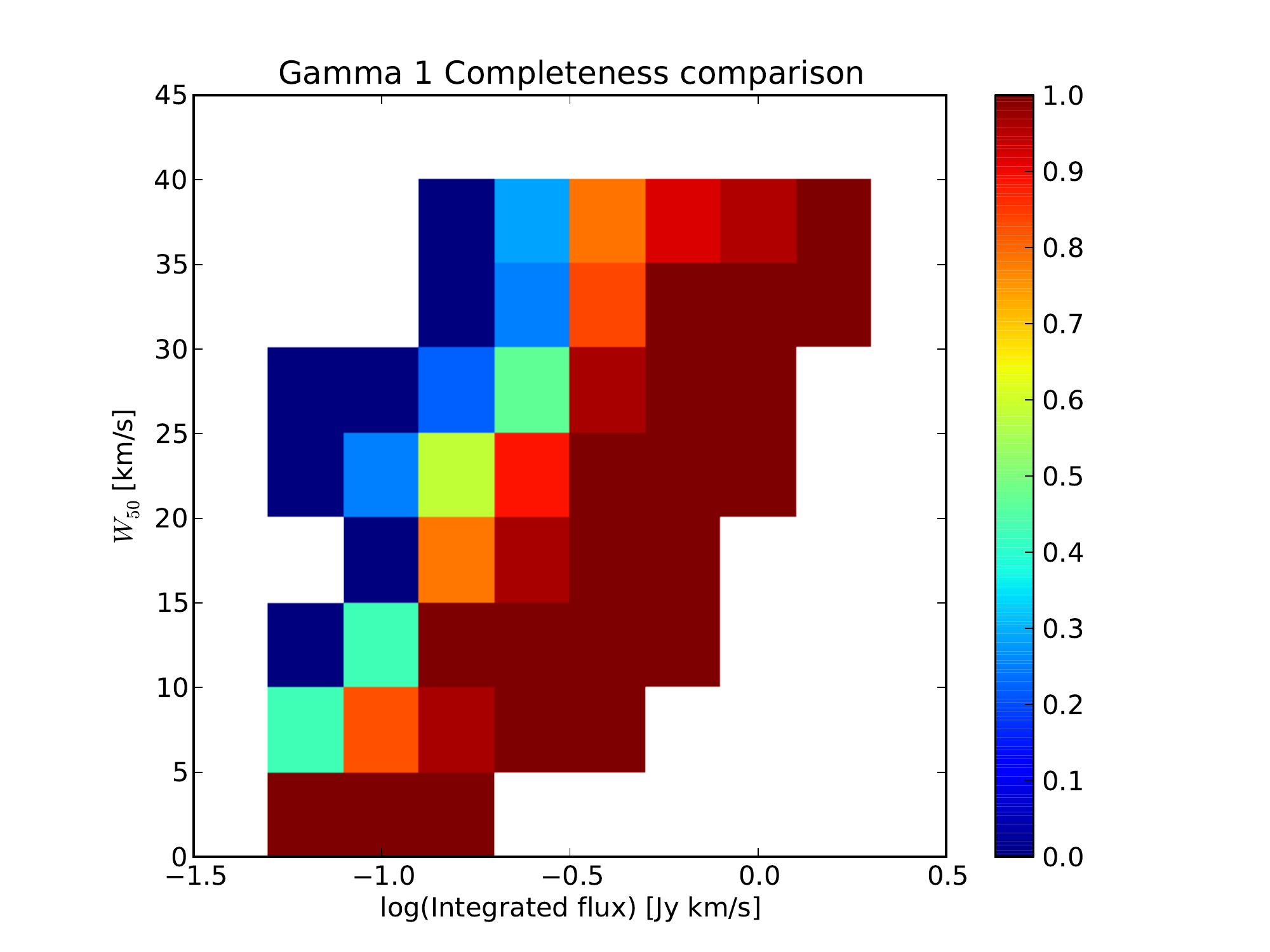}
\includegraphics[width=0.48\textwidth, angle=0]{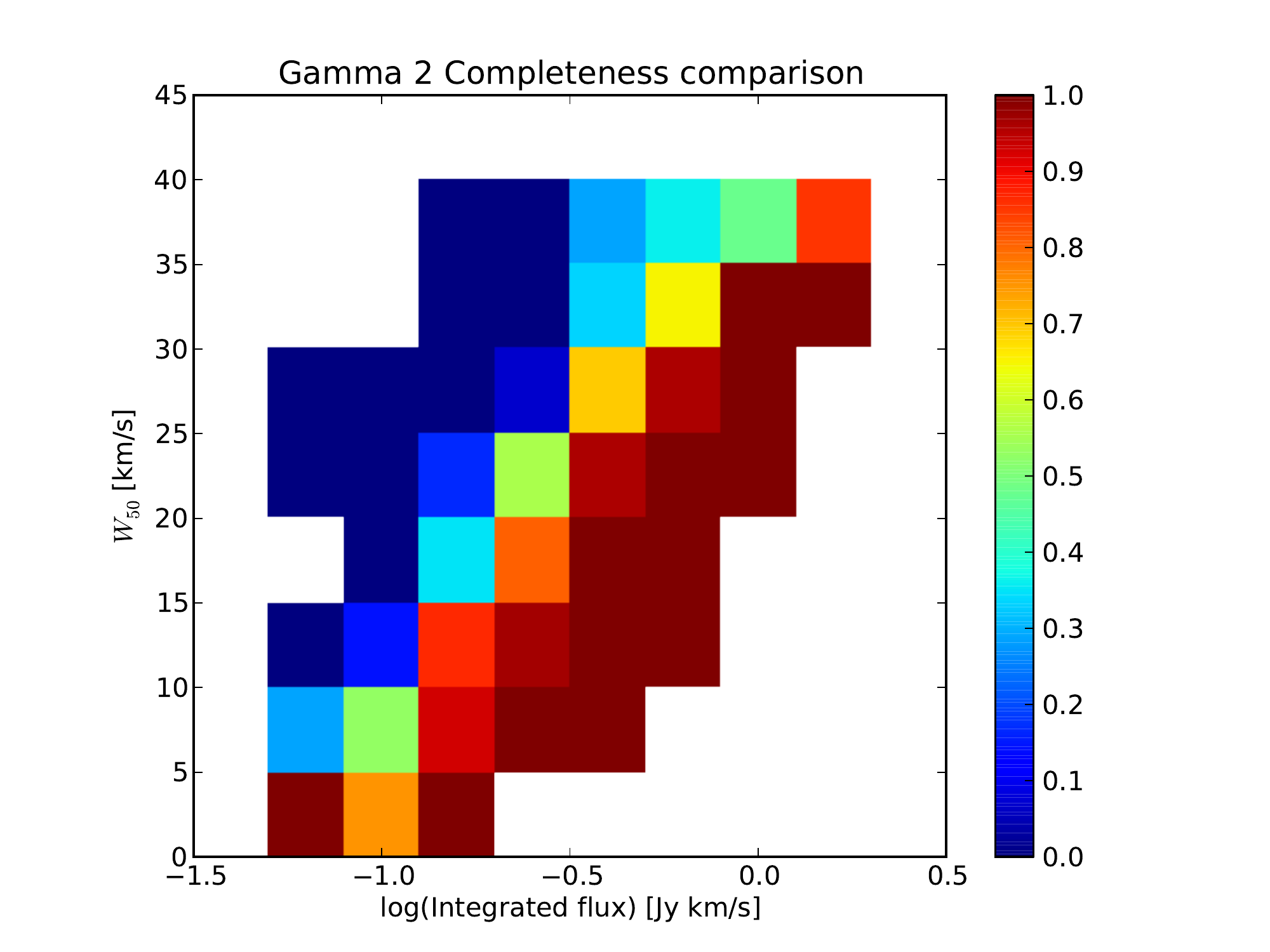}

\end{center}
\end{figure*}


\begin{figure*}[t]
\begin{center}

\includegraphics[width=0.48\textwidth, angle=0]{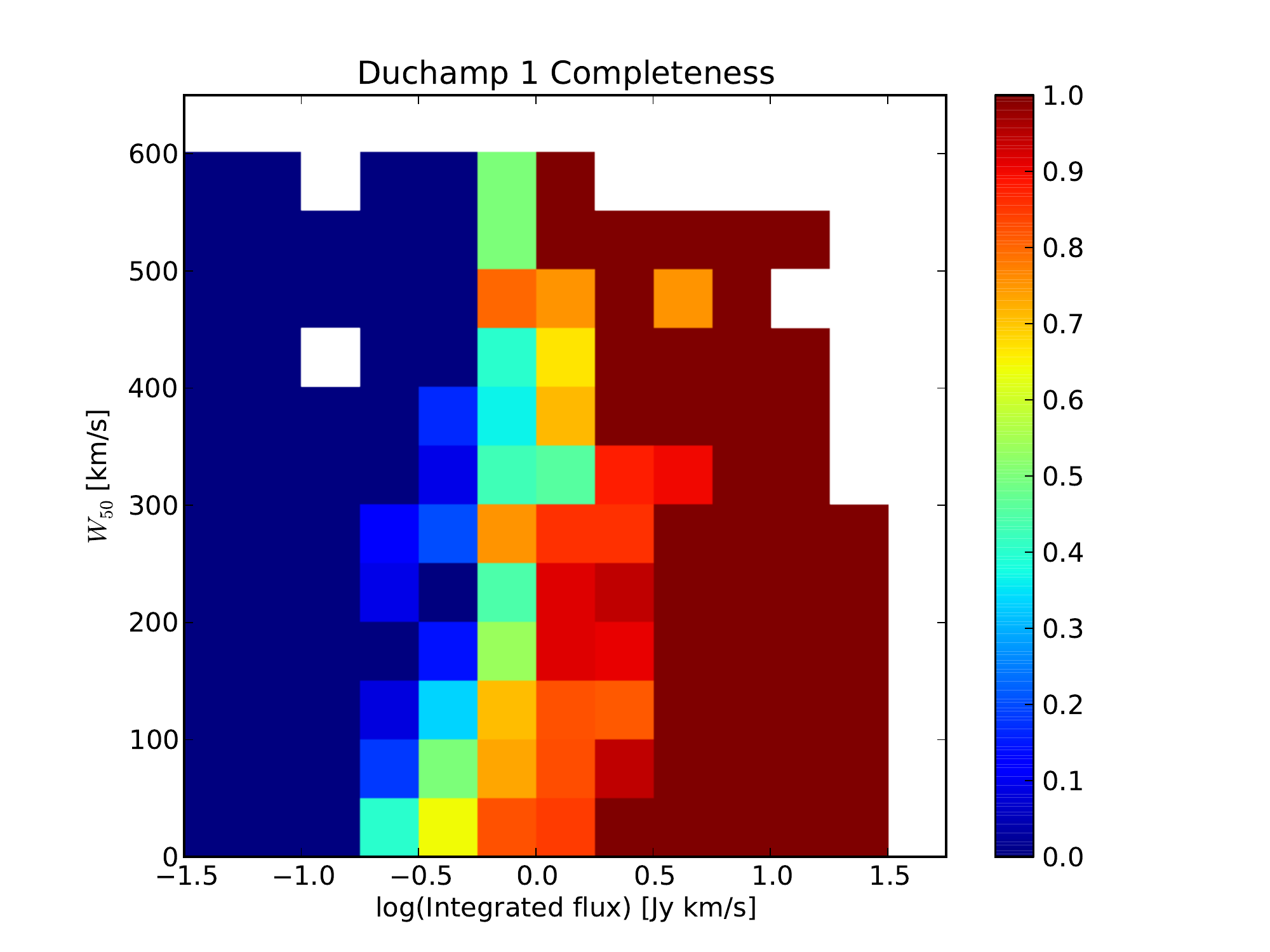}
\includegraphics[width=0.48\textwidth, angle=0]{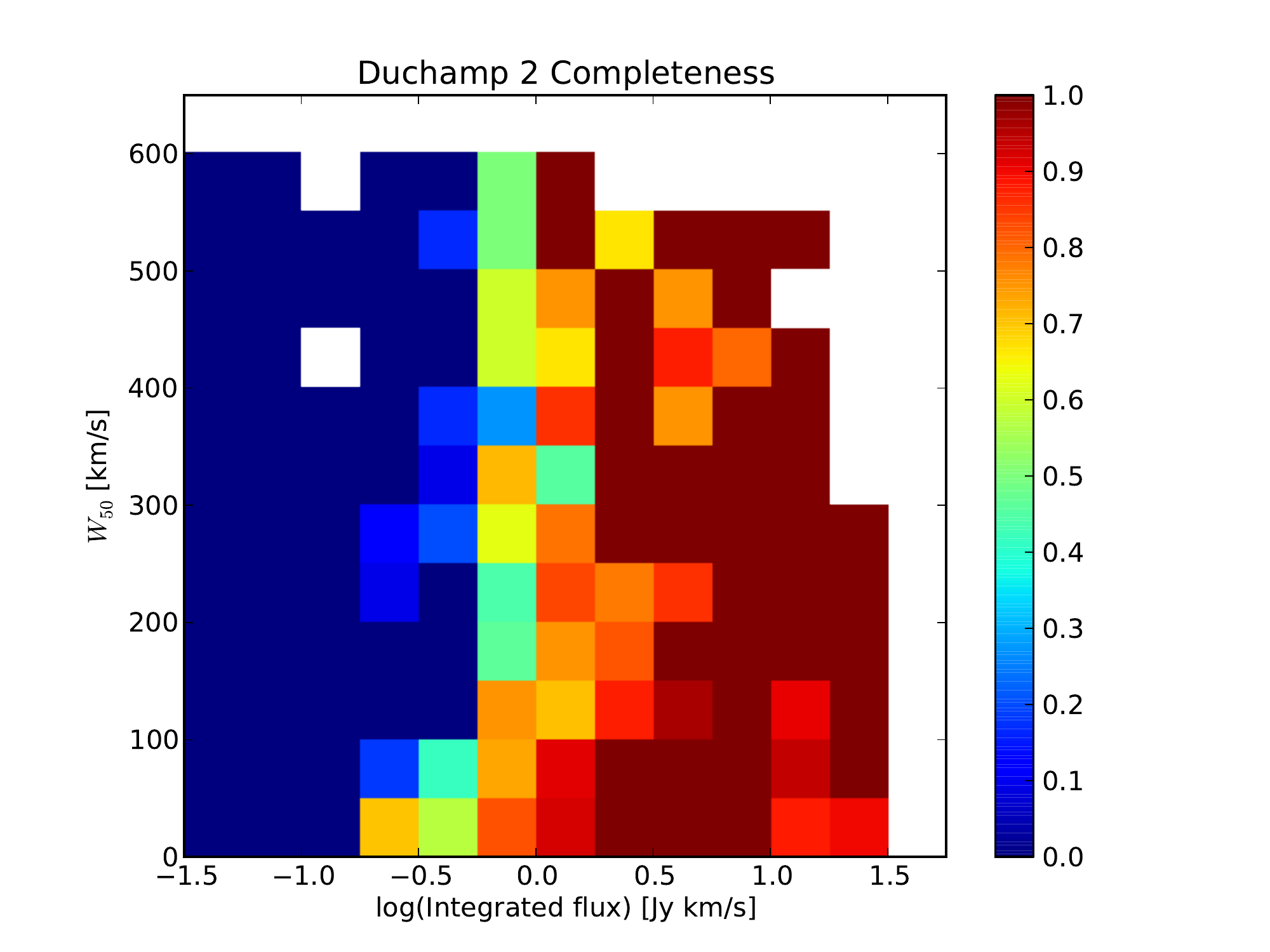}
\includegraphics[width=0.48\textwidth, angle=0]{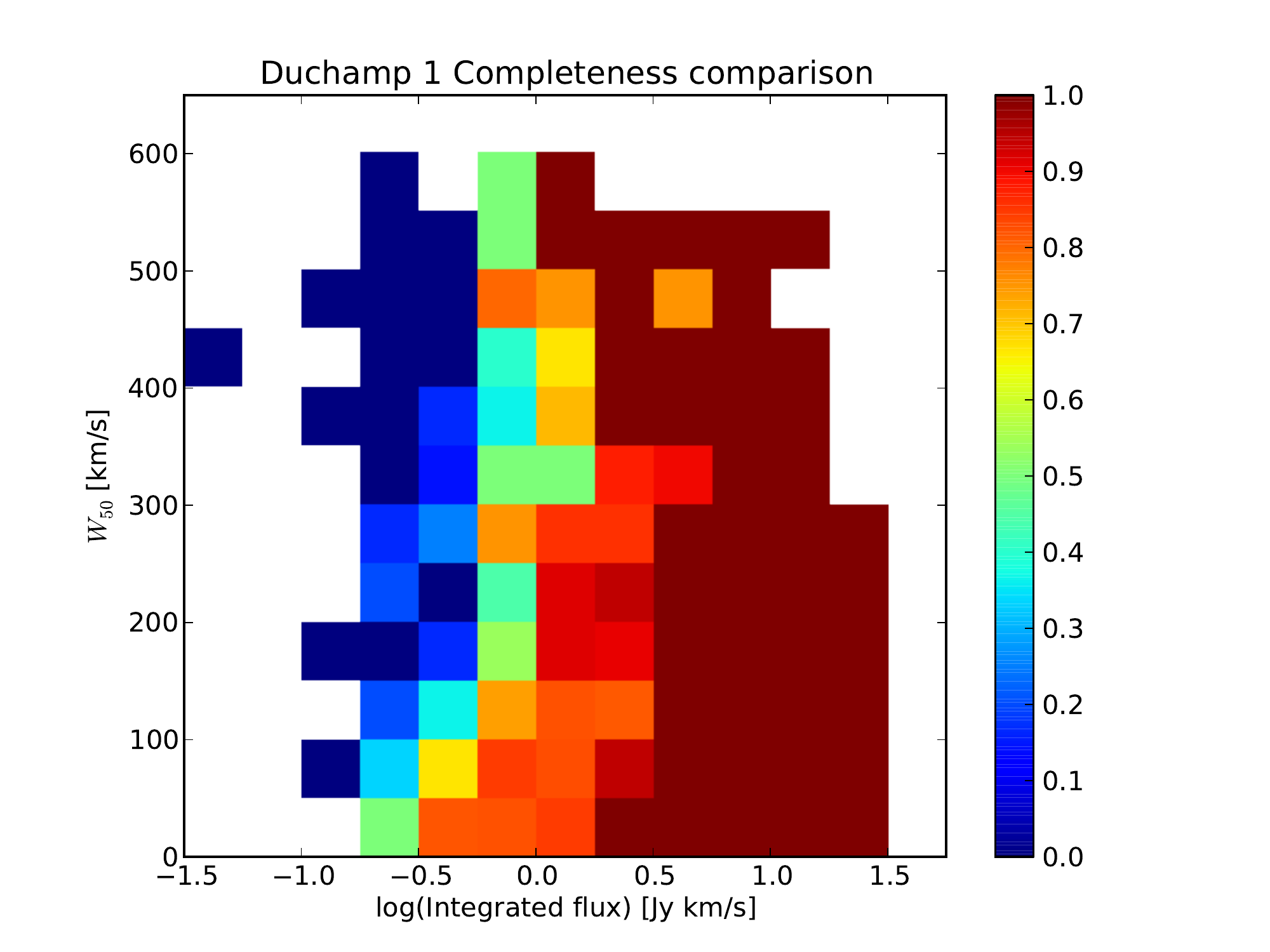}
\includegraphics[width=0.48\textwidth, angle=0]{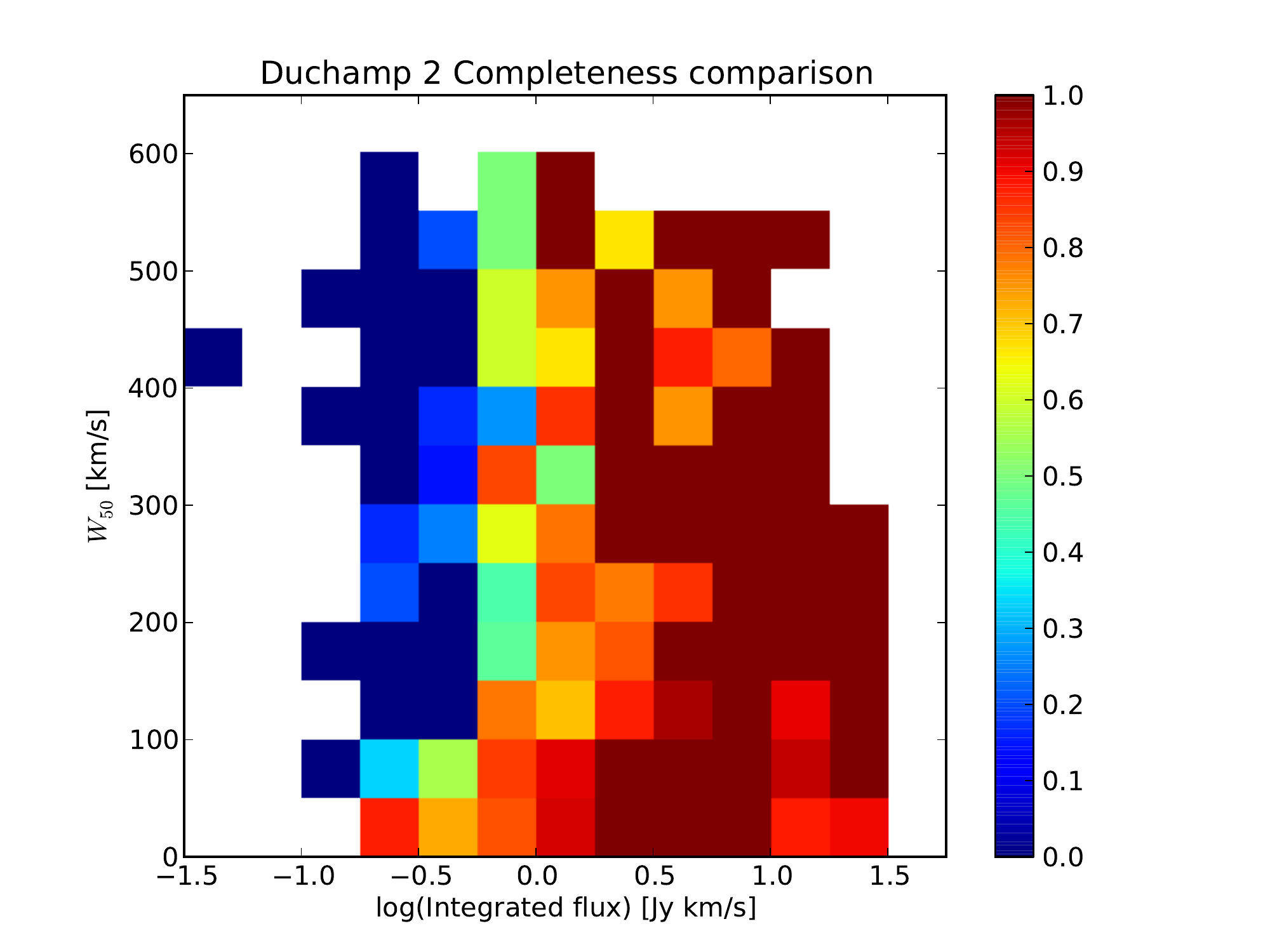}
\includegraphics[width=0.48\textwidth, angle=0]{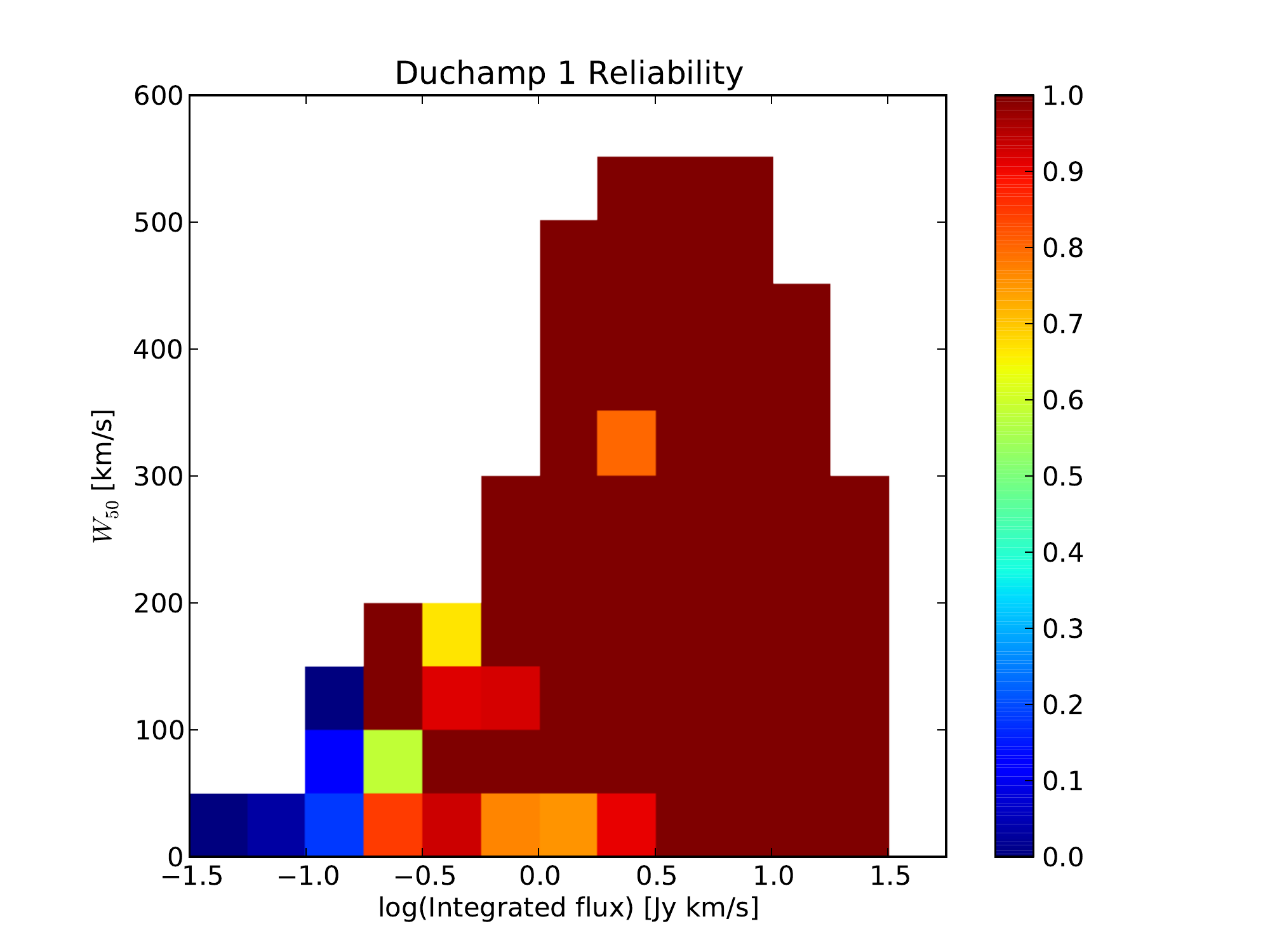}
\includegraphics[width=0.48\textwidth, angle=0]{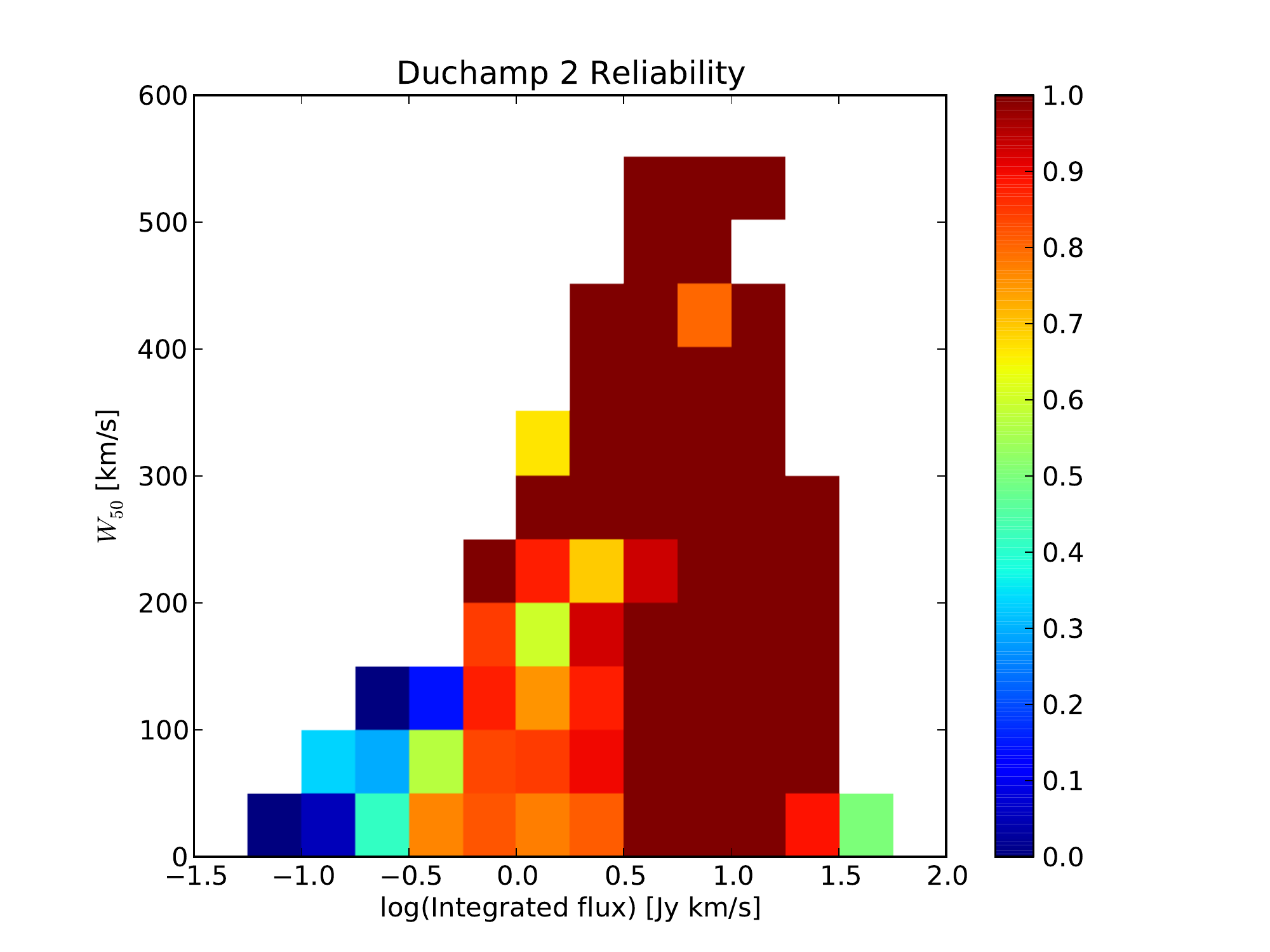}

\caption{Similar as Fig.~\ref{point_2D}, but now completeness and
  reliability is plotted for the model galaxies.}
\label{model_2D}
\end{center}
\end{figure*}

\begin{figure*}[t]
\begin{center}

\includegraphics[width=0.48\textwidth, angle=0]{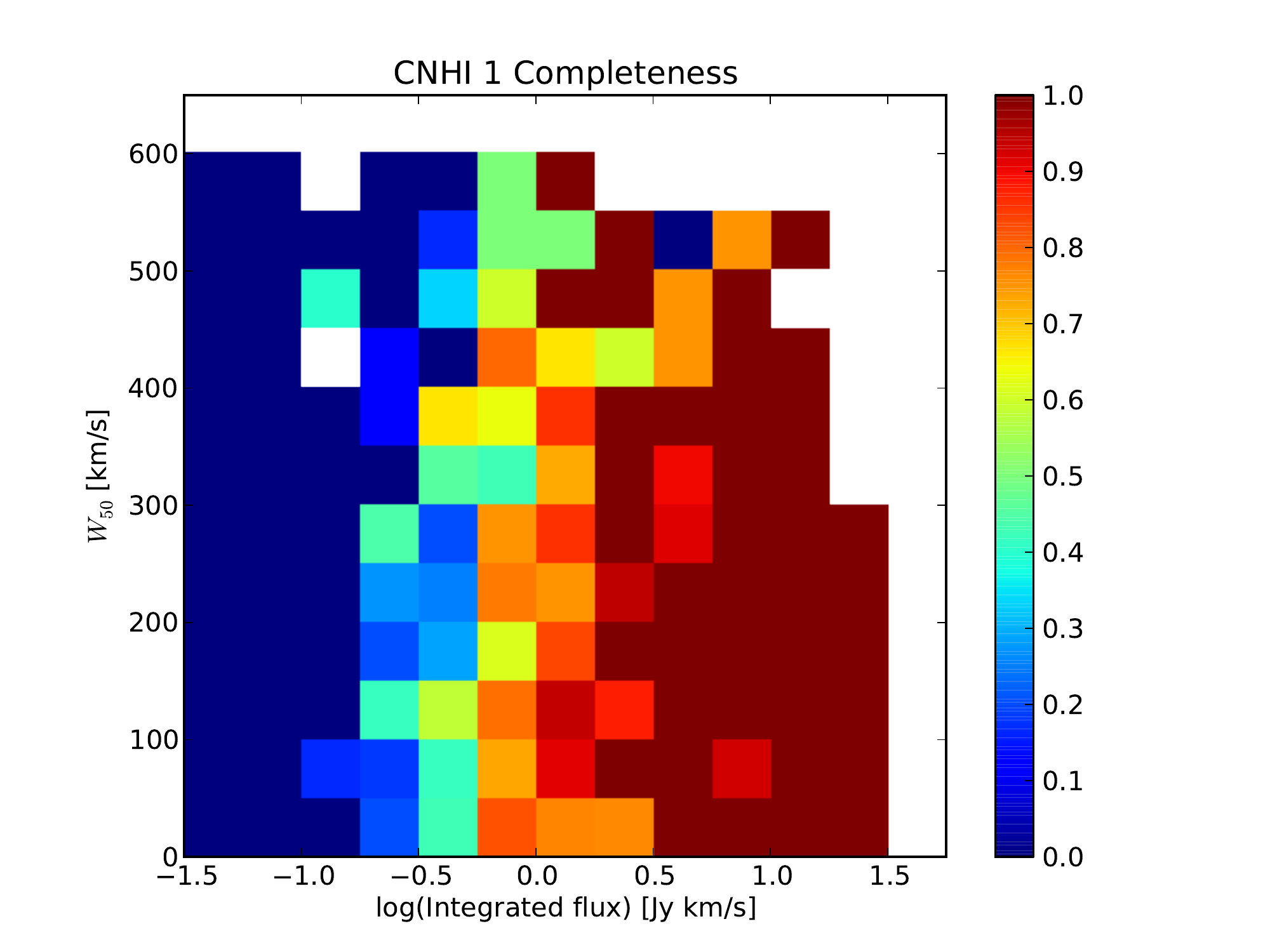}
\includegraphics[width=0.48\textwidth, angle=0]{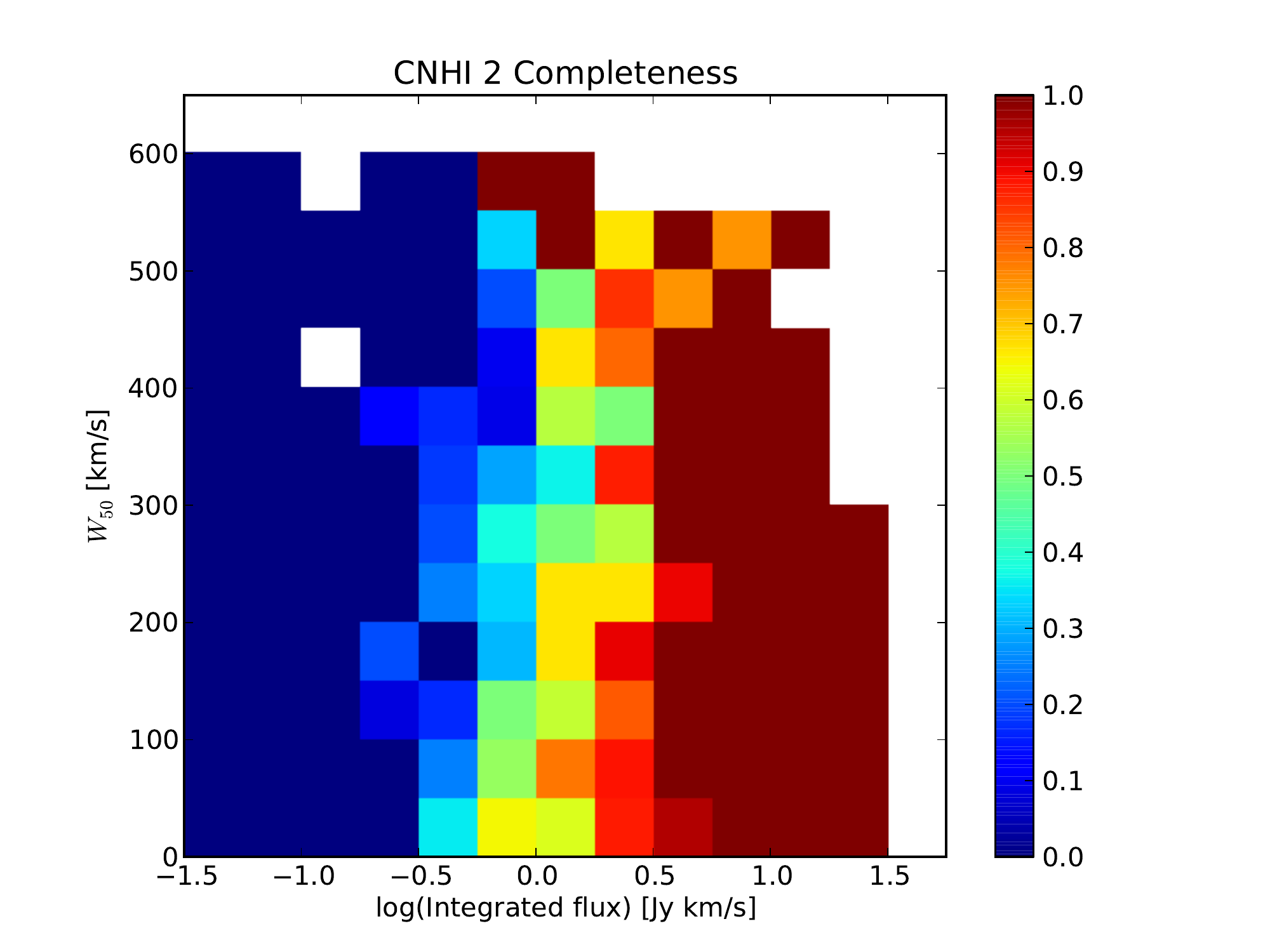}
\includegraphics[width=0.48\textwidth, angle=0]{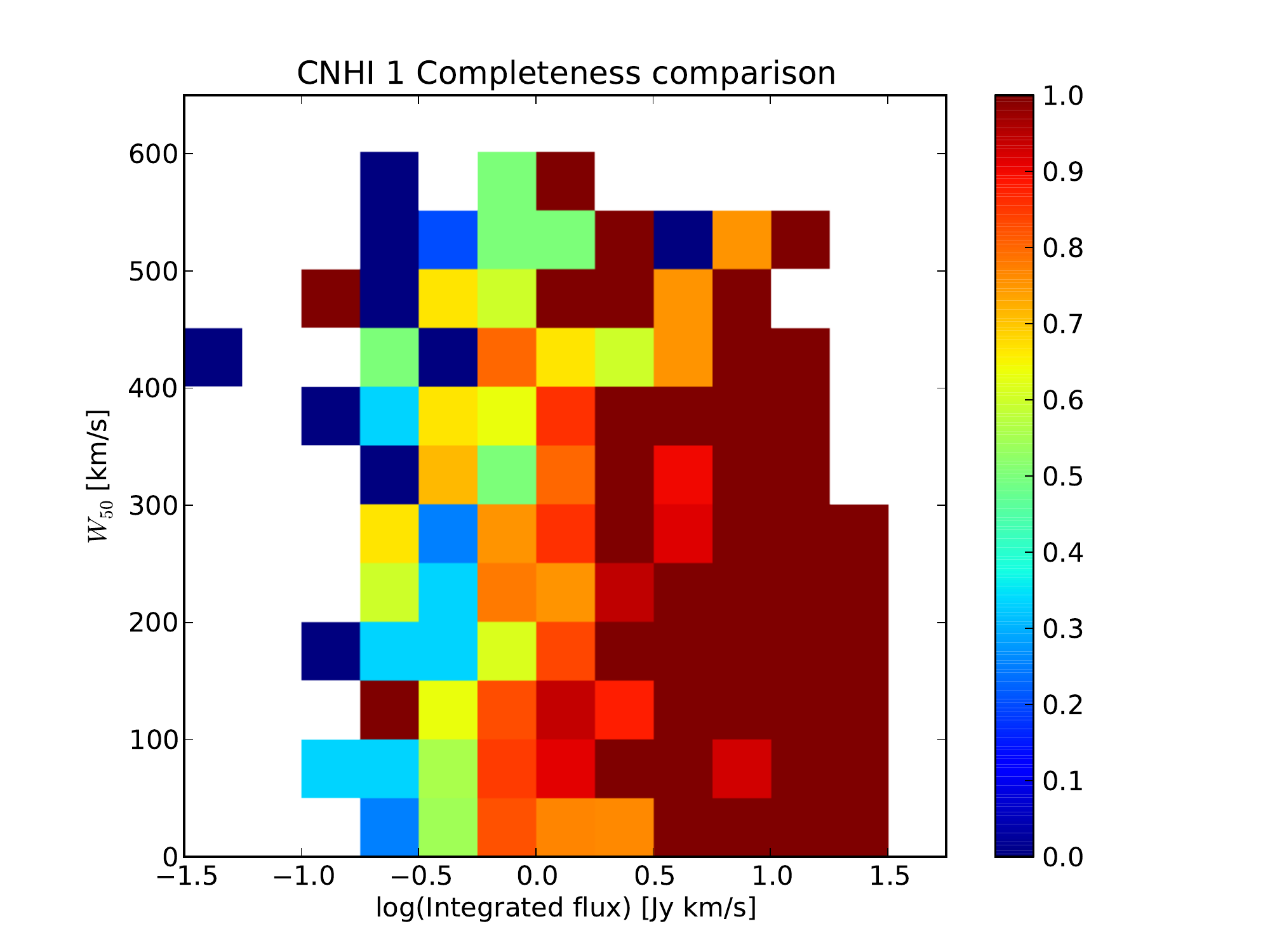}
\includegraphics[width=0.48\textwidth, angle=0]{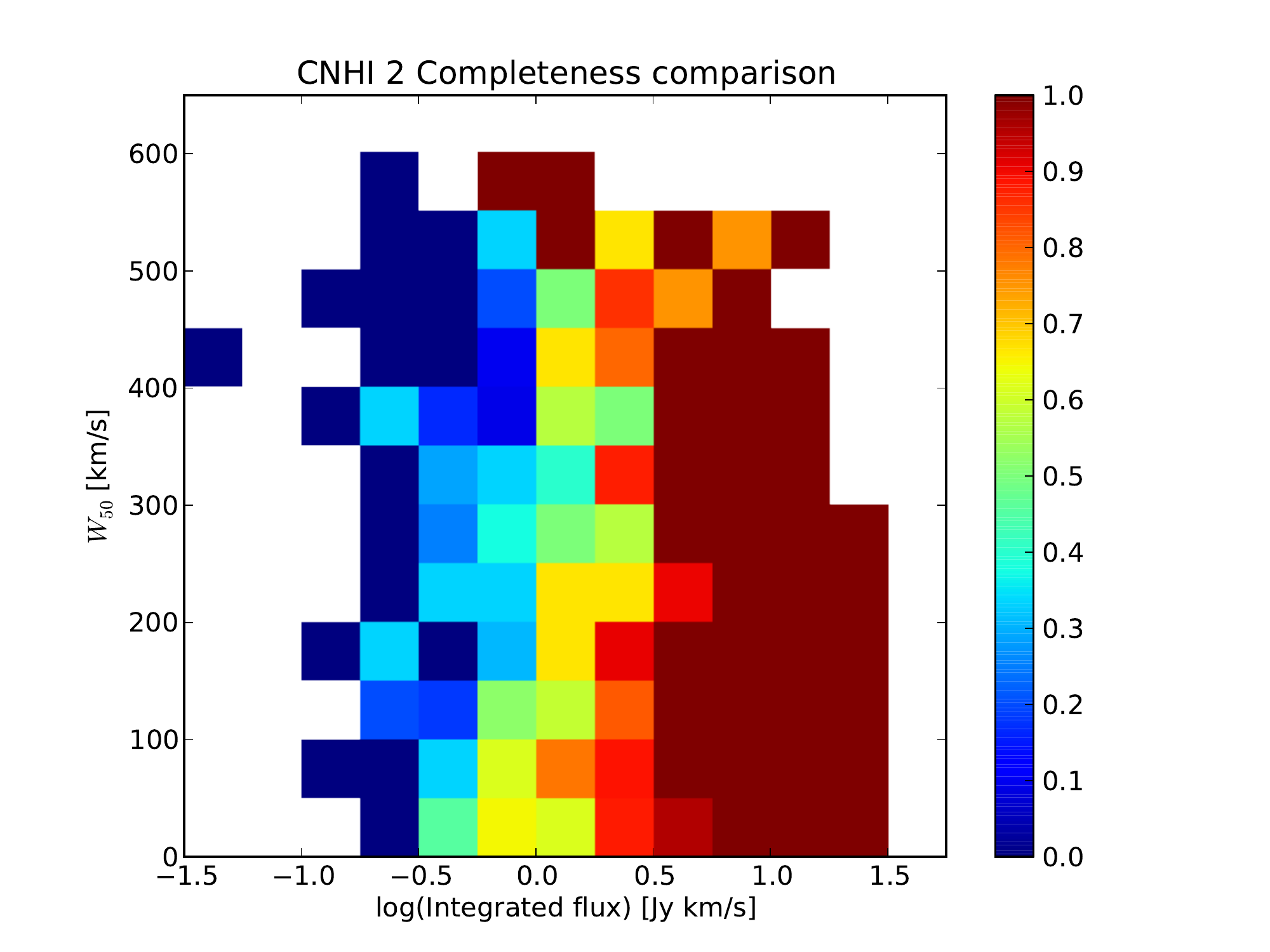}
\includegraphics[width=0.48\textwidth, angle=0]{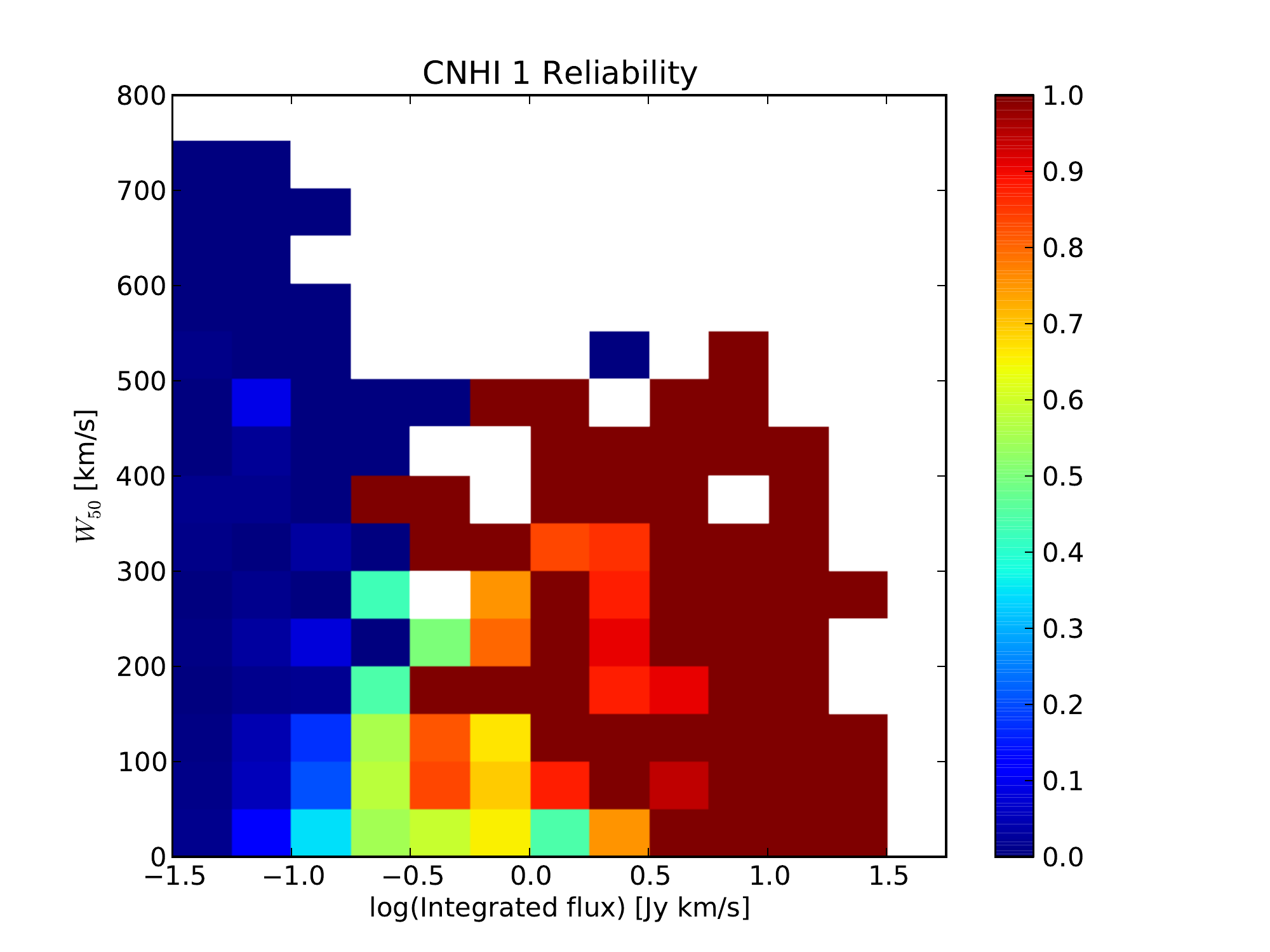}
\includegraphics[width=0.48\textwidth, angle=0]{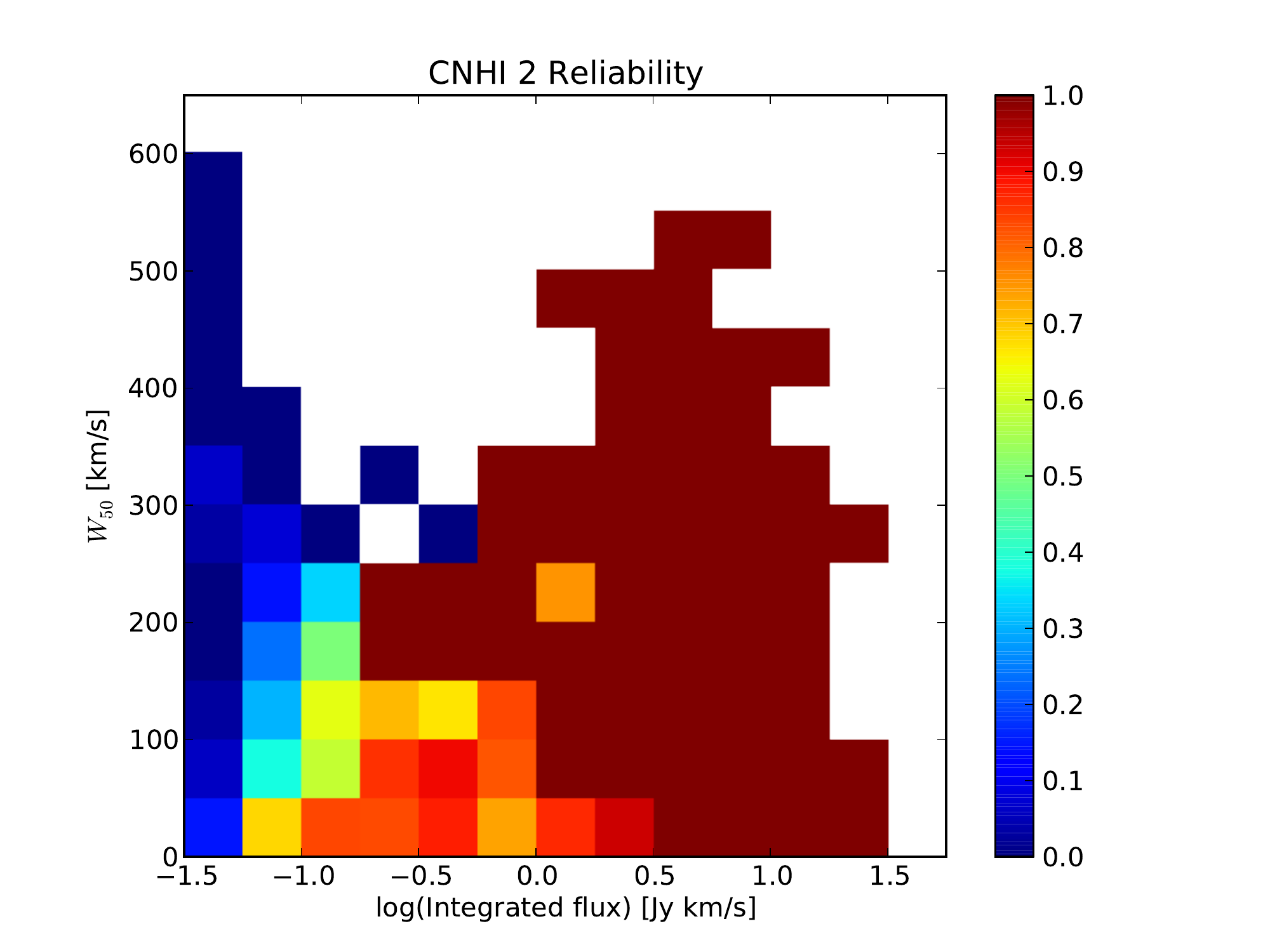}

\end{center}
\end{figure*}

\begin{figure*}[t]
\begin{center}

\includegraphics[width=0.48\textwidth, angle=0]{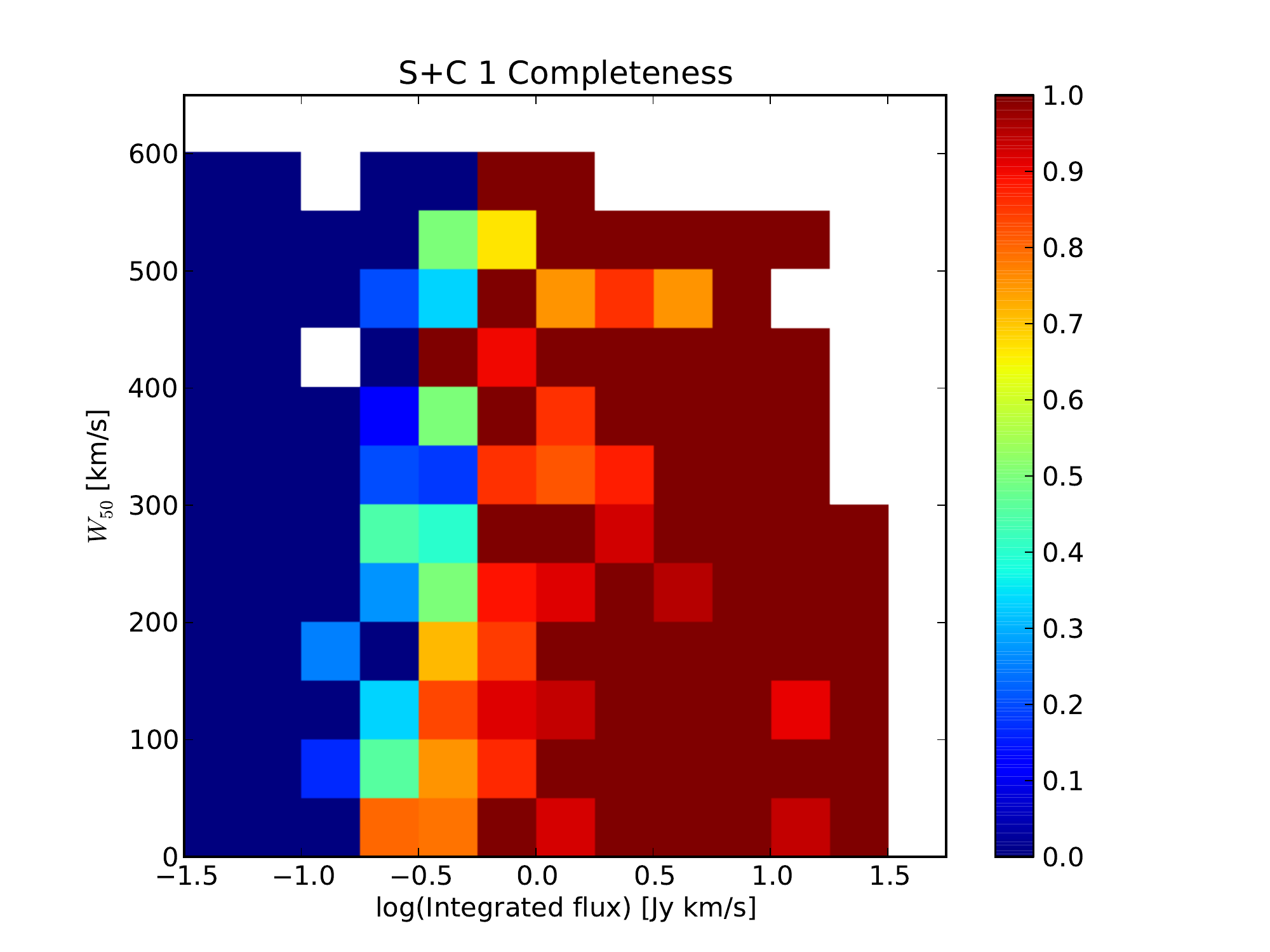}
\includegraphics[width=0.48\textwidth, angle=0]{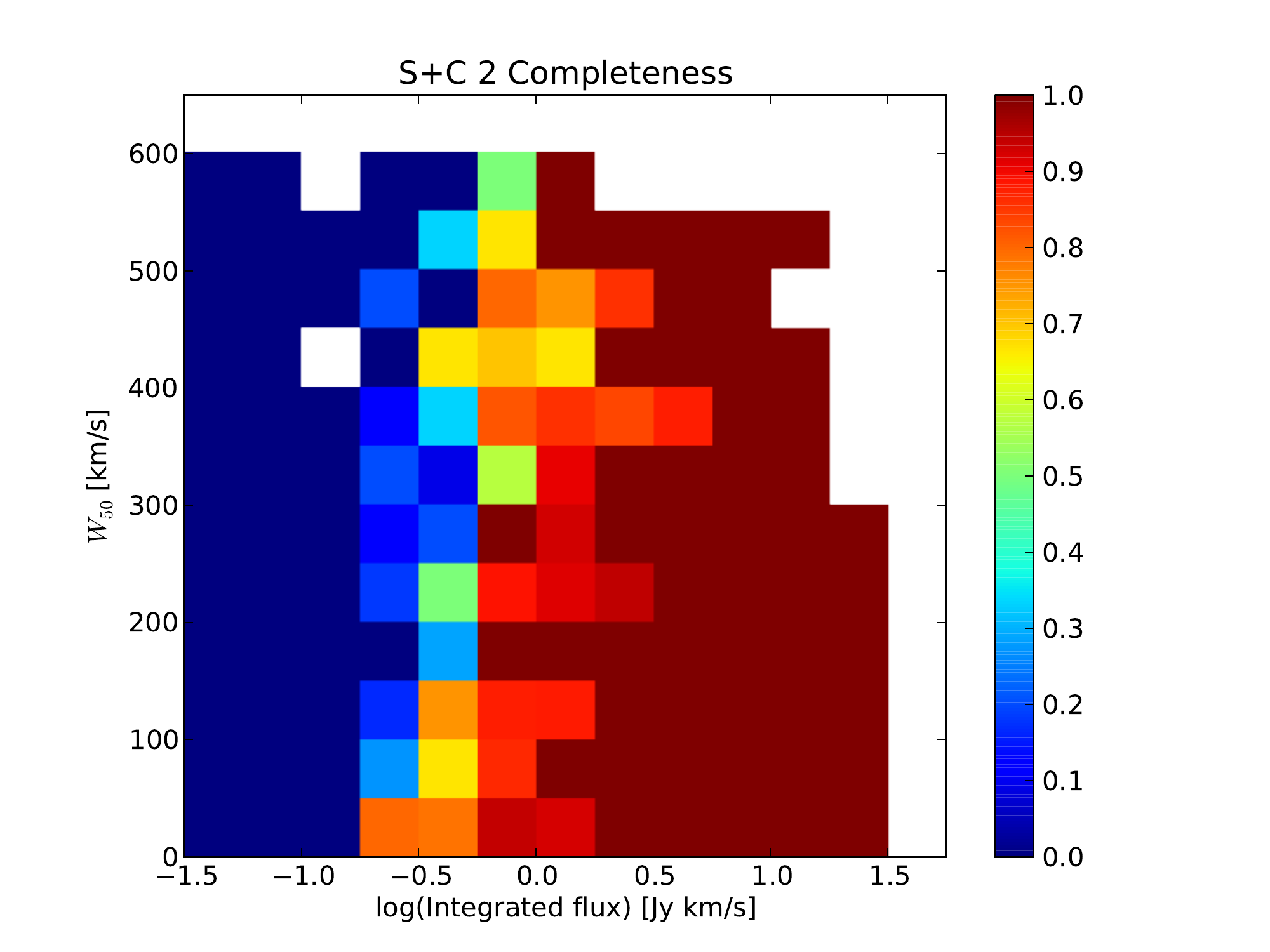}
\includegraphics[width=0.48\textwidth, angle=0]{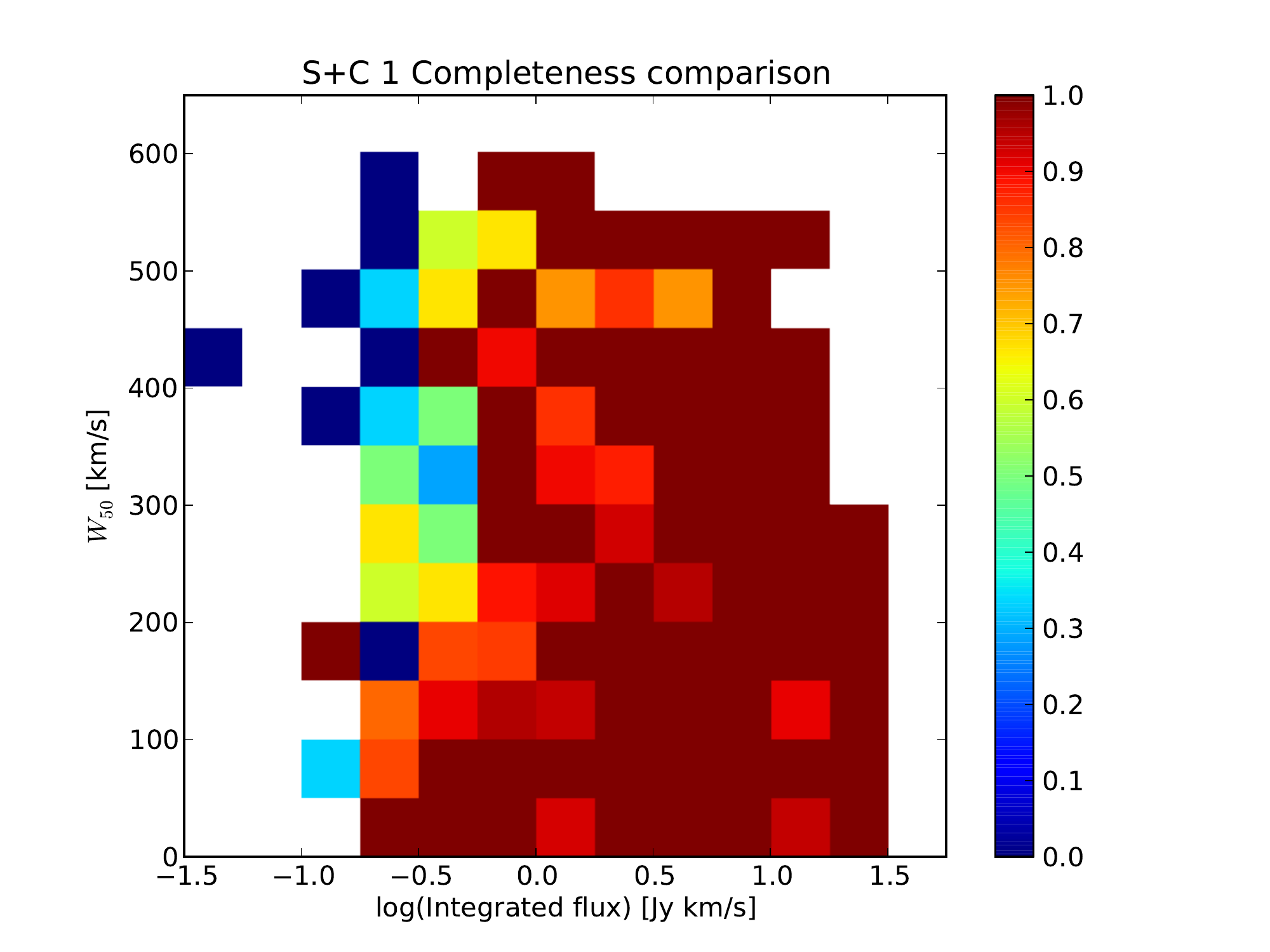}
\includegraphics[width=0.48\textwidth, angle=0]{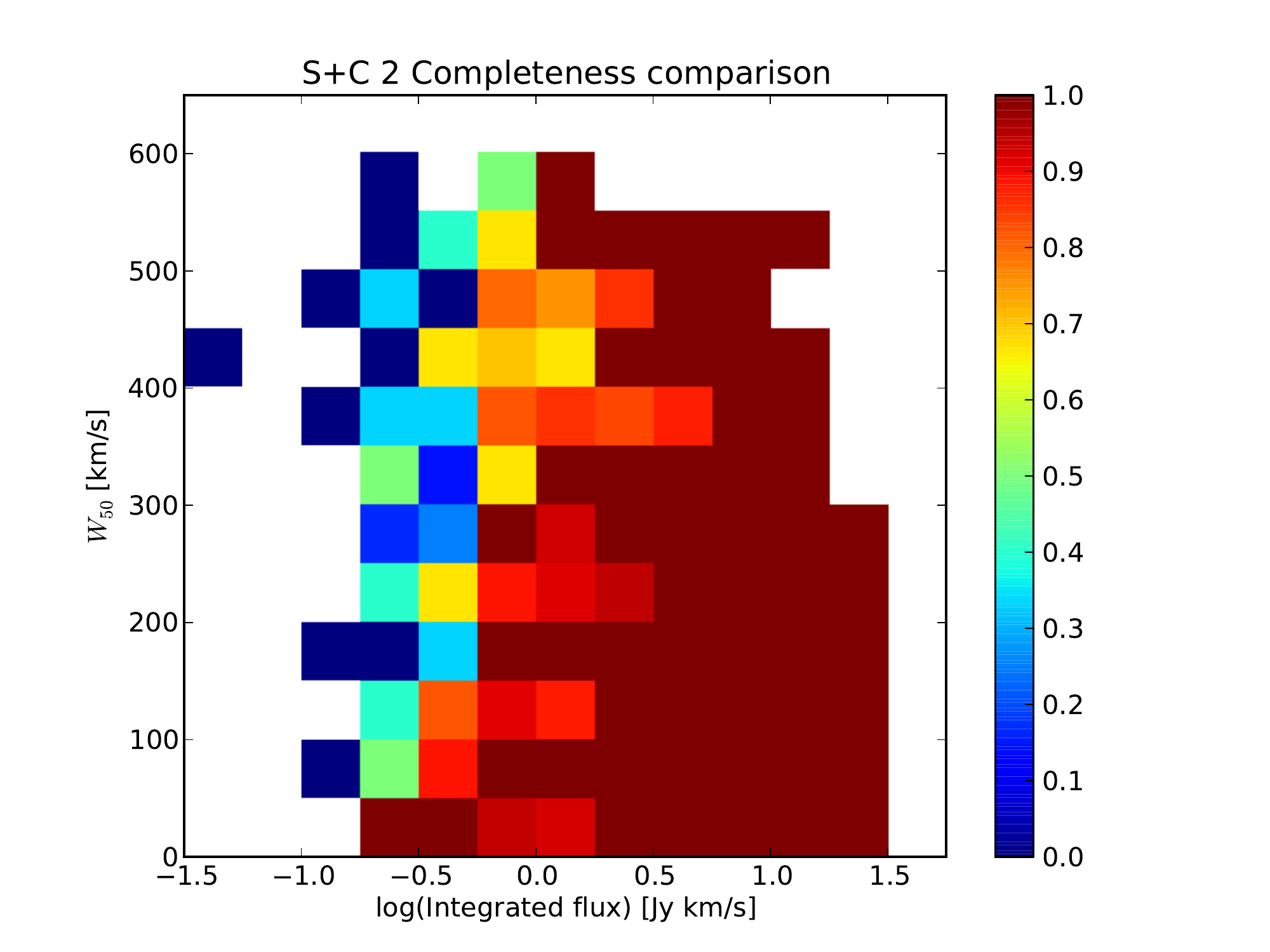}
\includegraphics[width=0.48\textwidth, angle=0]{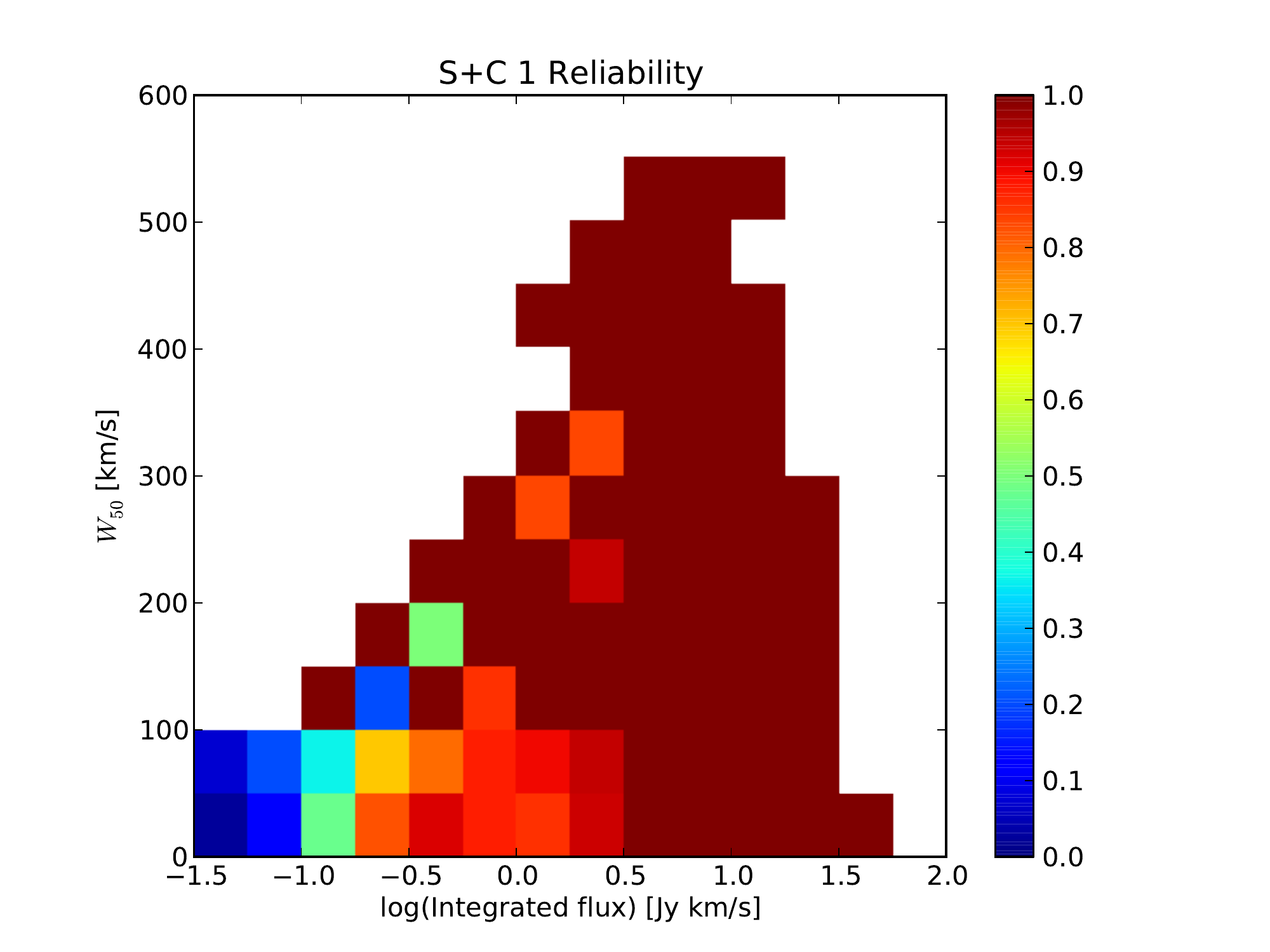}
\includegraphics[width=0.48\textwidth, angle=0]{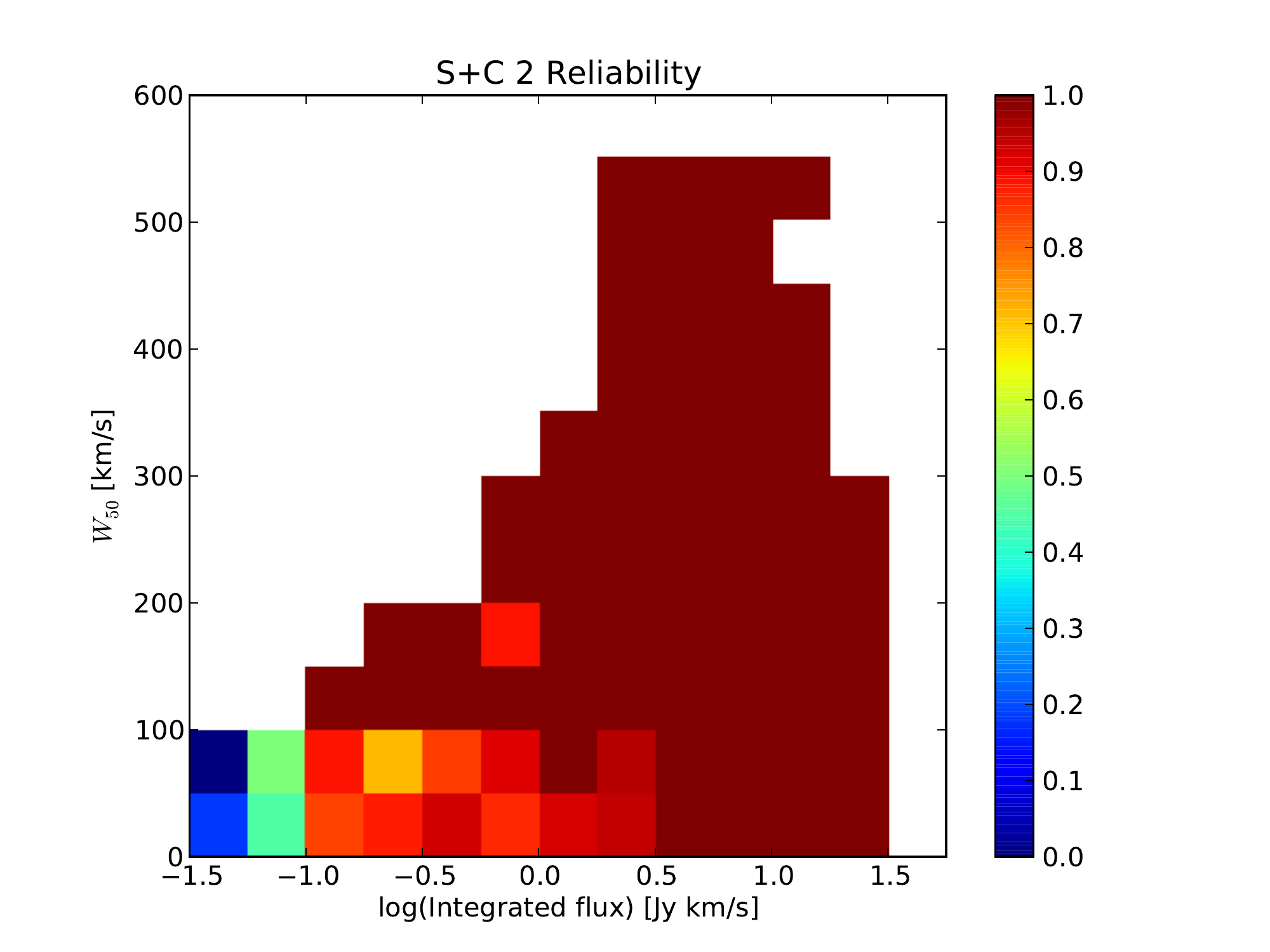}

\end{center}
\end{figure*}

\begin{figure*}[t]
\begin{center}

\includegraphics[width=0.48\textwidth, angle=0]{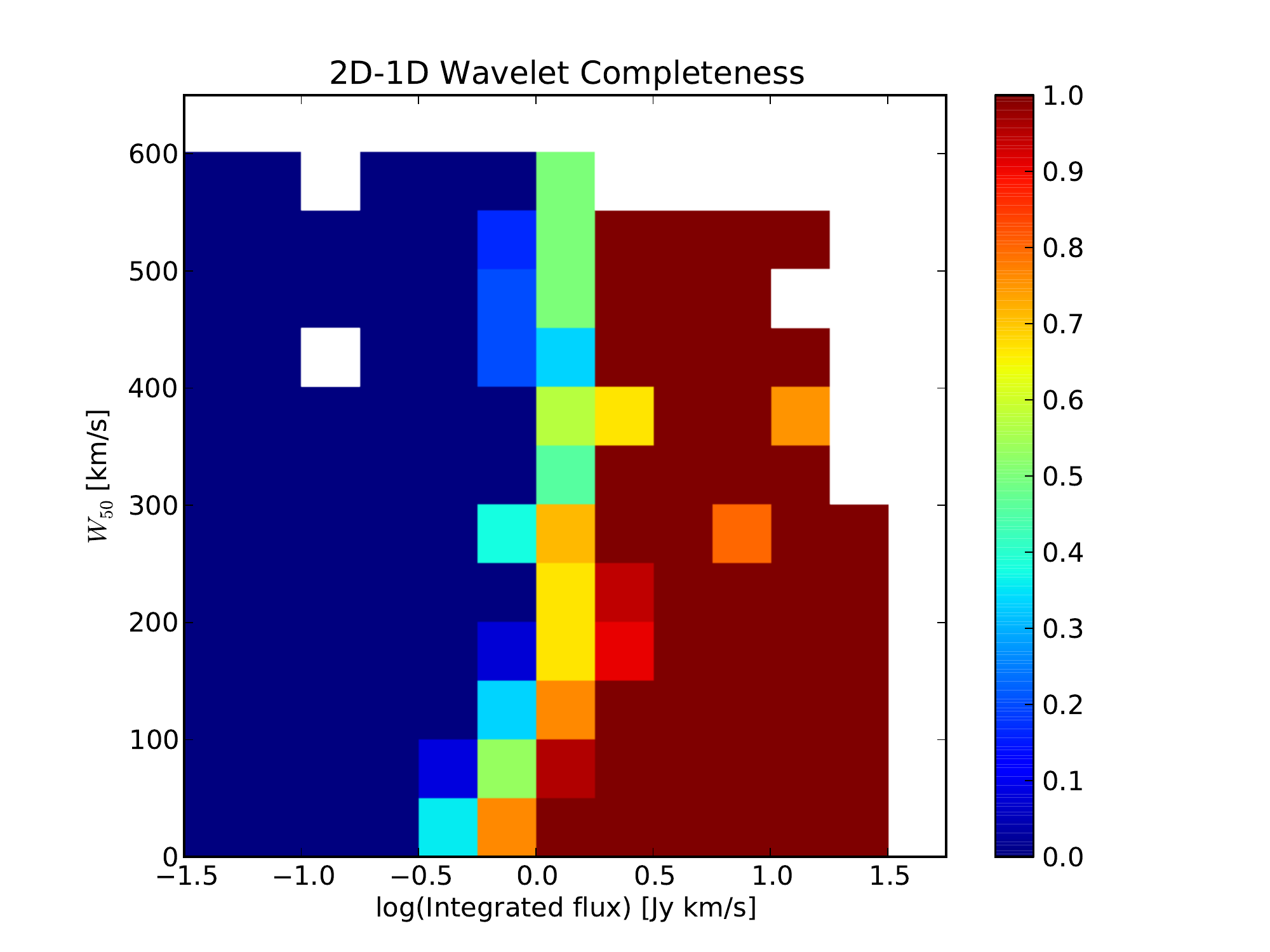}
\\
\includegraphics[width=0.48\textwidth, angle=0]{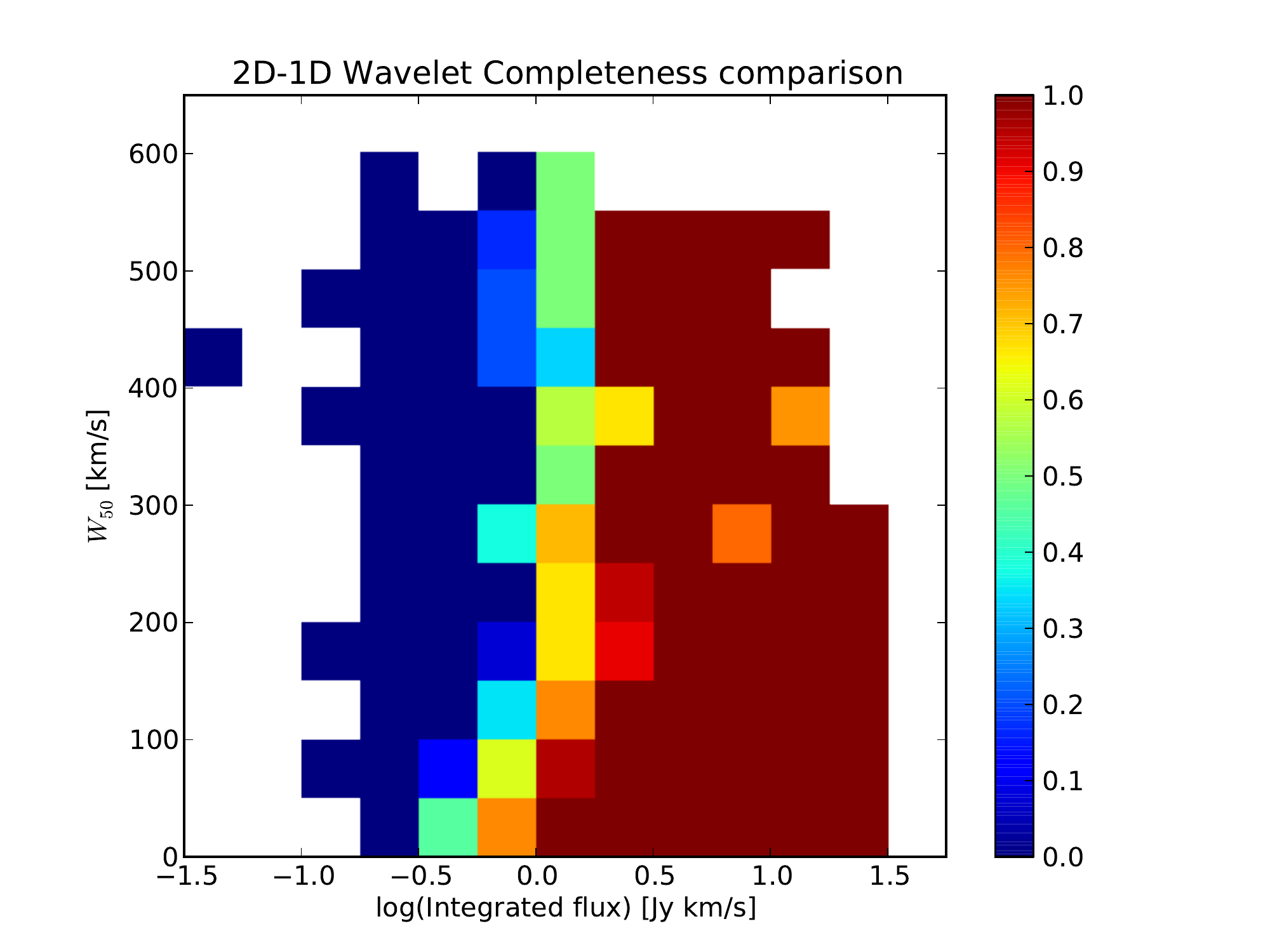}
\\
\includegraphics[width=0.48\textwidth, angle=0]{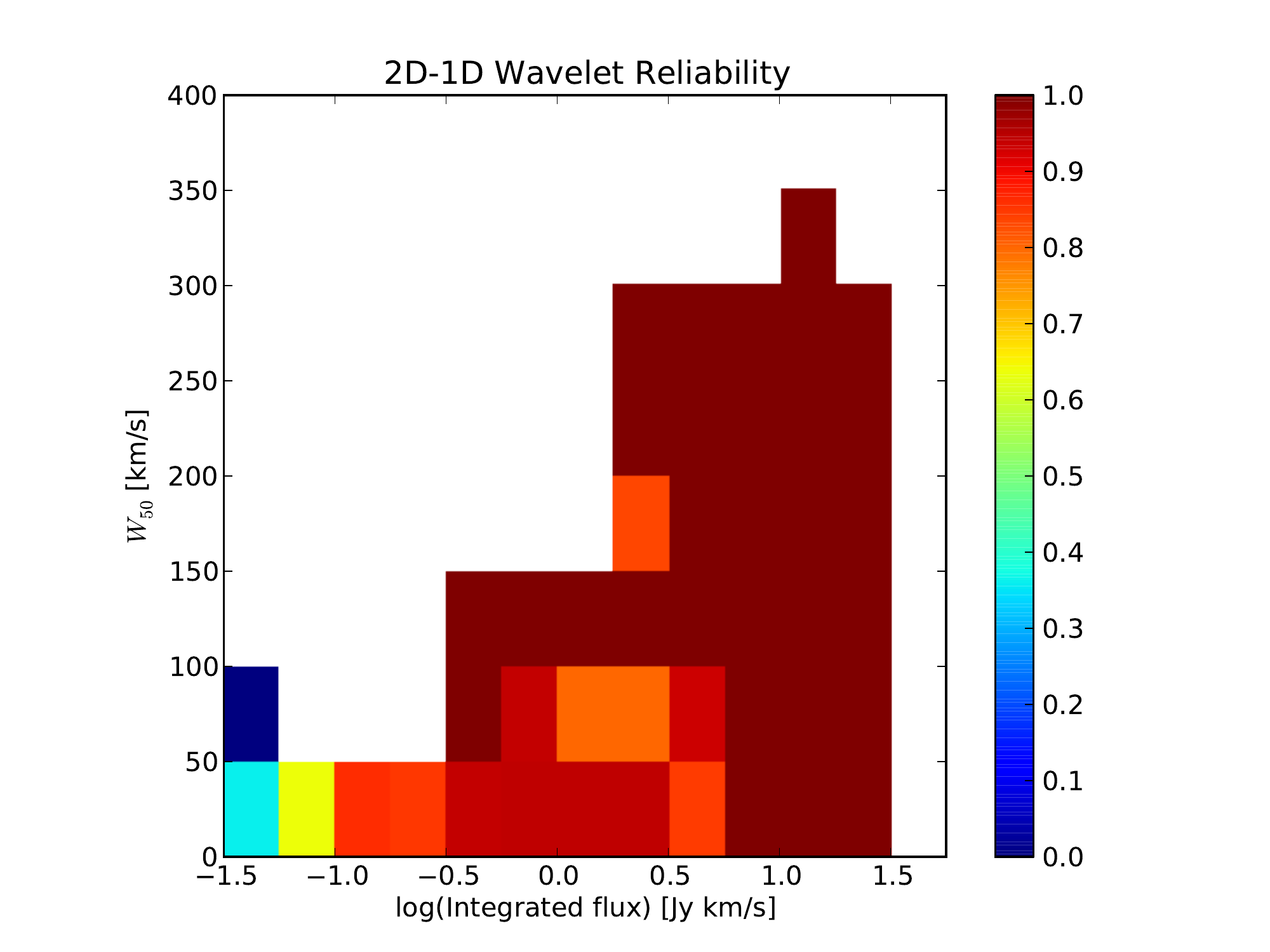}
\\

\end{center}
\end{figure*}

\begin{figure*}[t]
\begin{center}

\includegraphics[width=0.48\textwidth, angle=0]{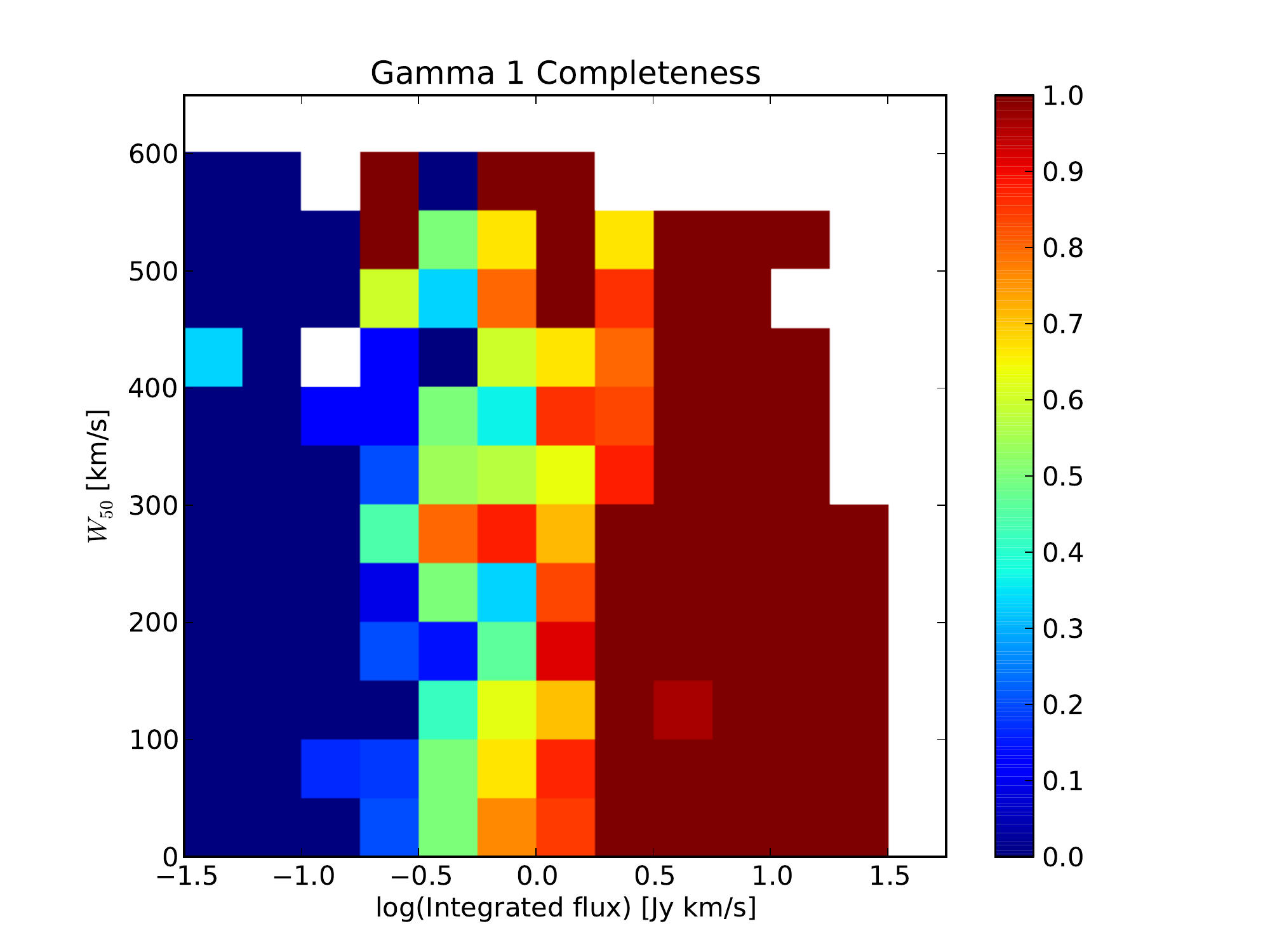}
\includegraphics[width=0.48\textwidth, angle=0]{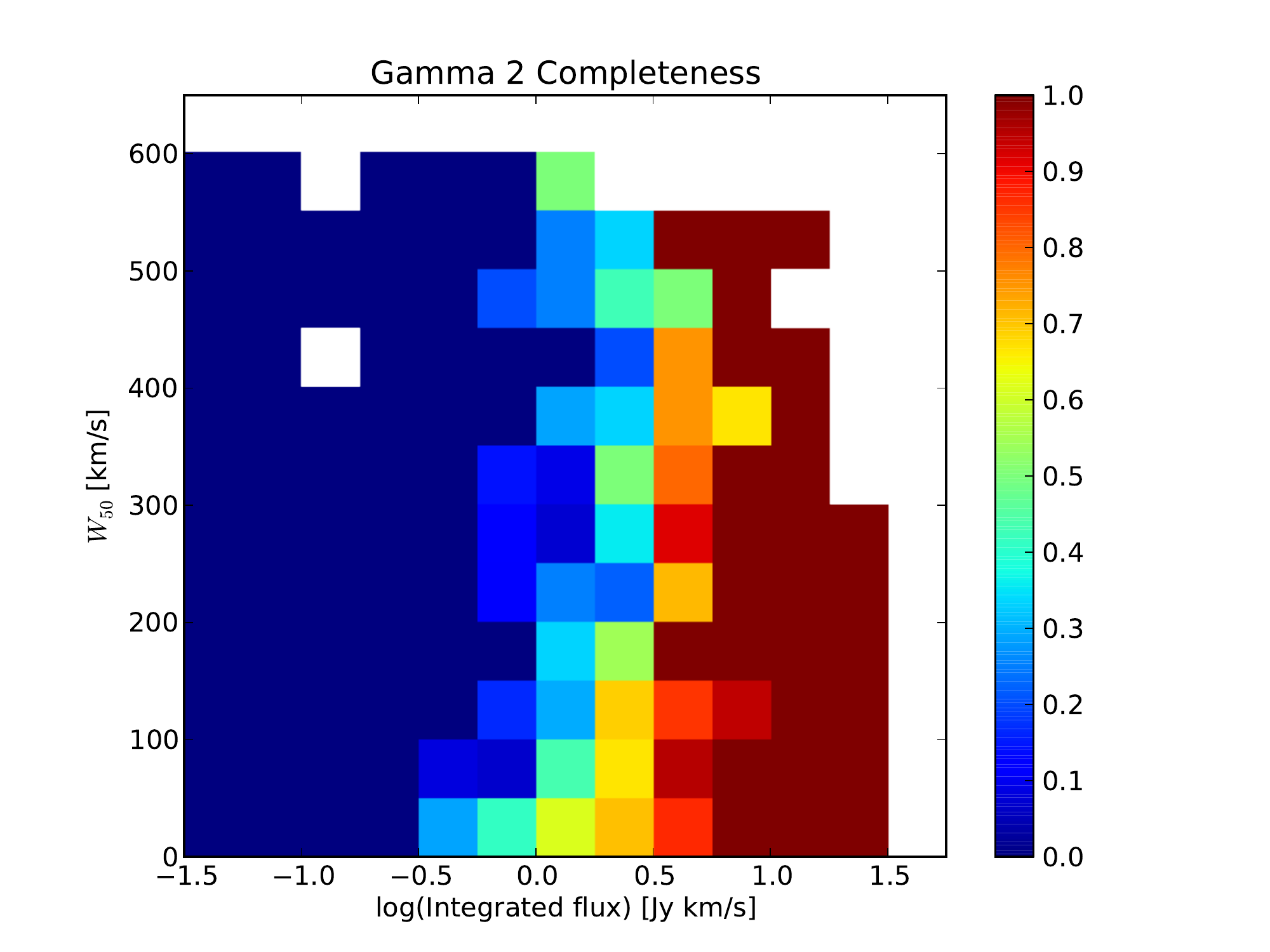}
\includegraphics[width=0.48\textwidth, angle=0]{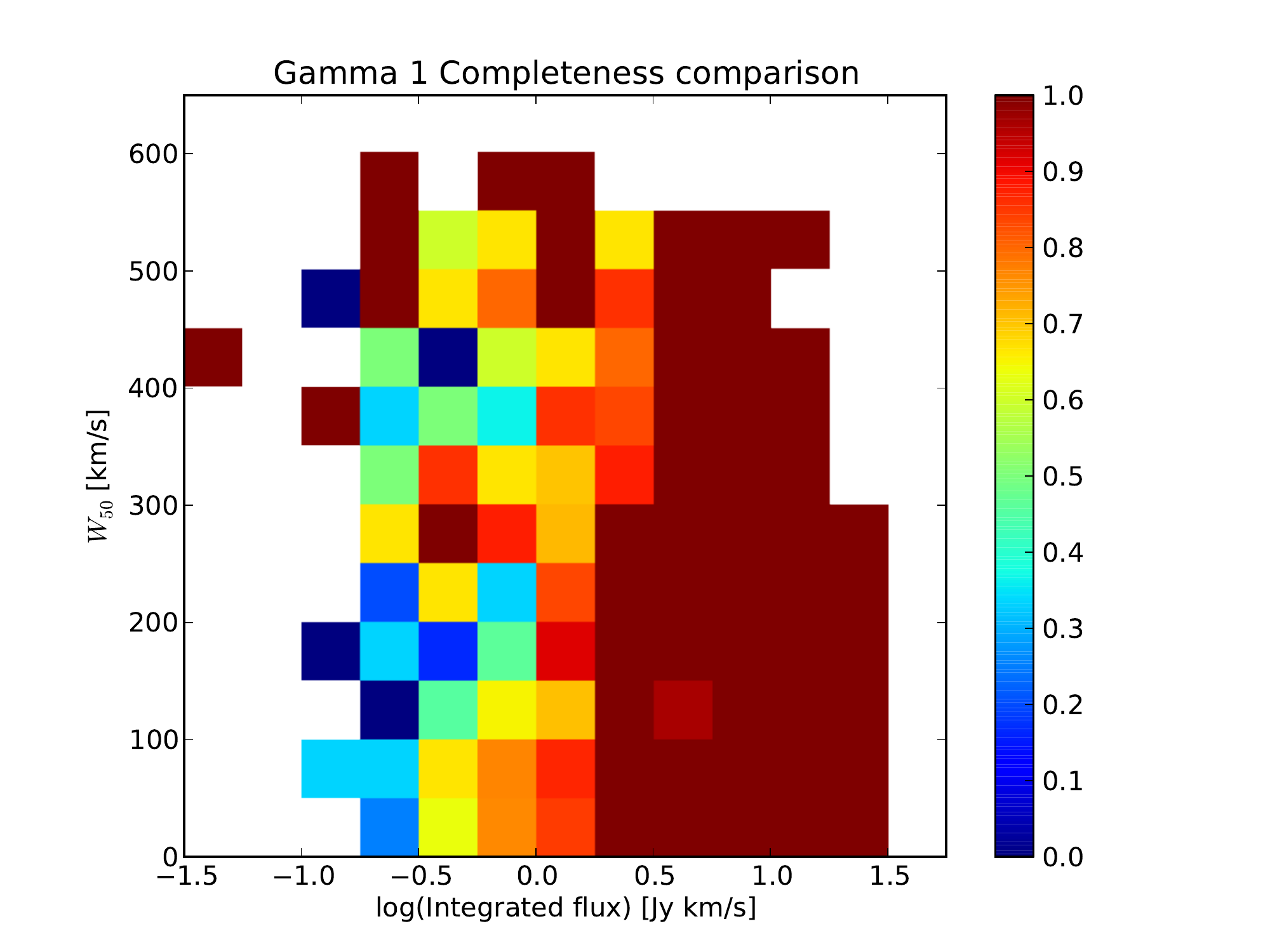}
\includegraphics[width=0.48\textwidth, angle=0]{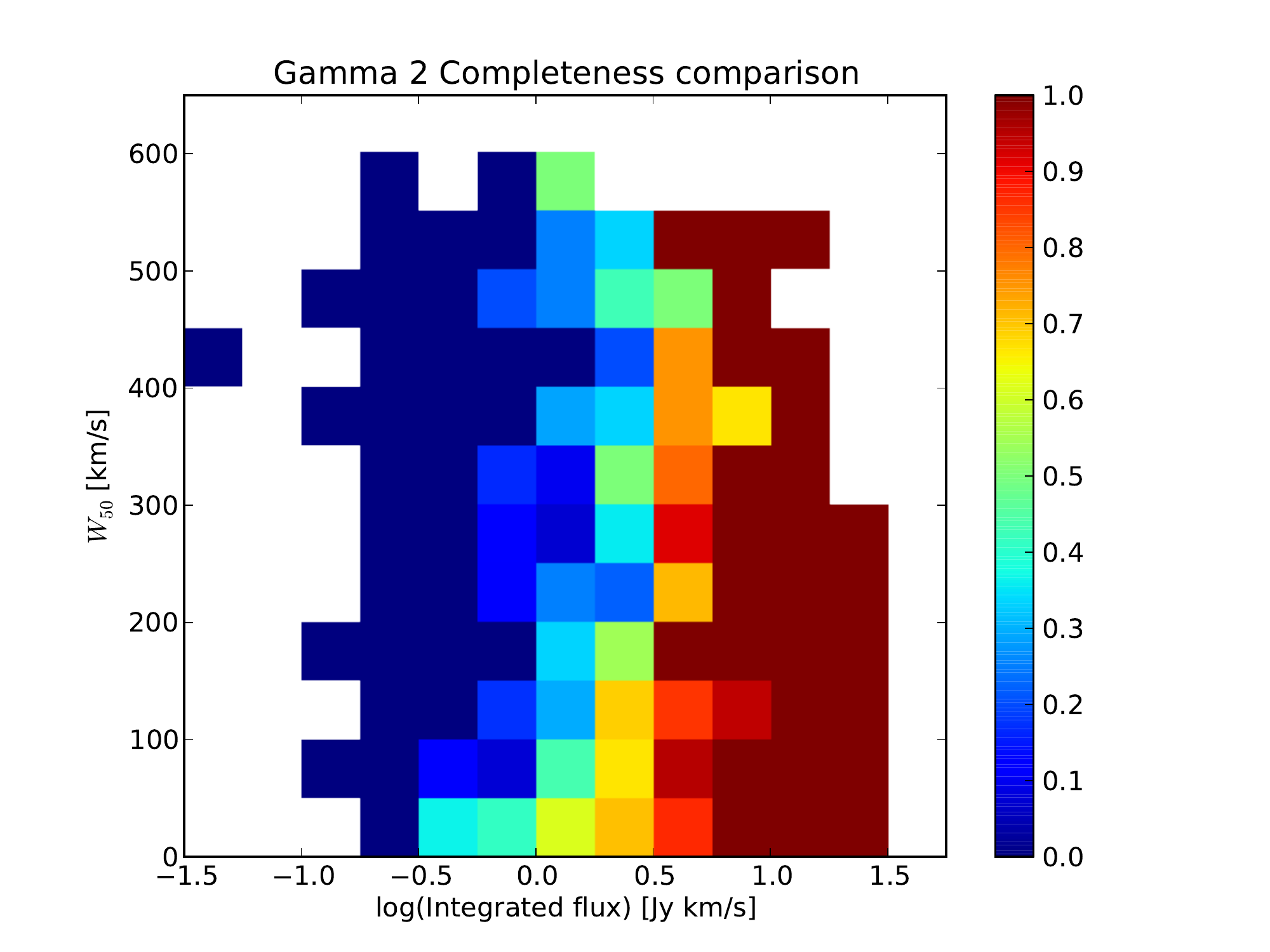}

\end{center}
\end{figure*}

\section{Discussion}

A different way of demonstrating the performance of the source finders
is by plotting the completeness of the source finders on a two dimensional
plot as a function of integrated flux and velocity width.  As a
reference the total number of objects in both cubes is shown on this
grid in Fig.~\ref{total_detections_2D}. 

\subsection{Point sources}

In Fig.~\ref{point_2D} we plot the completeness and the reliability
results of the different source finders when applied to the point
sources on a two-dimensional grid. For each result, completeness is
plotted as a function of integrated flux and velocity width
(represented by FWHM ($W_{50}$)) of the modelled point sources in the
top panels. In the middle panels the ratio is shown between number of
objects detected by the tested source finder and the number of sources
detected by any source finder. Instead of showing the overall
completeness this plot shows how a particular source finder performs
compared to the other source finding results. Regions in the parameter
space that appear blue in this plot are regions that can be improved
upon, as other source finders do detect objects within this parameter
space. Apart from showing how one source finder performs
  compared to the others, this plot also shows the parameter space
  that is covered by all the source finders combined.

  In the bottom panel of Fig.~\ref{point_2D} reliability is plotted as
  function of measured integrated flux and velocity width
  $(W_{50})$. These panels are not included for the {\sc Gamma-finder}
  as this source finder does not parameterise sources. The
  completeness plots in the top two panels all have the same scale as
  the parameters are based on the intrinsic parameters of the input
  catalogue. The scaling of the reliability panels is different in
  each plot as this is determined by the measured parameters of the
  different source finders. We have to emphasise here that the
  measured parameters are not by definition correct values as this
  depends on the capability to parameterise sources
  properly. Different parameterisation algorithms are used by the
  different source finders. We have not compared the parameters
  obtained from the source finders, but a possible difference has to
  be taken into account when comparing the plots.

{\sc Duchamp} is incomplete for small integrated fluxes, but is
basically 100\% complete for fluxes above 0.3 Jy~km s$^{-1}$. It is
expected that very low flux values are difficult to detect, however in
quite a large area of the parameter space sources are detected which
are not recovered by {\sc Duchamp}. This indicates that although in
Fig.~\ref{point_comp_flux} {\sc Duchamp} appears to be the best
performing source finder, another source finder is needed to detect
the very low fluxes, or {\sc Duchamp} has to be improved here. For
both {\sc Duchamp} tests the reliability is reasonable as most
detections are true detections and the false detections are especially
concentrated at very small fluxes.

The {\sc CNHI} source finder does not perform very well on the tested
point sources, it misses almost all sources with a FWHM velocity
width below 12 km s$^{-1}$. Apart from that this source finder also
misses a very significant fraction of the bright sources.  The number
of false detections is relatively large and spread over the whole
parameter range. Many of the false detections have low fluxes and very
broad line widths, much broader than any of the real line widths.

The {\sc S+C finder} detects sources down to very low integrated
fluxes, lower than most of the other source finders. As can be seen in
the middle panel of the first {\sc S+C finder} results, some of the
sources with a low integrated flux are only recovered by this source
finder. On the opposite side, the {\sc S+C finder} is not 100\%
complete at either large fluxes or large line widths. False detections
are quite difficult to distinguish when using this source finder, as
the false detections are not clustered in a narrow range of the parameter
space. For a large region in the plot the reliability fluctuates
around 50\%, indicating that the determined parameters of false
detections are very similar to that of true detections.

For the 2D-1D wavelet finder there seems to be a clear trend from 0\%
completeness at low fluxes to almost full completeness at high
integrated flux values, very similar to the {\sc Duchamp} results. In
the parameter space covering the largest fluxes and line widths, the
finder is not 100\% complete. This could be caused by the fact that
our model cube is very dense with many sources, and for the largest
wavelet scales these sources start to merge. The wavelet finder can be
improved here, as {\sc Duchamp} also uses wavelet reconstructions, but
appears to be less sensitive to this problem. The reliability of the
2D-1D wavelet finder is very good and 100\% in most of the parameter
space, although there are some false detections with a high integrated
flux, we have no good explanation for why the reliability decreases
here.

The {\sc Gamma-finder} seems to perform well on sources with a strong
integrated flux and narrow line width. In fact it is the best finder
for objects with a narrow line width below 5 km/s, although we have to
question how realistic such sources are when observing real
galaxies. As the {\sc Gamma-finder} does not give a mask or parameters
of the detected sources, we cannot make reliability plots for this
source finder.

\subsection{Model galaxies}

In Fig.~\ref{model_2D} we show very similar plots as in the previous
figure, but now for the model galaxies. In the top panels the
completeness of the different source finders is plotted, while the
middle panels compare the completeness of the source finders with
respect to each other. In the bottom panels the reliability of the
source finders is plotted. These modelled galaxies have more complex
structures compared to the point sources, and the completeness and
reliability results are very different. Note the different scales in
both integrated flux and velocity compared to the point sources in
Fig.~\ref{point_2D}.

{\sc Duchamp} is complete for objects with high flux in the first run,
but in the second run misses a few sources that should be easy to
detect due to their high flux. The only difference between the two
{\sc Duchamp} runs is the {\it growth} parameter, which has merged
some of the extended sources. As can be seen in the plot, the missed
sources have a large integrated flux but relatively narrow line width,
which indicates that they are spatially extended. As the objects were
all placed at a similar radial velocity in the cube, there is a high
risk of merging. There is a clear transition phase between
non-detected and detected objects and {\sc Duchamp} misses objects
with low integrated fluxes that are detected by at least one other
source finder. The reliability of {\sc Duchamp} looks very good, as
almost all false detections are clustered in a limited area of the
parameter space at small fluxes and narrow line widths.

The CNHI finder also shows a transition phase from non-detected
objects with a low flux to detected objects with high fluxes, however
the transition is much broader than for {\sc Duchamp}. The CNHI finder
is less likely here to miss sources with a narrow line width as the
velocity profiles of the model galaxies are much broader and more
realistic than for the point sources. When compared to the other
source finders, this finder detects a significant fraction of the
objects at low integrated flux. The reliability is worse than for the
other source finders, but a large fraction of the detected objects have very low fluxes,
covering a large range in line width.
 
For the {\sc S+C finder} the results are very impressive as it even
detects many of the sources with small flux and narrow line
width. This finder also has a very small number of false detections
that appear to be concentrated in a rather limited range of the
parameter space. Although currently this appears to be the best source
finder on the tested cube with model galaxies, it is not the best
source finder on the full parameter range. In particular, objects with a
small integrated flux and broad line width are missing, which in
  some cases are detected by the Gamma-finder.

The 2D-1D wavelet source finder has a very narrow transition between
detected and non-detected sources where almost all objects with a flux
below 0.5 Jy km/s are missed, while almost all objects with an
integrated flux above 1.5 Jy km/s are detected. The reliability of
this source finder is very good and the completeness can probably
be improved upon by decreasing the clipping threshold used on the
reconstructed wavelet scales. The parameter space covered in the
reliability plots is very different to the other source finders, the
2D-1D wavelet method seems to detect higher fluxes and smaller line-widths.

The {\sc Gamma-finder} has a relatively good performance in
completeness as it detects the objects with high fluxes, but also a
significant number of objects with low flux values. Interesting to see
is that the first {\sc Gamma-finder} results gives the best result for
objects with a low flux and broad line width. Although not plotted in
this figure, this good performance in completeness probably comes at
the cost of reliability as the reliability of the first run is very
low at 12\%.

\begin{figure*}[t]
\begin{center}
\includegraphics[width=0.48\textwidth, angle=0]{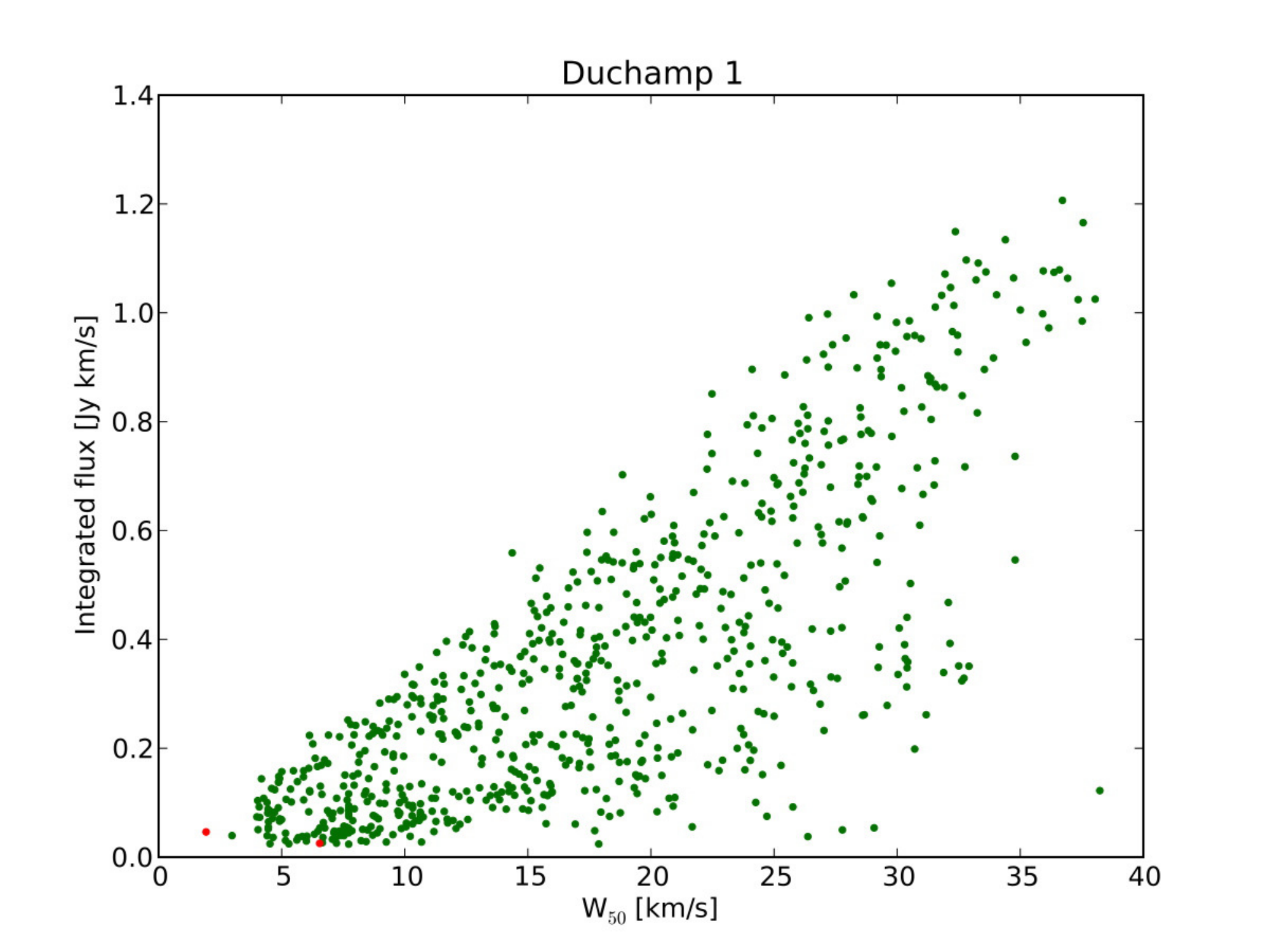}
\includegraphics[width=0.48\textwidth, angle=0]{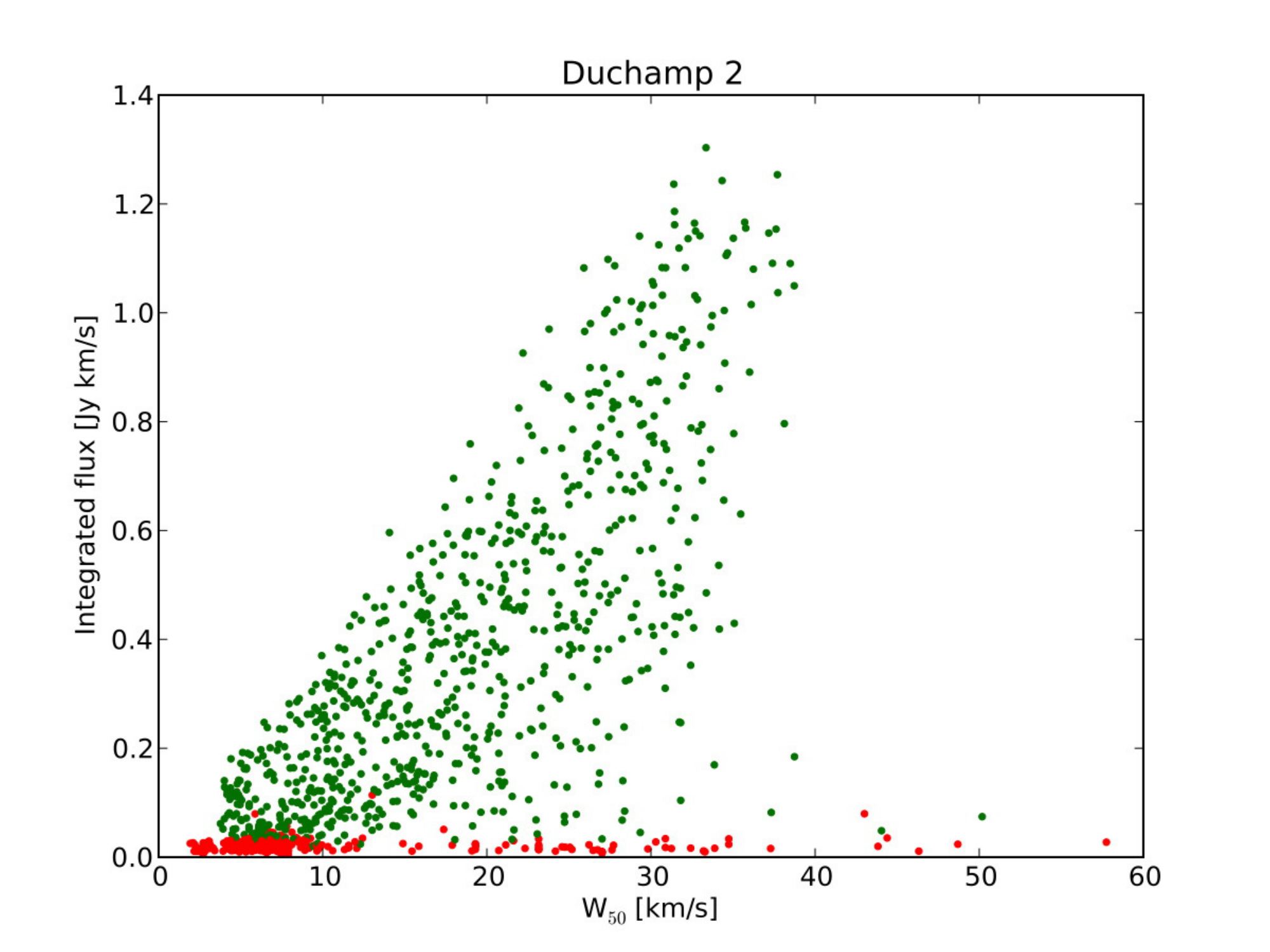}
\includegraphics[width=0.48\textwidth, angle=0]{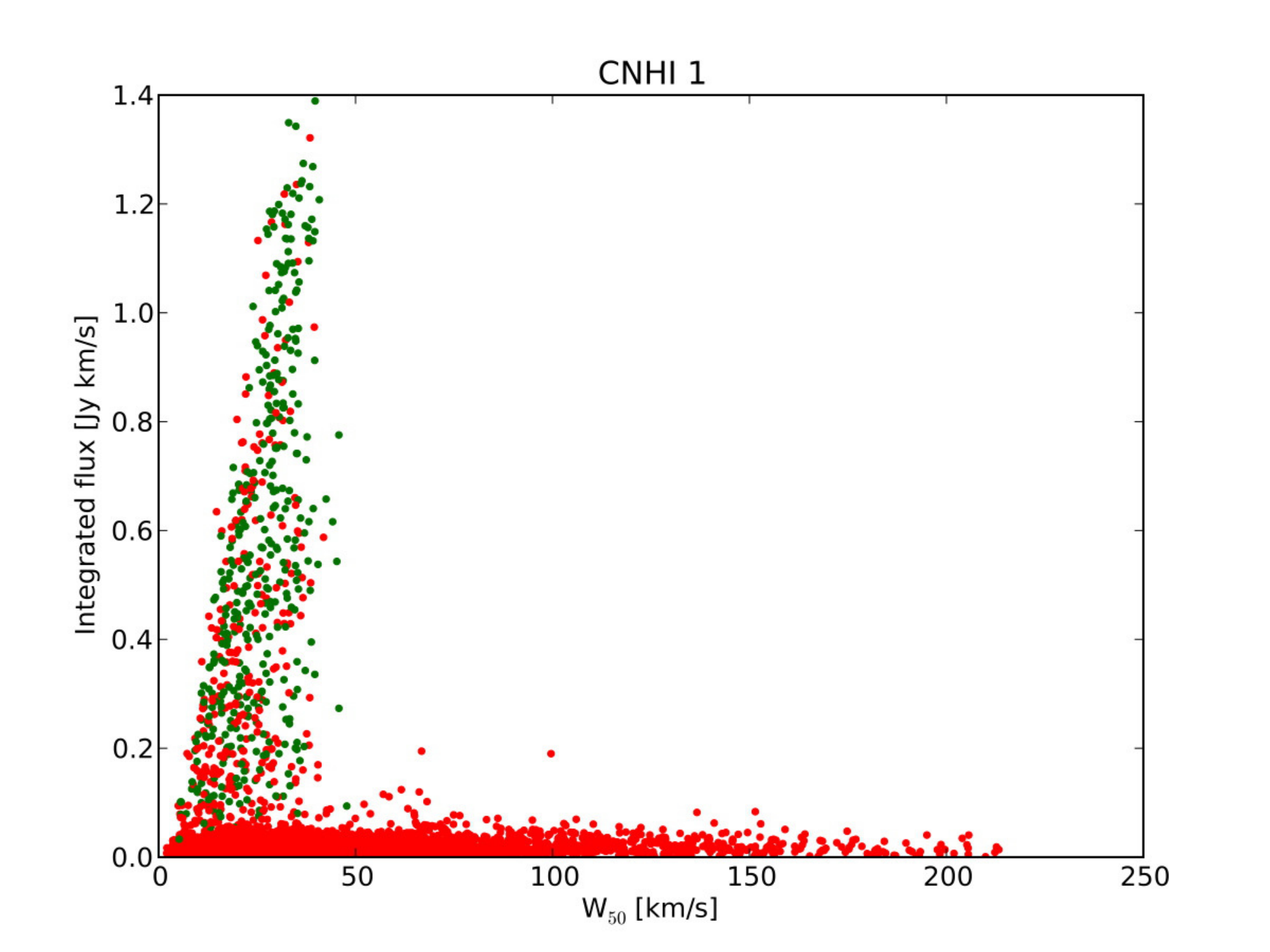}
\includegraphics[width=0.48\textwidth, angle=0]{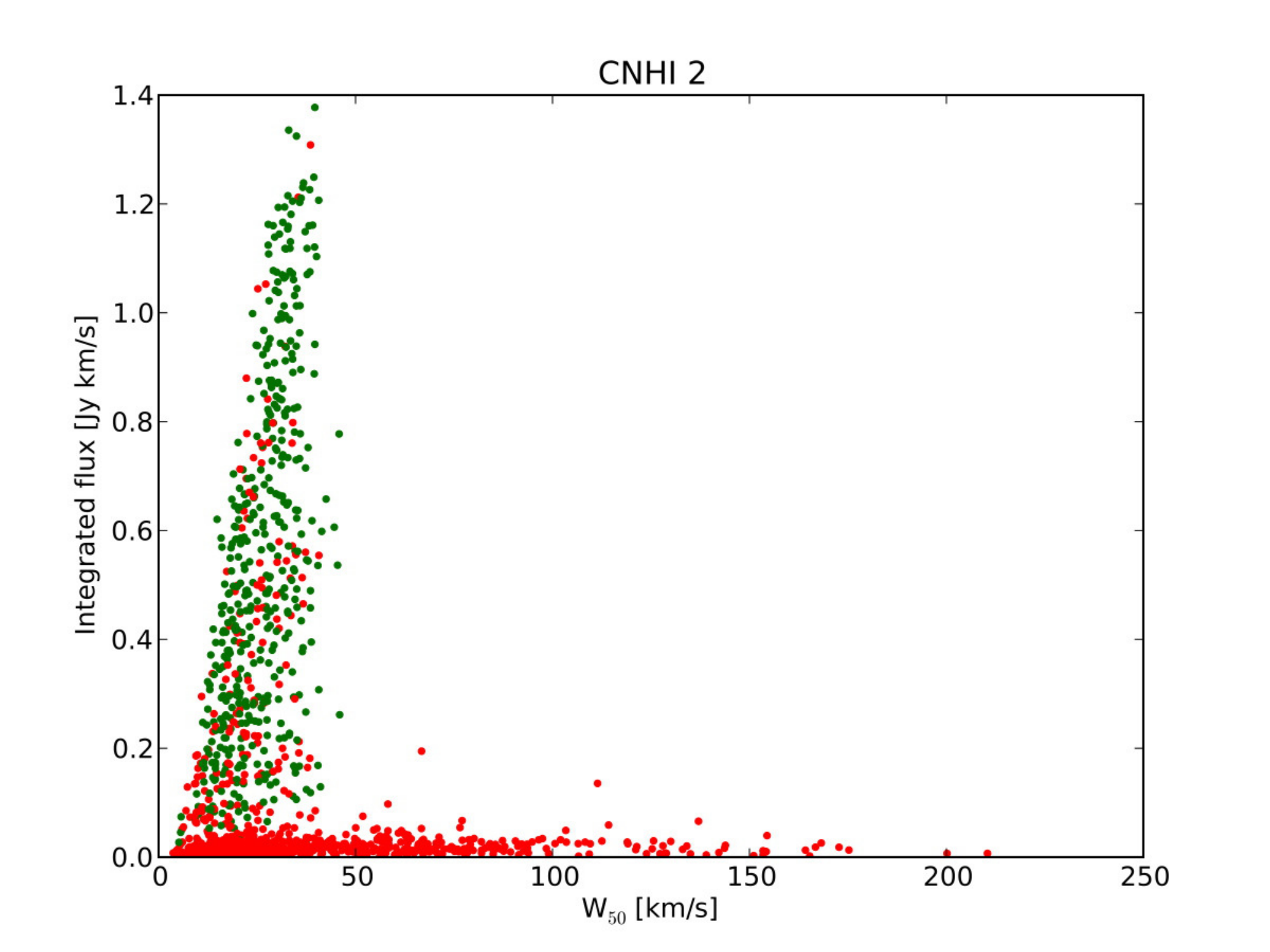}
\includegraphics[width=0.48\textwidth, angle=0]{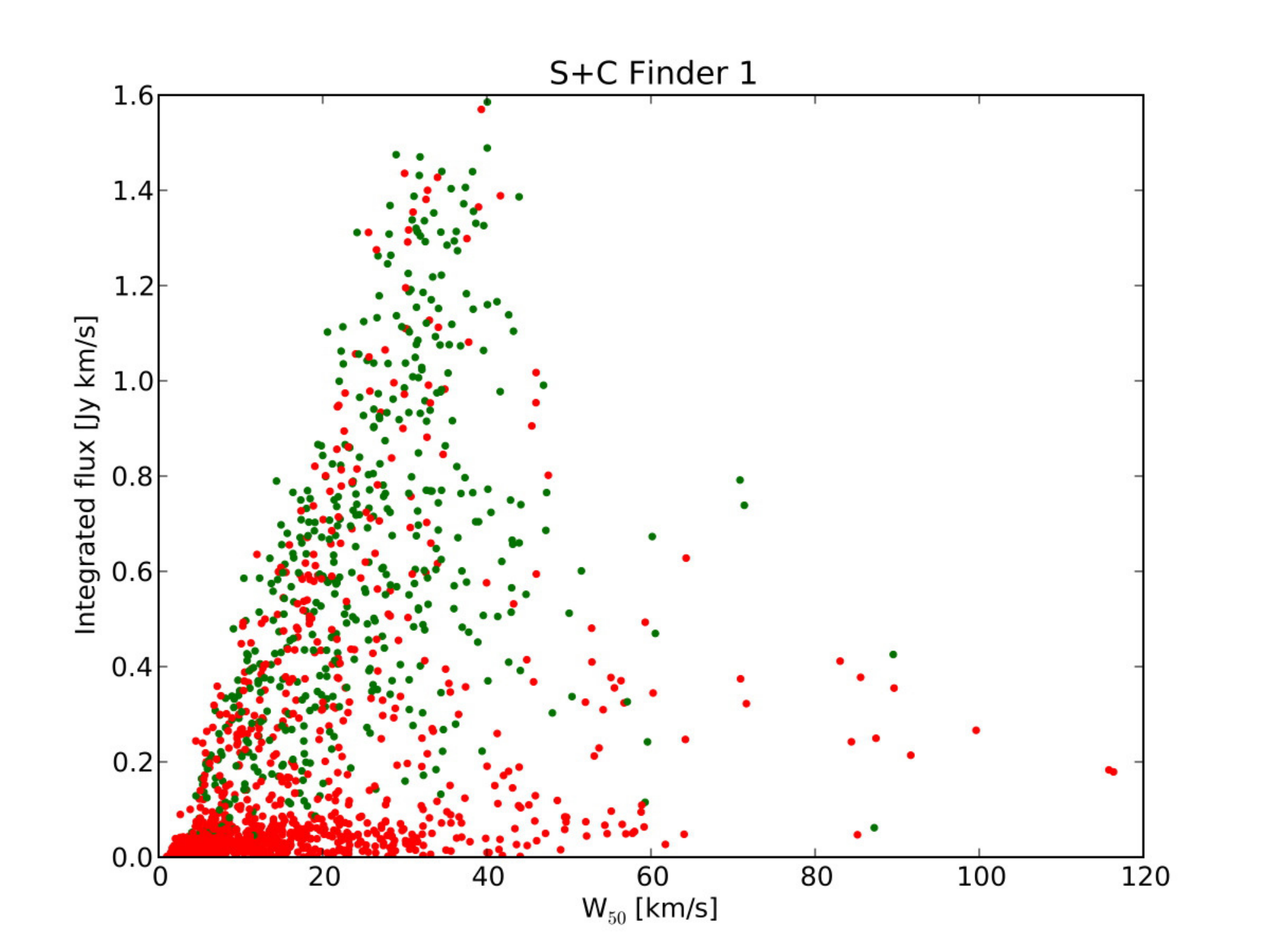}
\includegraphics[width=0.48\textwidth, angle=0]{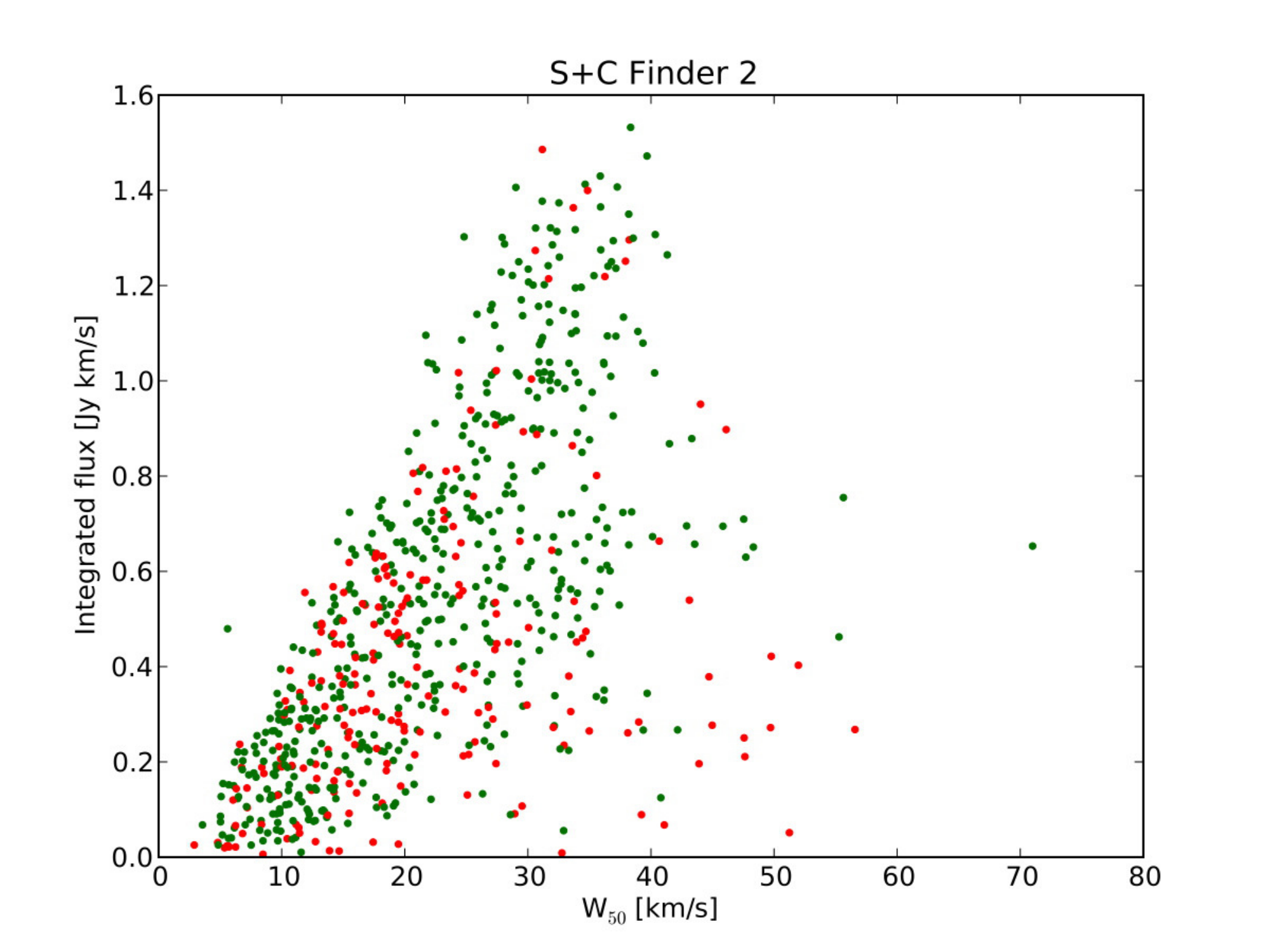}

\caption{True detections (green) and false detections (red) for all the source finders when applied on the
data cube with point sources. Detections are plotted as function of
integrated flux [Jy km/s] against $(W_{50})$ [km/s].}
\label{point_scatter}
\end{center}
\end{figure*}

\begin{figure*}[t]
\begin{center}

\includegraphics[width=0.48\textwidth, angle=0]{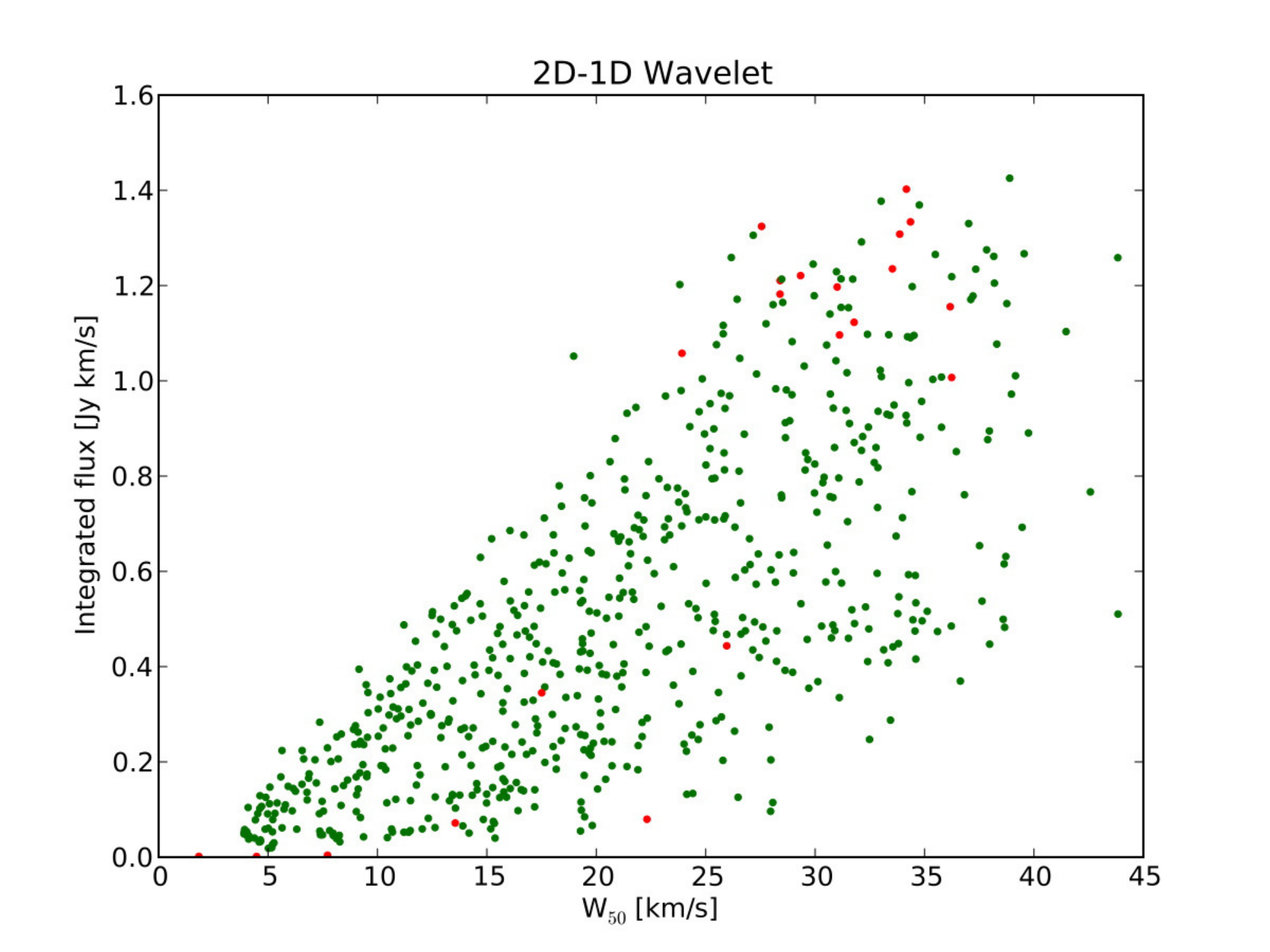}

Figure~\ref{point_scatter}: Continued
\end{center}
\end{figure*}

\begin{figure*}[t]
\begin{center}
\includegraphics[width=0.48\textwidth, angle=0]{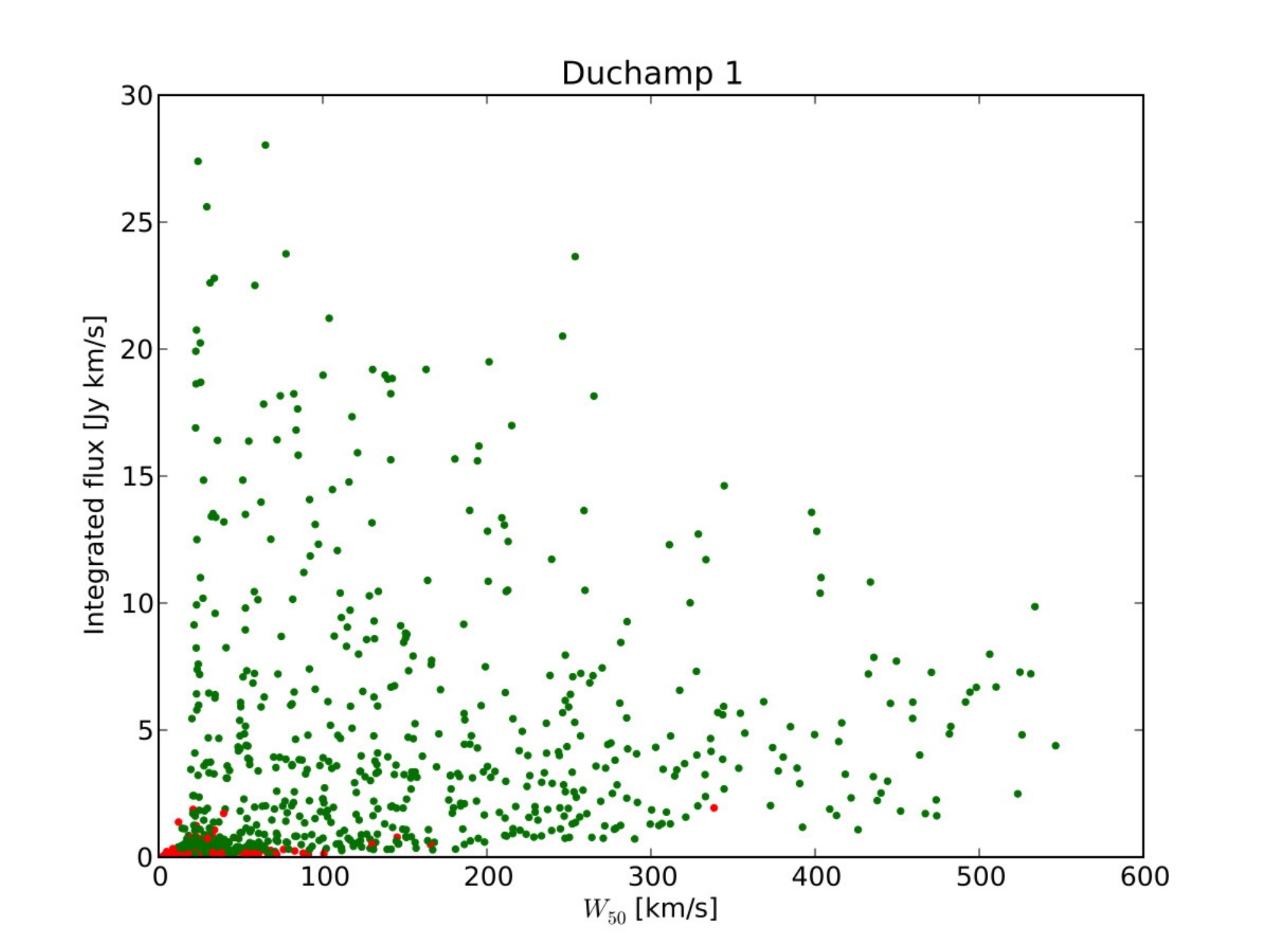}
\includegraphics[width=0.48\textwidth, angle=0]{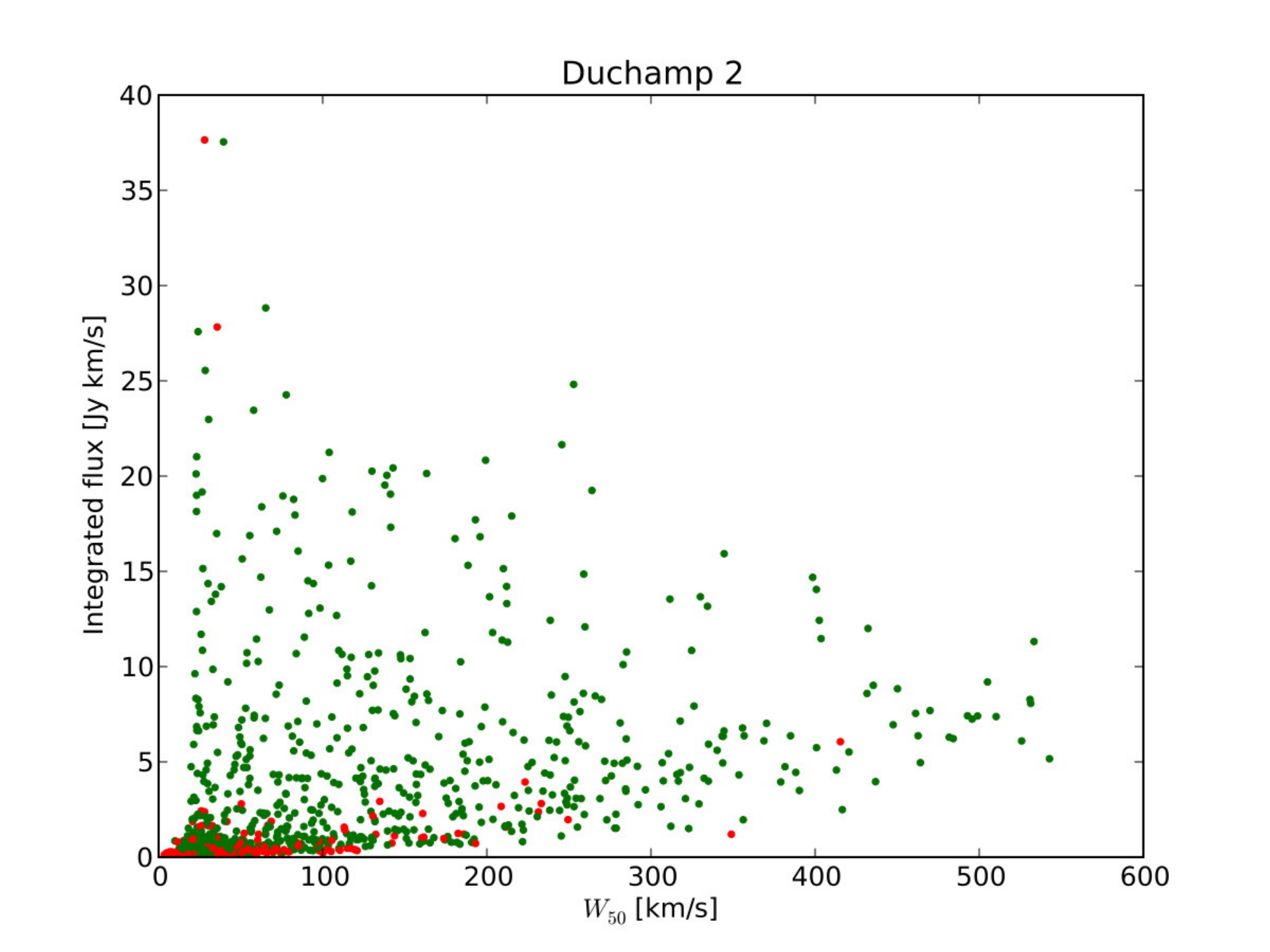}
\includegraphics[width=0.48\textwidth, angle=0]{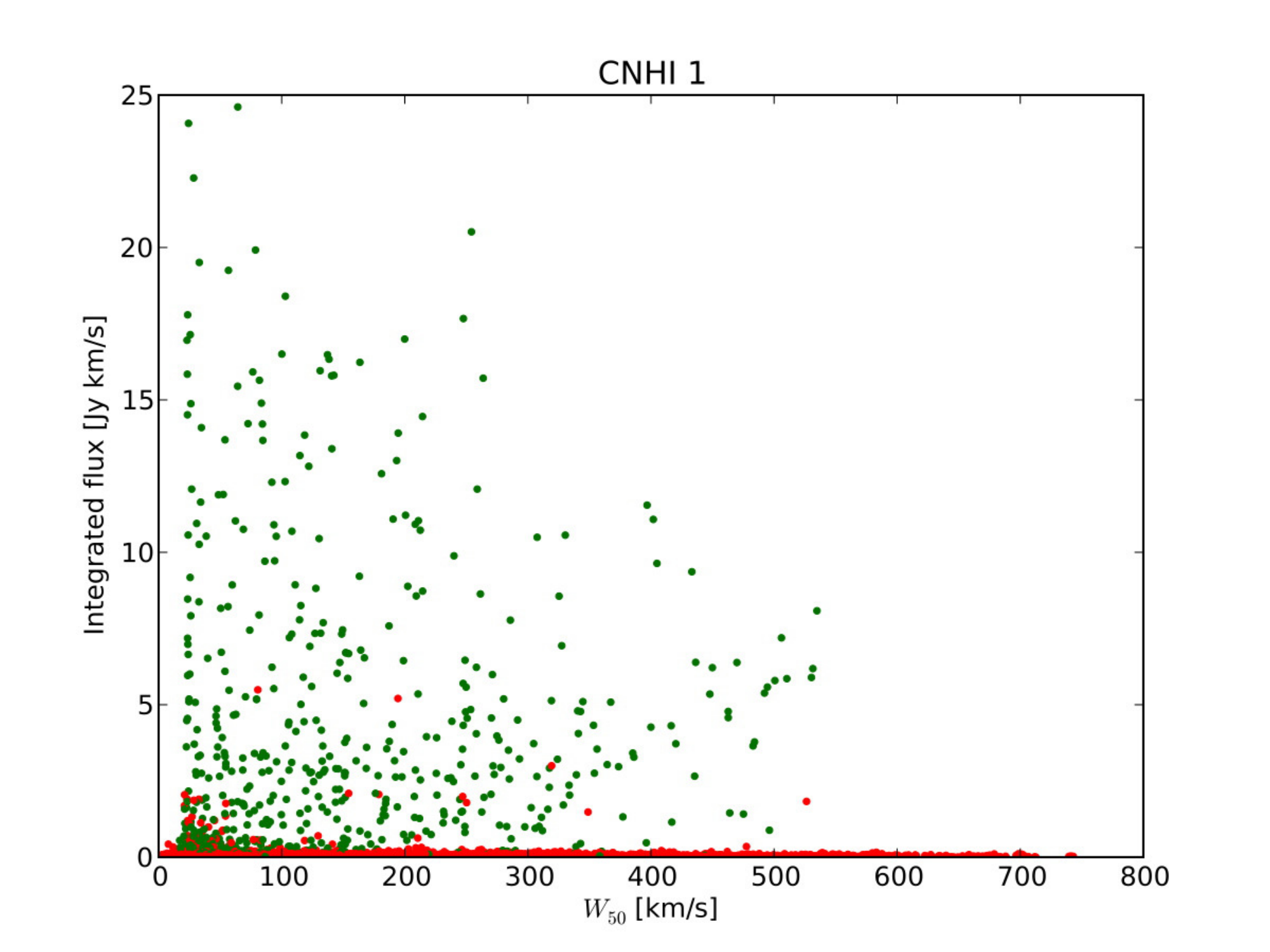}
\includegraphics[width=0.48\textwidth, angle=0]{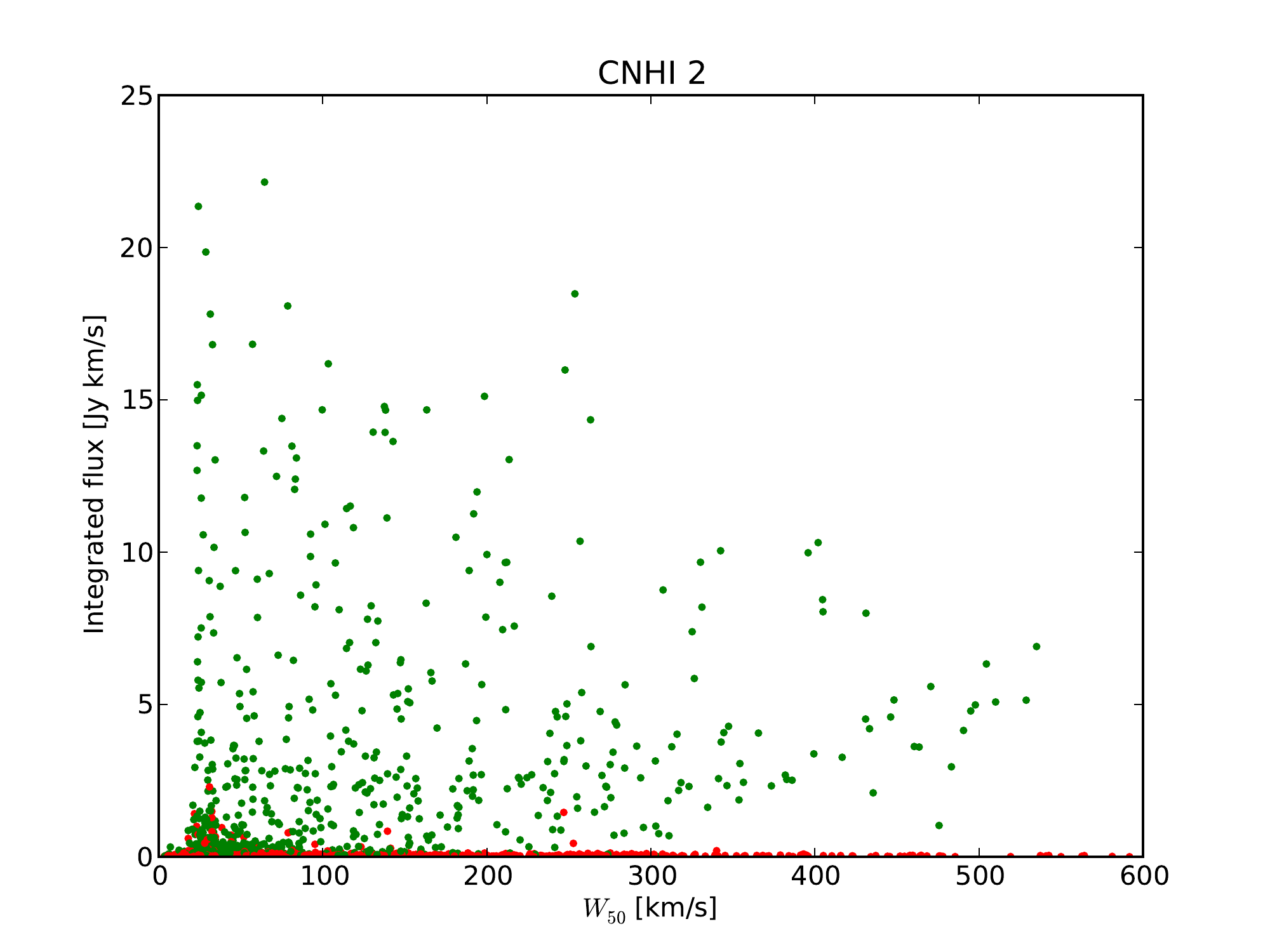}
\includegraphics[width=0.48\textwidth, angle=0]{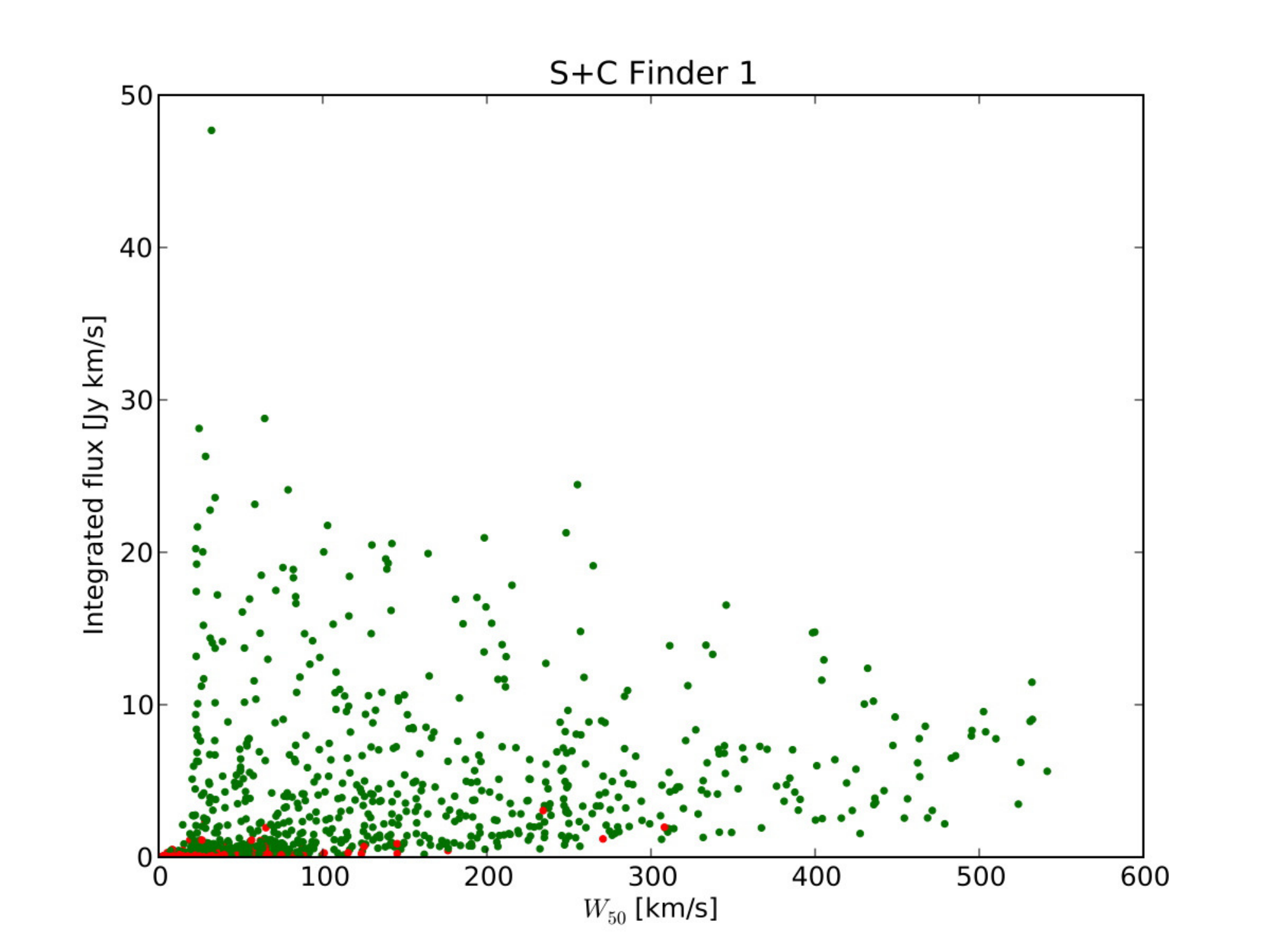}
\includegraphics[width=0.48\textwidth, angle=0]{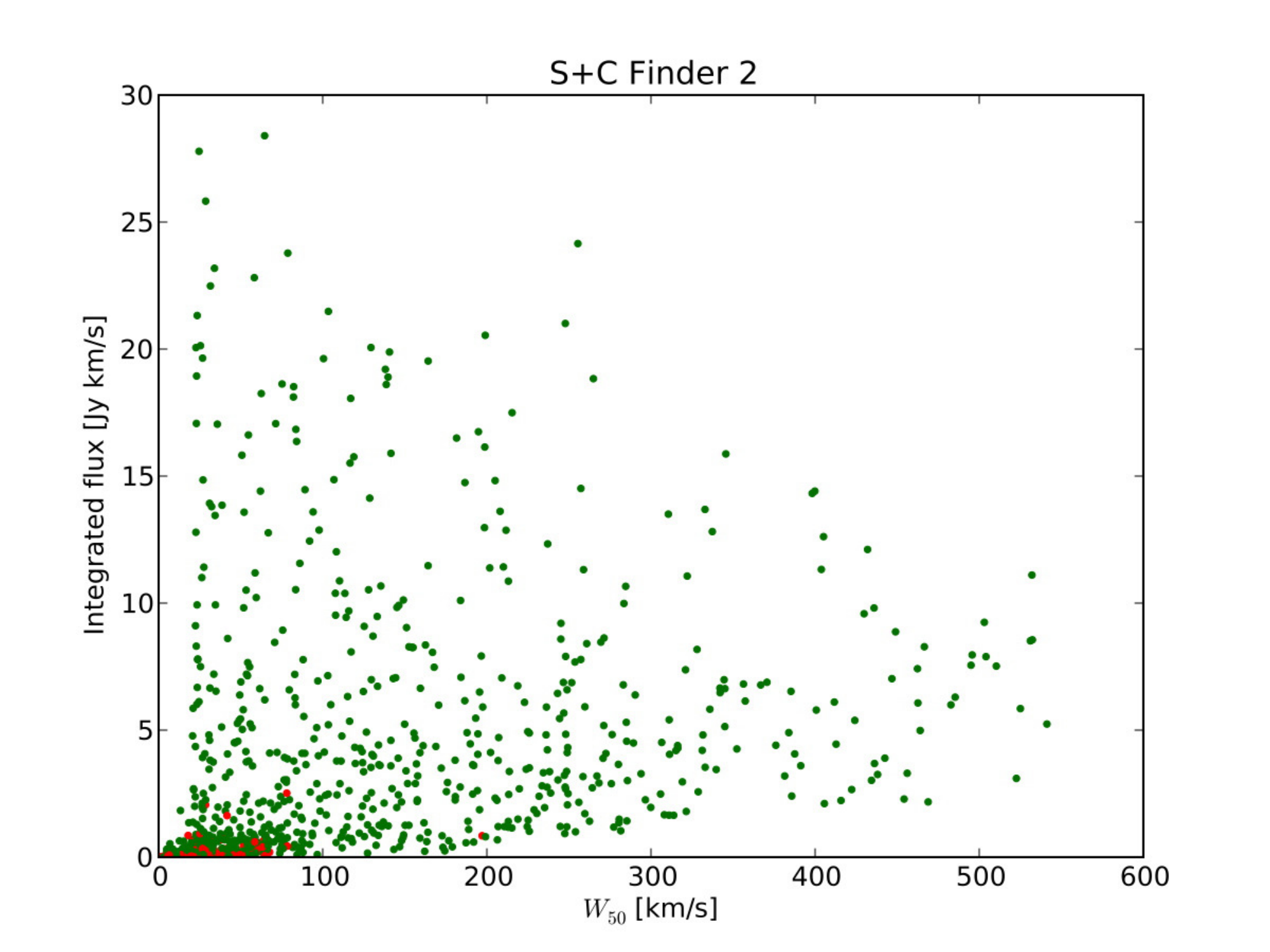}

\caption{Similar as Fig.~\ref{point_scatter}, but now the source
  finders are applied on the cube with model galaxies.}
\label{model_scatter}
\end{center}
\end{figure*}

\begin{figure*}[t]
\begin{center}

\includegraphics[width=0.48\textwidth, angle=0]{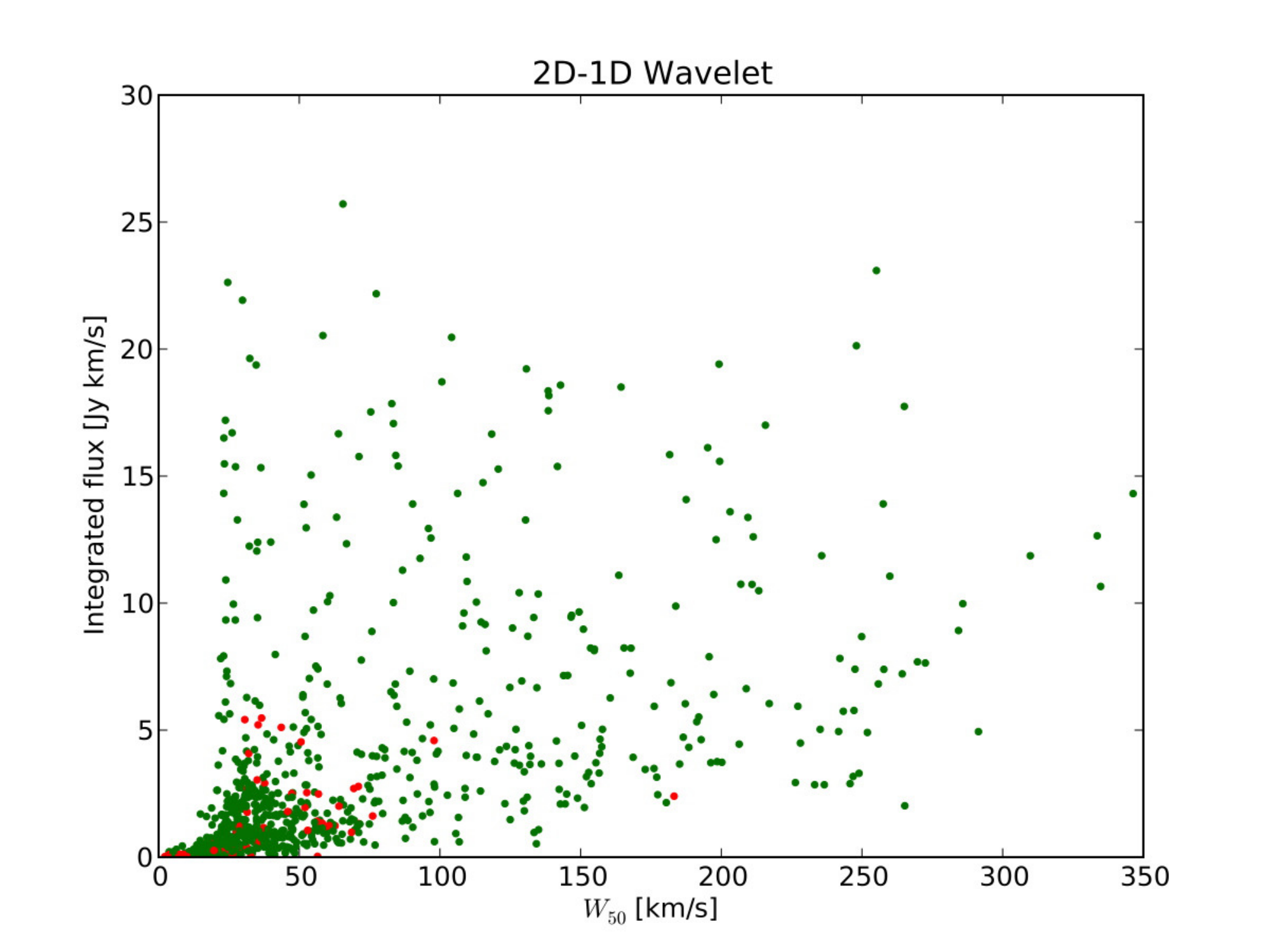}\\
Figure~\ref{model_scatter}: Continued
\end{center}
\end{figure*}

\subsection{Reliability of Source Finders}
In Fig.~\ref{point_comp_flux} and \ref{model_completeness} the
reliability of the different source finders is given by a single
number, which can be misleading. This number could be completely
dominated by a large number of false detections at a very low
threshold, while the source finder is very reliable for high flux
values.

To better understand where the bulk of the false detections are, all
detections are plotted in a scatter plot in Fig.~\ref{point_scatter}
for the point sources and Fig.~\ref{model_scatter} for the model
galaxies. The detections are again plotted as a function of velocity
width and integrated flux, where true detections are plotted in green
and false detections are plotted in red.

In the {\sc Duchamp} results for the point sources shown in the top
panels of Fig.~\ref{point_scatter} there are barely any false
detections in the first run. In the second run, all false detections
have low fluxes. A possibility that can improve the reliability is
to apply a cut in integrated flux after the parametrisation of
detections. In this example a cut a 0.05 Jy km/s would increase the
reliability to $\sim$100\% while the number of missed real detections
is still limited.

For the {\sc CNHI} finder the difference between true and false
detections is not so obvious. False detections are not clustered in a
clearly defined parameter space, but rather mixed with real
detections, making it more difficult to eliminate them after
post-processing. There is however a very large bulge of detections
with a low flux and broad line width.

The {\sc S+C finder} has a large number of false detections in the
first run, however a very large fraction can be eliminated by applying
a cut in integrated flux. In the second run the number of false
detections is much lower, however they are very well mixed with true
detections and difficult to eliminate. Although not shown in the plot,
particular for this source finder is that it also reports negative
fluxes. These are by default all considered false detections. Assuming
that the noise is symmetric, the reliability of positive detections
can be determined based on the properties of the negative
detections. This method is further explored and explained in
\citet{serra2011b}.

The 2D-1D wavelet source finder is very reliable for point sources as shown
before, with barely any false detections. The false detections are
however difficult to eliminate as they are concentrated toward high
fluxes and line widths. As mentioned this could be a consequence of
the used test cube which has a very high source density. Especially in
the case of strong sources the largest wavelet scales will merge sources,
decreasing the number of detected objects and hence the completeness.

\ \\

A very similar set of plots is given in Fig.~\ref{model_scatter},
where true and false detections of the model galaxies are plotted for
all the source finders apart from the {\sc Gamma-finder}. The
behaviour of the different algorithms is very similar to before, where
the false detections of {\sc Duchamp} tend to have a low integrated
flux, although it is difficult to completely isolate them. The CNHI
finder has a very large number of false detection with low flux and
broad line-width, many of which can be rejected to refine the
reliability. The performance of the {\sc S+C finder} is very good when
it comes to reliability as the number of false detections is
relatively low. Also the reliability of the 2D-1D wavelet finder is
very good, however the false detections are mixed with true
detections.

\ \\

\begin{figure*}[t]
\begin{center}
\includegraphics[width=0.48\textwidth, angle=0]{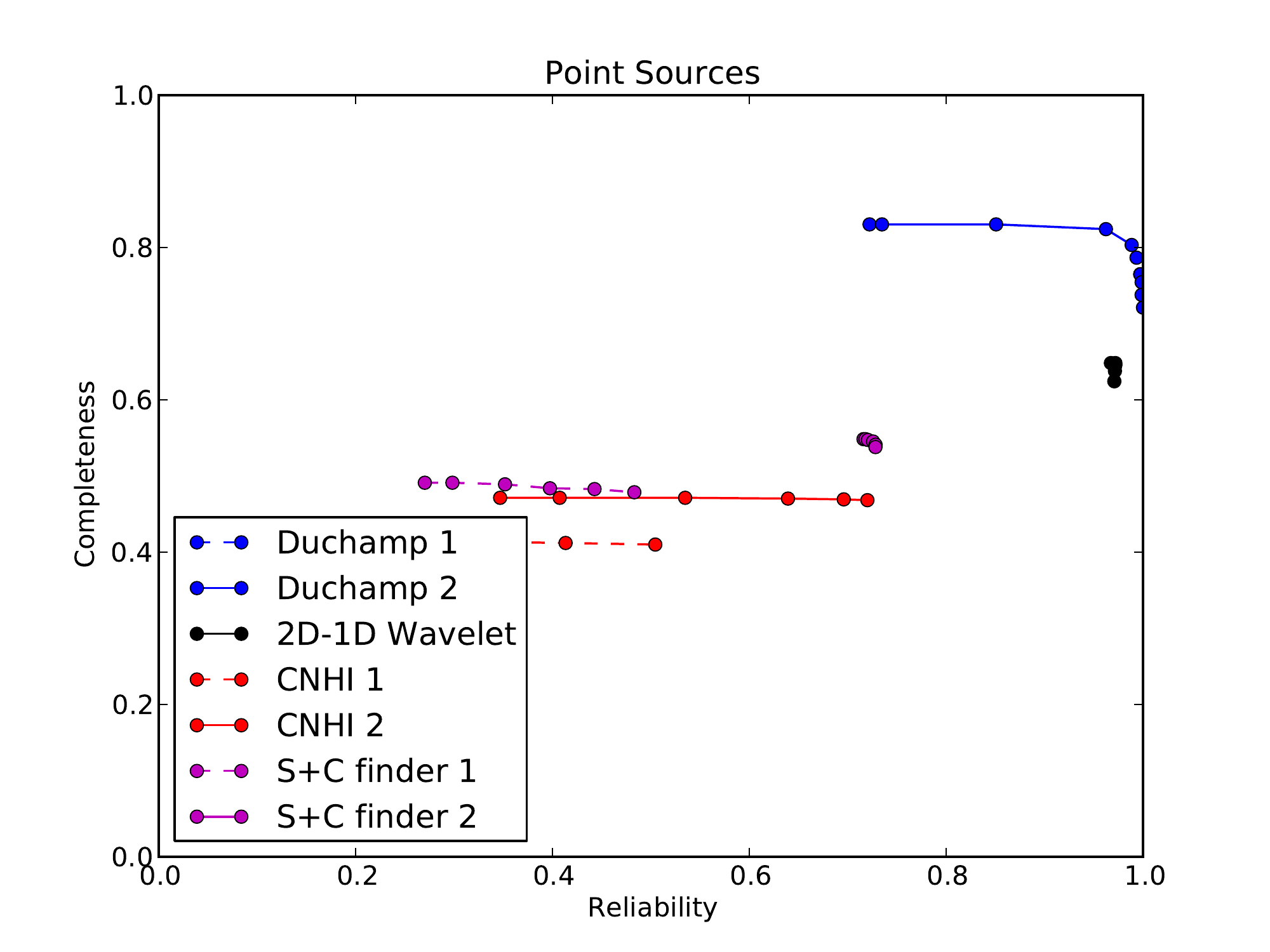}
\includegraphics[width=0.48\textwidth, angle=0]{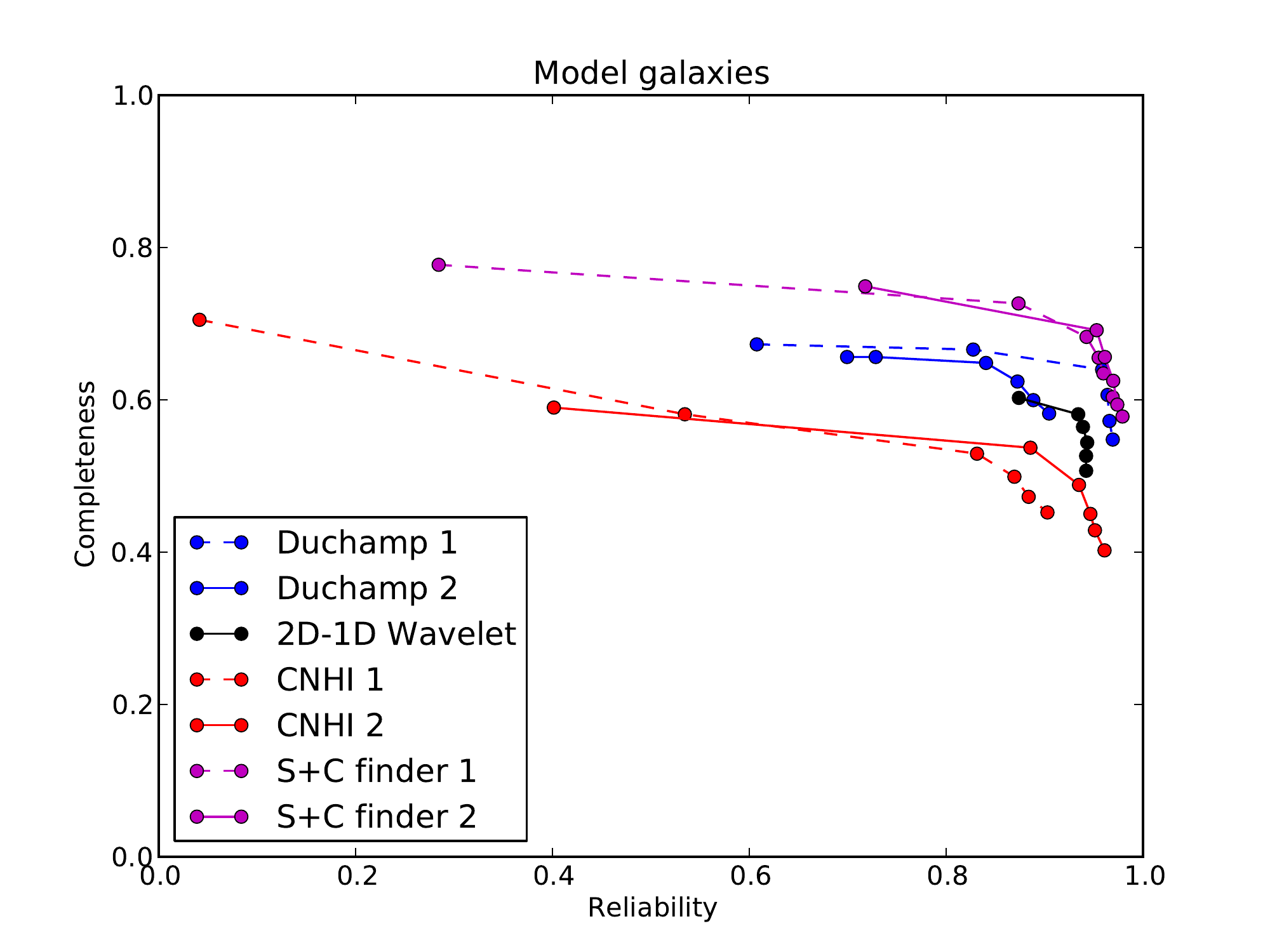}
\caption{Completeness as function of reliability for the point sources
  (left panel) and model galaxies (right panel) after applying a cut
  on the integrated flux ($F_{int}$). For the point sources cuts are
  applied at 0.0, 0.01, 0.02, 0.03, 0.04 and 0.05 Jy km s$^{-1}$,
  while for the model galaxies cuts are applied at 0.0, 0.1, 0.2, 0.3,
  0.4 and 0.5 Jy km s$^{-1}$. The points in the curve corresponds to a
  different cut, where the results move from high completeness and low
  reliability to low completeness and high reliability.}
\label{cut_results}
\end{center}
\end{figure*}

The reliability of the source finders can be dramatically
  improved upon through simple cuts in parameter space. To be able to
  do this, it is crucial to properly parameterise the detections which
  has not been done sufficiently at this stage. Nevertheless, to
  illustrate the concept, we applied a cut on the detections at
  different integrated flux levels. In Fig.~\ref{cut_results} the
  results are shown, where completeness is plotted as function of
  reliability for the different source finders after applying cuts at
  different flux levels. For the point sources cuts have been applied
  at $F_{int}$ = 0.0, 0.01, 0.02, 0.03, 0.04 and 0.05 Jy km s$^{-1}$
  while for the model galaxies at $F_{int}$ = 0.0, 0.1, 0.2, 0.3, 0.4
  and 0.5 Jy km s$^{-1}$. The results move from high completeness and
  low reliabiliy when not applying a cut to low completeness and high
  realiability when applying the most extreme cut.  Although the
  improvements in reliability vary amongst the different source
  finders, for each of them the raw reliability can be improved by
  tens of percent, while only losing a few percent in completeness. In
  the case of the second {\sc Duchamp} test on the point sources the
  reliability increases from 72\% to 96\%, while the completeness
  drops by only 0.6\% from 83\% to 82\% at the fourth data point. On
  the model galaxies the most impressive result is achieved with the
  {\sc S+C finder} where in the second run the reliability increases
  to above 95\%, while still maintaining a completeness of almost
  70\%.

\section{Conclusions}
In this paper we have compared the performance of five potential ASKAP
{\HI} source finders. The tested source finders are 1)~the {\sc
  Duchamp} source finder, 2)~the {\sc Gamma-finder}, 3)~the {\sc CNHI}
source finder, 4)~2D-1D Wavelet reconstruction source finder and
5)~the {\sc S+C finder}, a source finder based on sigma clipping of
smoothed versions of the original data cube. The source finders have
been applied to two data cubes with model sources, the first
containing point sources with a relatively narrow Gaussian line
profile and the second containing extended galaxies with inclinations
and rotation curves.

We have to stress that apart from the {\sc Gamma-finder} the
tested source finders are not final products but are still under
active development. In this paper we want to present the current
status of the different source finders, however there is significant
room for improvement as is also discussed in other papers in this
issue describing some of the tested source finders individually.

The testing of different source finding algorithms on different data
cubes has proven that it is very difficult to find a good source
finder which is reliable for many types of objects. Source finders
perform very differently depending on the type of object that is
detected.

An important feature of a source finder is its reliability, which has
not yet been fully explored. Although a number for the raw reliability
can be given, in many cases the false detections are clustered within
a certain range of flux and line width. We are confident that a large
fraction of the false detections can be rejected through simple cuts
in parameter space as has been demonstrated in the discussion,
however to be able to do this properly all detections have to be
parameterised accurately which has not been done yet.

For the current source finders and datasets, we find that for point
sources 50\% completeness can be achieved at an integrated
signal-to-noise ratio of $\sim$4-5 sigma, and 100\% completeness can
be achieved around an integrated signal-to-noise ratio of
$\sim$10. For the extended sources the completeness estimates are very
similar: for the best results 50\% completeness is achieved at an integrated
signal-to-noise ratio of $\sim$4-6 and 100\% completeness is achieved
at an integrated signal-to-noise ratio of $\sim$10.

It is interesting to see that the different source finders achieve a
different performance, depending on the type of object. Currently none
of the source finders excels at being able to achieve the best result
in the full parameter space when looking at integrated flux and line
width. Nevertheless we have pointed out the strong and weak points of
the different source finders, which provides input for future
development and testing.

For the tested parameters, currently {\sc Duchamp} gives the best
results on point sources, while the {\sc S+C finder} gives the best
result for extended objects when looking at the completeness. Due to
the different smoothing levels that have been applied in the {\sc S+C
  finder}, this algorithm is best capable of matching the true shape
of an object. As the {\sc S+C finder} concept is simple yet powerful,
we recommend that the other source finders improve their performance
by incorporating smoothing on multiple scales.

Currently all the tested source finders perform reasonably well,
however there is significant room for improvement to meet our goals.
All of the source finders have a certain area in parameter space where
they perform best and we will combine the algorithms of different
source finders to optimise the result.

\citet{duffy2011} give predictions of the number of objects that will
be detected with WALLABY and DINGO. They predict that at an angular
resolution of $30''$, 14\% of the WALLABY sources will be unresolved
and the bulk of the remainder will be marginally resolved, while for
DINGO 93.3\% of the sources will be unresolved.  This means that
  many of the unresolved sources in DINGO will have very different
  profiles to the ones tested in this paper. At an angular resolution
of $10''$ these numbers change dramatically, as for WALLABY none of
the sources will be point sources as all sources will be larger than
one beam, and for DINGO 7.4\% of the objects will be smaller than one
beam.

Although the two cubes that have been used for testing cover a large
area in parameter space, they do not sample the full signal-to-noise
ratio range properly. We have started efforts to test the source
finders on models covering a large range of parameters, keeping
integrated signal-to-noise values constant. These tests should give
accurate estimates of how many sources can be detected by WALLABY and
DINGO.

We have a fairly good understanding of the different source finders on
simulated objects as presented in this paper. The cubes that have been
tested are ideal cubes in the sense that the noise is Gaussian and
does not have any systematic artefacts caused by continuum sources,
solar ripples, phase errors, radio frequency interference, etc. These
contributions have not been taken into account but will have a very
significant effect on the performance of source finders, especially in
terms of reliability. The simulated model sources are perfectly
symmetric sources without any weird or unexpected shapes or extended
tails. To have a better understanding of the performance of the source
finders, the next step will be to test the source finders on a cube
containing data from real galaxies as they occur in the Universe.





\end{document}